\documentclass[10pt,a4paper]{article}

\usepackage{lineno}
\usepackage{subfig}
\usepackage{algorithm}
\usepackage{algpseudocode}
\usepackage[square, sort, numbers]{natbib}
\usepackage{cancel}
\usepackage{graphicx}
\usepackage{siunitx}

\usepackage[a4paper, left=2cm, right=2cm, top=2cm]{geometry}
\usepackage{authblk}
\usepackage{amsmath, amsthm, amssymb, amsfonts}

\usepackage{tikz}
\usetikzlibrary{external}
\usepackage{etoolbox}  
\providetoggle{tikzExternal}
\settoggle{tikzExternal}{false}  
\usepackage{hyperref}

\title{Incremental Model Order Reduction of Smoothed-Particle Hydrodynamic Simulations}

\author[1,3]{Eduardo Di Costanzo\thanks{eduardo.dicostanzo@andritz.com}}
\author[2]{Niklas K\"uhl}
\author[1]{Jean-Christophe Marongiu}
\author[3]{Thomas Rung}

\affil[1]{R\&D Department, ANDRITZ Hydro, Rue des Deux-Gares 6, 1800 Vevey, Switzerland}
\affil[2]{Hamburg Ship Model Basin, Bramfelder Strasse 164, D-22305 Hamburg, Germany}
\affil[3]{Hamburg University of Technology, Institute for Fluid Dynamics and Ship Theory, Am Schwarzenberg-Campus 4, D-21075 Hamburg, Germany}

\begin{document}

\maketitle

\begin{abstract} 
Engineering simulations are usually based on complex, grid-based, or mesh-free methods for solving partial differential equations. The results of these methods cover large fields of physical quantities at very many discrete spatial locations and temporal points. Efficient compression methods can be helpful for processing and reusing such large amounts of data.
A compression technique is attractive if it causes only a small additional effort and the loss of information with strong compression is low.
The paper presents the development of an incremental Singular Value Decomposition (SVD) strategy for compressing time-dependent particle simulation results. The approach is based on an algorithm that was previously  developed for grid-based, regular snapshot data matrices. It is further developed here to process highly irregular data matrices generated by particle simulation methods during simulation.
Various aspects important for information loss, computational effort and storage requirements are discussed, and corresponding solution techniques are investigated. These include the development of an adaptive rank truncation approach, the assessment of imputation strategies to close snapshot matrix gaps caused by temporarily inactive particles, a suggestion for sequencing the data history into temporal windows as well as bundling the SVD updates.
The simulation-accompanying method is embedded in a parallel, industrialized Smoothed-Particle Hydrodynamics  software and applied to several 2D and 3D test cases. The proposed approach reduces the memory requirement by about 90\% and increases the computational effort by about 10\%,  while preserving the required accuracy. For the final application of a water turbine, the temporal evolution of the force and torque values for the compressed and simulated data is in excellent agreement.

\end{abstract}

\begin{flushleft}
\end{flushleft}

\section{Introduction}
Current and future computational engineering must support more than just the mere analysis of a given design. Instead, the procedures should also optimize the design, enhance its robustness against changing operating conditions, and facilitate the reuse of computed (field) data as a database for machine learning of various design-relevant input/output relationships. 

The analysis of an engineering design is usually based on complex, grid-based, or mesh-free first-principle simulation methods, the results of which cover large fields of physical quantities in many million (discrete) spatial locations and many thousand temporal points. Processing, i.e., reusing such large amounts of data, which typically result from the numerical solution of Partial Differential Equations (PDEs), is a challenge. In this case, data compression of the results, using methods such as Proper Orthogonal Decomposition (POD), can effectively save storage space for large-scale but low-rank data sets. The POD technique \cite{sirovich1987turbulence} is particularly attractive when it incurs only a small computational overhead and the loss of information for strong compression is minor. 

POD is generally considered a model order reduction approach that approximates a given dataset optimally on a low-dimensional basis \cite{benner2015survey}. Aiming at data compression, the strategy usually involves a Singular Value Decomposition (SVD) of a snapshot data matrix to derive a reduced basis spanned by the POD modes. Due to the often very good approximation by a small number of consecutive initial POD modes, only a few are usually considered at the expense of a small loss of information. Linearity assumptions limit the use of POD strategies. Nevertheless, the solution manifold is often sufficiently well approximated by a low-rank subspace, which recommends the use of the POD  for model order reduction of nonlinear systems \cite{NAKAMURA2024117340, lassila2014model, kunisch2002galerkin, willcox2002balanced, XIAO2024117099,Abbaszadeh2020, Ballarin2016}. This is especially true for unsteady systems where the data is generated progressively in time, and the snapshot matrix usually consists of many more discrete spatial points (defining the length of the matrix column)  than temporal points (defining the number of matrix columns). SVD procedures are traditionally offline approaches and thus require storing the entire snapshot matrix before performing the SVD, which still demands prohibitive computer memory. To solve this issue, efficient algorithmic alternatives dedicated to the incremental computation of the SVD (iSVD) that accompanies the data growth over time have been developed \cite{FAREED20181942, FAREED2019223, Brand2002, Brand2006, KUHL2024109022, KUHL2025106579}.

Focusing on unsteady gradient- and thus adjoint-based optimization methods raises related issues due to the oppositely directed temporal information transport of the primal and adjoint PDEs involved. To obtain the gradient of the objective function with respect to the optimization parameters, one usually has to solve the forward time-dependent primal problem PDEs and store all forward solutions for later use when integrating the backward-directed adjoint PDEs. Naively storing and subsequently reading the forward solutions to and from the disk at runtime is prohibitive in large-scale industrial applications. It calls for an incrementally processed model order reduction during the integration of the primal flow.

This paper investigates iSVD as a data compression technique for particle simulation methods. The goal is to reduce the storage requirements for future time-dependent, PDE-constrained optimization as much as possible while minimizing the required computational overhead. Though incrementally constructed ROMs have recently been used to handle memory limitations for optimization studies \cite{vezyris2019incremental, margetis2021lossy, margetis2022reducing, KUHL2024109022, li2021towards, li2024incremental}, the iSVD has never been utilized for dynamic meshes and meshless methods such as Smoothed Particle Hydrodynamics (SPH). The challenges and innovations are considerable and relate to applying an iSVD to particle data obtained from massively parallel SPH simulations. Three aspects are particularly important. In contrast to grid-based methods, (a) the spatial positions of the data vary over time; (b) the snapshot matrix is irregularly populated because the number of particles within the domain typically varies over time; (c) all particles are only active during a partial period when simulating open boundary domains.

The paper is organized as follows: Sections 2 and 3 are devoted to the Full-Order Model (FOM) and the numerical method used to solve the governing equations. Section 4 briefly outlines the model reduction strategy and the iSVD. Section 5 addresses the specifics and challenges of the iSVD in conjunction with an SPH-based snapshot matrix. Verification and validation examples are reported in Sec. 6, emphasizing accuracy and eﬃciency aspects for 2D and 3D benchmark flows. They refer to (a) a simple 2D sloshing in a closed box, (b) a 2D impinging jet featuring open boundaries, and (c) an unsteady 3D Pelton turbine runner. Note that although only hydrodynamic applications are shown, the strategy presented here can also be used for other particle simulation applications. Conclusions are drawn in Sec. 7.  

In the remainder of the paper, reference properties are usually marked by an index ''0''. Field quantities are defined regarding Cartesian coordinates denoted by Greek superscripts, and Latin subscripts distinguish the particles and their spatial position. Moreover, we employ lower- and upper-case bold letters to denote vectors and tensors, respectively.

\section{Governing Equations}
The section briefly recalls the governing equations for a weakly compressible, inviscid fluid. The chosen flow model is best suited for the application of particle methods to the pressure-driven flow around a Pelton turbine at high Reynolds numbers, whereby friction losses on the bucket's surfaces, which are responsible for a reduction in the machine's efficiency, are deliberately not taken into account. 

A standard compressible formulation for conserving mass, volume, and momentum is employed for an inviscid single-phase flow. With regard to a numerical particle approach,  the density $\rho_i$, velocity $\mathbf{v_i}$ and volume $\omega_i$ of a fluid particle currently located in the position $\mathbf{x}_i$ and transported with the velocity $\mathbf{v}^0_i$ follow from the system of advection, space conservation, mass and momentum equations, viz.
\begin{equation}
\frac{d\mathbf{x}_i}{dt} = \mathbf{v}_i^0 \, , 
\label{POSE} 
\end{equation}
\begin{equation}
\frac{d{\omega}_i}{dt}=
{\omega}_i \, \nabla \cdot \mathbf{v}_i^0 
  \, ,
\label{VOLE} 
\end{equation}
\begin{equation}
\frac{d\rho_i}{dt} = -\rho_i \nabla \cdot \mathbf{v}_i \quad \to \quad
\frac{d (\omega_i \rho_i)}{dt} =
-\rho_i \omega_i  \nabla \cdot (\mathbf{v}_i - \mathbf{v}^0_i) 
\, ,
\label{CE} 
\end{equation}
\begin{equation}
\frac{d\mathbf{v}_i}{dt}=  
 -  (\mathbf{v}_i - \mathbf{v}^0_i) \cdot \left[\nabla \mathbf{v}_i \right]
- \left(\frac{\nabla p}{\rho} \right)_i  \
 + \mathbf{g} 
 \quad \to \quad
\frac{d (\rho_i \omega_i \mathbf{v_i})}{dt} = - \rho_i \omega_i \nabla \cdot \left[  (\mathbf{v}_i - \mathbf{v}^0_i) \mathbf{v}_i \right] 
 - \omega_i \nabla p_i + \rho_i \omega_i \mathbf{g}
 \, .
\label{ME} 
\end{equation}
Here $p$ is the pressure, $\mathbf{g}$ denotes a gravitational body force per unit mass, $\nabla$ is the (left) spatial gradient, and $d \phi /dt$ indicates the temporal changes of the focal particle properties. Equations (\ref{POSE})-(\ref{ME}) refer to an \textit{Arbitrary Langragian-Eulerian} (ALE) form. For $\mathbf{v}^0 = 0$, the particle does not move, and one obtains the classical Eulerian description with $d/dt \to \partial/\partial t$. Moving the particle with the flow, i.e., $\mathbf{v}^0 = \mathbf{v}$, a Lagrangian description is recovered with $d/dt \to D/D t$.

\subsection{Constitutive Equation}
Using a weakly compressible fluid model, the pressure is derived from a heuristic polytropic pressure-density relation (cf. \cite{monaghan1994simulating}), aka. Tait's relation,
\begin{equation}
p = p_0 \left[ \left(\frac{\rho}{\rho_0} \right)^{\gamma} - 1 \right] \, \quad {\rm with} \quad \gamma = 7 , 
\label{eq:Tait}
\end{equation}
where $\rho_0$ is a reference density, and $p_0$ is linked to an artificial reference speed of sound using  $p_0  = \rho_0  c_0^2/\gamma$. To keep the density variations below 1$\%$, the speed of sound value is usually obtained from an estimated maximum flow speed using \emph{ $c_0 \ge 10 |\mathbf{v}|_{\mathrm{max}}$} which ensures a Mach number $M \le 0.1$. The fluid considered in this work has a reference density of $\rho_0 = 1000 \, \text{kg/m}^3$.

\section{Computational Model}
The Lagrangian SPH method displays advantages over classical Eulerian flow simulation methods for problems with substantial temporal changes of the wetted domain. An example is violent free surface flows, such as those encountered in hydraulic machinery simulations. In these cases,  grid-based methods often require dynamic grid refinement and coarsening, which is not trivial in 3D parallel frameworks. On the contrary, SPH approaches are mesh-free and discretize the material, i.e., it does not require the use or adaption of the computational grid to a free surface. Nowadays, SPH strategies are successfully applied to a broad range of engineering investigations, including --for example-- free-surface flows \cite{Cola09, Zheng2020, Ehigiamusoe2018}, multi-continua applications \cite{Ulr13}, fluid-structure interaction \cite{ZhangColagrossi2017, Cummins2012, SALIS2024103885}, geotechnical \cite{Bui24, Manenti2024}, energy harvesting from water waves and wind offshore \cite{YANG2024104049, CAPASSO2025124508}, and turbine applications \cite{Marongiu2010, Rentschler2016}. Recent advancements include addressing volume conservation in sloshing flows and free-surface conditions \cite{PILLOTON2024116640}, improving consistency and transport-velocity formulations via reverse kernel gradient correction \cite{ZHANG2025117484}, implicit shifting methods \cite{RASTELLI2023116159}, Quasi-Lagrangian formulations, and energy balance techniques with Riemann solvers have also been introduced \cite{MICHEL2023116015}. Results of the FOM employed in this work were obtained from the in-house Andritz solver, ASPHODEL. This section provides a brief overview of the implemented methods. A more detailed introduction can be found in \cite{Marongiu2007, Leduc2010, andritzoverview, neuhauser}.

\subsection{Numerical Method} 
Equations (\ref{POSE}-\ref{ME}) are discretized in space and time and subsequently explicitly integrated in time. For this purpose, the flow field is first discretized spatially by $N$ particles of equal mass $m$ and initially equal density $\rho_i (t_0) = \rho_0$ and volume $\omega_i (t_0) = m/\rho_i (t_0)$. 

\subsubsection{Smoothed-Particle Hydrodynamics Approximation}

Smooth Particle Hydrodynamics rely on two fundamental approximation steps: kernel-based mollification and discrete particle approximation of integrals. The starting point is the identity
\begin{equation}
f(\mathbf{x}_i) = \int f(\mathbf{x}_j) \delta_D(\mathbf{x}_i-\mathbf{x}_j) dx_j , 
\end{equation}
with $\delta_D(\mathbf{x}_i-\mathbf{x}_j)$ being the Dirac function, that vanishes everywhere but for the case $\mathbf{x}_i=\mathbf{x}_j$. Any physical quantity \( f(\mathbf{x}) \), such as the density or the velocity, is subsequently approximated by the convolution of \( f(\mathbf{x}) \) with a smooth, kernel function \(  W(\|\mathbf{x}_i - \mathbf{x}_j\|, h) = W_{ij} \) that mollifies the Dirac function and is numerically integrated with the help of neighboring particles $j$, i.e.
\begin{equation}
 \int f(\mathbf{x}_j) W (\|\mathbf{x}_i - \mathbf{x}_j\|, h) 
 dx_j \approx \sum_{j \in D_i} \omega_j f_j W_{ij} \, .
\end{equation}
The isotropic bell-shaped kernel function $W$ is usually assumed to be positive and symmetric to mimic the properties of the Dirac function. The most crucial feature of $W$ is its compact support \( D \), which is characterized by the (homogeneous) length \( h \). 
Similarly, the SPH approximation of the gradient \( \nabla f(\mathbf{x}_i) \) is derived using integration by parts and the symmetry relation $\nabla_j W_{ij} = - \nabla_i W_{ij}$, which yields
\begin{equation}
\nabla f(\mathbf{x}_i) \approx \sum_{j \in D_i} \omega_j f(\mathbf{x}_j) \nabla_i W_{ij} + \sum_{k \in \partial D_i} \partial\omega_k  f(\mathbf{x}_j) \mathbf{n}_k W_{ij}.
 \label{eq:gradsph}
\end{equation}
The normal vector and the area of the discretized boundary are denoted by $\mathbf{n}_k$, and $\partial \omega_k $. Due to the compact support of $W$, the second term in (\ref{eq:gradsph}) is only non-zero if the kernel support intersects the domain's boundary. Since the kernel function is a smooth analytical function, its gradient can be computed analytically, greatly simplifying the procedure.

The paper adopts an ALE formulation of the SPH method as initially suggested in \cite{Vila1999} and described in \cite{andritzoverview}.
The corresponding spatially discretized version of Eqns. (\ref{POSE}-\ref{ME}) reads
\begin{equation}\label{eq:motion}
  \frac{d \mathbf{x}_i}{d t} = \mathbf{v}^0_i 
\end{equation}
\begin{equation}\label{eq:volume}
  \frac{d\omega_i}{dt} = \omega_i \sum_{j \in \mathcal{D}_i} \omega_j \left[\mathbf{v}^0_j - \mathbf{v}^0_i\right] \cdot \boldsymbol{\nabla}_i W_{ij}    
\end{equation}
\begin{equation}\label{eq:mass}
  \frac{d(\omega_i \rho_i)}{dt} = - \omega_i \sum_{j \in \mathcal{D}_i} 2 \omega_j \rho_{ij}^E \left[\mathbf{v}_{ij}^E - \mathbf{v}^0_{ij}\right] \cdot \boldsymbol{\nabla}_i W_{ij} 
\end{equation}
\begin{equation}\label{eq:momentum}
  \frac{d(\omega_i \rho_i \mathbf{v}_i)}{dt} = \omega_i \rho_i \mathbf{g}  - \omega_i \sum_{j \in \mathcal{D}_i} 2 \omega_j \left\{\rho_{ij}^E \mathbf{v}_{ij}^E \otimes [\mathbf{v}_{ij}^E - \mathbf{v}^0_{ij}] + p_{ij}^E \mathbf{I}\right\} \cdot \boldsymbol{\nabla}_i W_{ij} \, .
\end{equation}
For simplicity, the system of equations (\ref{eq:motion}-\ref{eq:momentum})  is presented without the inclusion of boundary terms in Eq. \ref{eq:gradsph}. Mind that the space conservation law (\ref{eq:volume}) has been augmented by the subtraction of a term proportional to $\sum \boldsymbol{\nabla}_i W_{ij}$ on the right-hand side (r.h.s.), which vanishes in continuous space, to improve the consistency of the discrete model. Similarly, the continuity equation (\ref{eq:mass}) and the momentum equation (\ref{eq:momentum}) have been augmented by the addition of a term proportional to $\sum \boldsymbol{\nabla}_i W_{ij}$ on the r.h.s., to support the conservation on the particle level. The superscript $E$ indicates interaction values between pairs of neighboring particles obtained by the solution of a Riemann problem, viz. 
\begin{equation}
\mathbf{v}_{ij}^E = \frac{\mathbf{v}_i + \mathbf{v}_j}{2} + \frac{p_j - p_i}{2\bar{\rho}_{ij}\bar{c}_{ij}} , 
\end{equation}
\begin{equation}
p_{ij}^E = \frac{p_i + p_j}{2} + \frac{\bar{\rho}_{ij}\bar{c}_{ij}(\mathbf{v}_j - \mathbf{v}_i)}{2} , 
\end{equation}
where the double subscripts in the expressions $\bar{\rho}_{ij}$ and $\bar{c}_{ij}$ refer to averaged values of the density and speed of sound of the particles $i$ and $j$. The interface velocity between these particles reads $\mathbf{v}_{ij}^0 = 0.5(\mathbf{v}_{i}^0+\mathbf{v}_{j}^0)$. The present study employs a constant speed of sound and a Wendland $C^4$ kernel \cite{Wendland1995}, viz.  
\begin{equation}\label{eq:WendlandC4}
    W(\mathbf{x},h) = \frac{\sigma}{h^d} \theta\left(\frac{\|\mathbf{x}\|}{2h}\right),
\end{equation}
\begin{equation}
    \theta(q) = (1-q)_{+}^{5}(8q^2 + 5q + 1),
\end{equation}
\begin{equation}
    x_{+}^{n} = \begin{cases}
    x^n & \text{if } x > 0, \\
    0 & \text{if } x \leq 0.
\end{cases}
\end{equation}

\noindent Here, $W(\mathbf{x}, h)$ represents the smoothing kernel function, $\mathbf{x}$ is the distance vector between two particles, and $h$ is the smoothing length. The parameter $d$ represents the number of spatial dimensions in the simulation, $\|\mathbf{x}\|$ denotes the Euclidean norm of the distance vector $\mathbf{x}$. The function $\theta(q)$ defines the shape of the kernel as a function of the normalized distance $q = \|\mathbf{x}\|/(2h)$. The subscript $()_+$ represents the positive part, meaning that the function goes to zero if the input is negative. The normalization factor $\sigma$ is given by $\sigma = \frac{3}{4\pi}$ in 2D and $\sigma = \frac{165}{256\pi}$ in 3D. 

\subsubsection{Temporal Discretization}
The system (\ref{eq:motion}-\ref{eq:momentum}) is discretized by adaptive time steps and explicitly integrated in time. In the scope of the present paper, we employ a third-order Runge-Kutta scheme and adjust the time step to comply with
\begin{equation}\label{eq:CFL}
\Delta t = K_{\text{CFL}} \min_{i \in D} \frac{h_i}{\|\mathbf{v}_i\| + c_i},
\end{equation}
where \(K_{\text{CFL}}\) denotes the maximum CFL number assigned to $K_{\text{CFL}}=0.5$ in the current 2D studies and $K_{\text{CFL}}=0.3$ in the 3D application.

\subsubsection{Initial and Boundary Conditions}
The initially realized particle distance reads $\Delta x$ and is usually related to the smoothing length of a kernel function $h$. The present studies are based upon $h/\Delta x =1.2$,  which works well in practice, cf. \cite{neuhauser}. The boundary fluxes are computed with partial Riemann solvers for wall boundaries, while open boundaries are managed according to a non-reflecting characteristic boundary condition derived from \cite{Selle}.

\section{Model Reduction Strategy}
\subsection{Singular Value Decomposition}
The \textit{Singular Value Decomposition} (SVD) is a unique and guaranteed to exist matrix factorization, which can provide the best low-rank approximation of a matrix in the least-squares sense \cite{eckart1936approximation}. A matrix $Y \in \mathbb{C}^{m \times n}$ is decomposed into two orthogonal matrices $U \in \mathbb{C}^{m \times m}$ and $V \in \mathbb{C}^{n \times n}$, along with a diagonal matrix $S = diag(s_i) \in \mathbb{R}^{m \times n}$, such that $Y=USV^*$, where $*$ is used to indicate the conjugate transpose. The columns of $U$ are called left-singular vectors or modes and describe patterns in the original data. Additionally, these columns are arranged based on their ability to capture the variance in the columns of $Y$. The matrix $S$ contains the singular values of $Y$, which are non-negative and arranged in decreasing order of magnitude. The magnitude of these singular values provides a measure of the relative importance of each mode. The columns of $V$ are called right-singular vectors. These vectors represent how the modes combine, scaled by the corresponding singular values, to reconstruct the columns of $Y$. The singular values of $Y$ are also the square roots of the eigenvalues of the correlation matrix $Y^* Y$ (or $Y Y^*$), while the singular vectors $U$ (or $V$) are its eigenvectors. This explains why the columns of $U$ ($V$) are ordered by how much correlation they capture in the columns (rows) of $Y$ \cite{Taira2017, Brunton_Kutz_2019}.

There can be at most $rank(Y) = r \leq min(m,n)$ non-zero singular values. In the case where \(m > n\) and \(Y\) is full rank (i.e., \(\text{rank}(Y) = n\)), the decomposition can be shown as:

\begin{center}
\iftoggle{tikzExternal}{
            \tikzsetnextfilename{svd-sketch}
            \begin{tikzpicture}
              \draw[thick] (0, 0) rectangle (2, 3);
              \node at (1, 1.5) {\(Y\)};
              \node at (2.5, 1.5) {\(=\)};
              \draw[<->] (-0.2, 3) -- (-0.2, 0);
              \node at (-.5, 1.5) {\(m\)};
              \draw[<->] (0, -.7) -- (2, -.7);
              \node at (1, -.5) {\(n\)};
            
              \draw[thick] (3.5, 0) rectangle (6.5, 3);
              \draw[dashed] (5.5, 0) -- (5.5, 3);
              \node at (5, 1.5) {\(U\)};
              \draw[<->] (3.3, 3) -- (3.3, 0);
              \node at (3, 1.5) {\(m\)};
              \draw[<->] (3.5, -.7) -- (6.5, -.7);
              \node at (5, -.5) {\(m\)};
              \draw[<->] (5.5, -.5) -- (6.5, -.5);
              \node at (6, -.3) {\(m-n\)};
            
              \draw[thick] (7.5, 0) rectangle (9.5, 3);
              \draw[dashed] (7.5, 1) -- (9.5, 1);
              \node at (8.5, 2) {\(S\)};
              \node at (8.5, 0.5) {\(0\)};
              \draw[<->] (7.3, 3) -- (7.3, 0);
              \node at (7, 1.5) {\(m\)};
              \draw[<->] (7.5, -.7) -- (9.5, -.7);
              \node at (8.5, -.5) {\(n\)};
            
              \draw[thick] (10.5, 1) rectangle (12.5, 3);
              \node at (11.5, 2) {\(V^*\)};
              \draw[<->] (10.3, 1) -- (10.3, 3);
              \node at (10.1, 2) {\(n\)};
              \draw[<->] (10.5, .8) -- (12.5, .8);
              \node at (11.5, .6) {\(n\)};
            \end{tikzpicture}
        }{
            \includegraphics{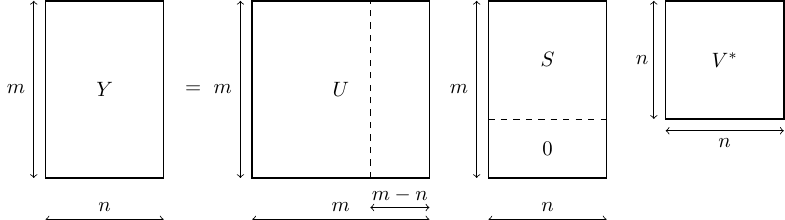}
        }
\end{center}
By excluding the rows of $S$ that contain only zeroes and the corresponding columns of $U$, one obtains the reduced SVD (also called thin SVD or economy-size SVD). For the rest of this work, it is assumed that data are real numbers arranged in matrices where $m>n$. Only the necessary components of the SVD are kept, hence: $Y=USV^T$, $Y \in \mathbb{R}^{m \times n}$, $U \in \mathbb{R}^{m \times n}$, $S \in \mathbb{R}^{n \times n}$, and $V \in \mathbb{R}^{n \times n}$. Thanks to the optimal truncation properties of the SVD, further reduction is possible by neglecting the smallest non-zero singular values according to a desired criterion and only keeping the $q$ most significant singular values. In this case, the SVD is known as truncated SVD: $U \in \mathbb{R}^{m \times q}$, $S \in \mathbb{R}^{q \times q}$, and $V \in \mathbb{R}^{n \times q}$.
\begin{center}
    \iftoggle{tikzExternal}{
        \tikzsetnextfilename{tsvd-sketch}
        \begin{tikzpicture}
          \draw[thick] (0, 0) rectangle (2, 3);
          \node at (1, 1.5) {\(Y\)};
          \node at (2.5, 1.5) {\(=\)};
          \draw[<->] (-0.2, 3) -- (-0.2, 0);
          \node at (-.5, 1.5) {\(m\)};
          \draw[<->] (0, -.7) -- (2, -.7);
          \node at (1, -.5) {\(n\)};
        
          \draw[thick] (3.5, 0) rectangle (5, 3);
          \node at (4.25, 1.5) {\(U\)};
          \draw[<->] (3.3, 3) -- (3.3, 0);
          \node at (3, 1.5) {\(m\)};
          \draw[<->] (3.5, -.7) -- (5, -.7);
          \node at (4.25, -.5) {\(q\)};
        
          \draw[thick] (6, 1.5) rectangle (7.5, 3);
          \node at (6.75, 2.25) {\(S\)};
          \draw[<->] (5.8, 3) -- (5.8, 1.5);
          \node at (5.5, 2.25) {\(q\)};
          \draw[<->] (6, 1.3) -- (7.5, 1.3);
          \node at (6.75, 1.1) {\(q\)};
        
          \draw[thick] (8.5, 1.5) rectangle (10.5, 3);
          \node at (9.5, 2.25) {\(V^T\)};
          \draw[<->] (8.3, 1.5) -- (8.3, 3);
          \node at (8.1, 2.25) {\(q\)};
          \draw[<->] (8.5, 1.3) -- (10.5, 1.3);
          \node at (9.5, 1.1) {\(n\)};
        \end{tikzpicture}
        }{
            \includegraphics{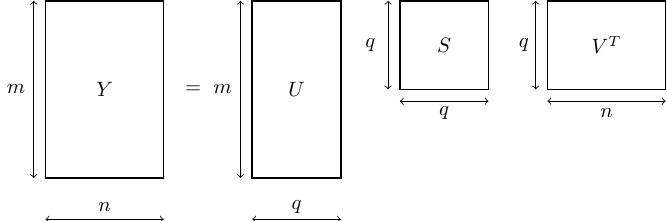} 
        }
\end{center}

The truncated SVD is the starting point of many algorithms utilized for dimensionality reduction, data compression, noise reduction, solving systems of algebraic equations, inverting rank-deficient matrices, and more \cite{Taira2017, Brunton_Kutz_2019, Kaur2020a, Kaur2020b, Tenekedjiev2021}. Nevertheless, since the degree of achievable compression depends on the specific data, no a-priori criteria are usually available to decide how many modes are necessary to keep. Commonly used methods for truncation include setting a hard threshold for singular values, retaining a specified percentage of the total energy, or analyzing the singular values to identify an "elbow" or "knee" in their distribution \cite{Gubisch, GraessleFEM, Brunton_Kutz_2019}. It was suggested in \cite{KUHL2024109022} that ignoring the first $o$ singular values might be beneficial while computing the retained energy to avoid machine accuracy issues.

\subsection{Incremental Singular Value Decomposition}
\label{sec:iSVD}
Typical SVD reduction methods require the complete flow field upfront, leading to significant memory consumption. Incremental SVD approaches have been developed, addressing this challenge and allowing the SVD to be updated incrementally with rank-b modifications. In this work, only the the additive modification to add columns to Y is considered. A brief summary of this specific incremental SVD approach is provided. For further details, the reader is referred to \cite{Brand2002,Brand2006, KUHL2024109022}. 

Let us assume that $Y=USV^T$ is known, and that new data becomes available. The original matrix $Y$ and the updated matrix $\Tilde{Y}$ are given by Eqn. \ref{eq:initial Y} and Eqn. \ref{eq:updated Y}, respectively. The objective is to compute the SVD of $\Tilde{Y}$, without having to reconstruct $Y$ first.

\begin{equation}
Y = 
\left[
\begin{matrix}
\mathbf{y_{1,1}} & \cdots & \mathbf{y_{1,n}} \\
\vdots & \ddots & \vdots \\
\mathbf{y_{m,1}} & \cdots & \mathbf{y_{m,n}} \\
\end{matrix}
\right]
\label{eq:initial Y}
\end{equation}

\begin{equation}
\Tilde{Y} = 
\left[
\begin{matrix}
\mathbf{y_{1,1}} & \cdots & \mathbf{y_{1,n}} & \mathbf{y_{1,n+1}} & \cdots & \mathbf{y_{1,n+b}} \\
\vdots & \ddots & \vdots & \vdots & \ddots & \vdots \\
\mathbf{y_{m,1}} & \cdots & \mathbf{y_{m,n}} & \mathbf{y_{m,n+1}} & \cdots & \mathbf{y_{m,n+b}} \\
\end{matrix}
\right]
\label{eq:updated Y}
\end{equation}

Let the new data be represented by $B$, the so called bunch-matrix, of width $b$ Eqn. (\ref{eq:B_matrix}).

\begin{equation}
B = 
\left[
\begin{matrix}
\mathbf{y_{1,n+1}} & \cdots & \mathbf{y_{1,n+b}} \\
\vdots & \ddots & \vdots \\
\mathbf{y_{m,n+1}} & \cdots & \mathbf{y_{m,n+b}} \\
\end{matrix}
\right]
\end{equation}

The SVD of $\tilde{Y}$ is obtained from the following steps, as derived in \cite{KUHL2024109022} from \cite{Brand2002,Brand2006}:
\begin{enumerate}
    \item The QR decomposition of the matrix $(I - UU^T)B$ is computed, where $I$ is the identity matrix. 
    \begin{equation}
    Q_B R_B = (I - UU^T)B
    \end{equation}

    \item The matrix $K$ is built:
    \begin{equation}
    K = \begin{bmatrix}
    S & U^T B \\
    0 & R_B
    \end{bmatrix}
    \end{equation}

    \item The SVD of $K$ is computed:
    \begin{equation}
    K = U' S' V'^T
    \end{equation}

    \item Finally, the existing SVD is updated:
        \begin{equation}
        \tilde{V}_q = \begin{bmatrix}
        V & 0 \\
        0 & I
        \end{bmatrix} V'_q
        \label{eq:tilde_V_q}
        \end{equation}

        \begin{equation}
        \tilde{S}_q = S'_q
        \label{eq:tilde_S_q}
        \end{equation}

        \begin{equation}
        \tilde{U}_q = [U \, Q_B] U'_q
        \label{eq:tilde_U_q}
        \end{equation}
\end{enumerate}

The index $q$ in equations \eqref{eq:tilde_V_q}, \eqref{eq:tilde_S_q}, and \eqref{eq:tilde_U_q} refers to the truncation rank of the SVD.

\section{Model Reduction of SPH-Data}
\label{sec:iSVD-SPH}

In the SPH framework, unique complexities arise from potential particle (a) deletion, (b) domain entering, or (c) exiting. In the latter case, the position of the calculation points remains unknown, resulting in additional fields --$\{x,y,z\}$-- to reduce and store via SVD. Similarly, particles may lack data prior to their injection or after their deletion, leading to bunch matrices of irregular row and column size.

Let $Y \in \mathbb{R}^{m_1 \times T}$ be the matrix containing the values $y_{i,j}$ of the field $y$ for particle $i \in \{1, \ldots, m_1\}$ and time $j \in \{0, \ldots, T\}$. The bunch matrix is $B \in \mathbb{R}^{m \times b}$ with $m = m_1 + m_2$, where $m_2$ is the number of particles injected since $T$ and $b$ refers to the bunch size. $B$ might present missing (non-existing) values, here indicated by \textbf{null}, due to the particle not being in the domain at the respective time, as shown in the following
\begin{equation}
B = 
\begin{bmatrix}
\mathbf{y_{1,T+1}} & \mathbf{y_{1,T+2}} & \cdots & \mathbf{null} & \mathbf{null} \\
\vdots & \vdots & \ddots & \vdots & \vdots \\
\mathbf{y_{m_1,T+1}} & \mathbf{y_{m_1,T+2}} & \cdots & \mathbf{y_{m_1,T+b-1}} & \mathbf{y_{m_1,T+b}} \\
\vdots & \vdots & \ddots & \vdots & \vdots \\
\mathbf{null} & \mathbf{null} & \cdots & \mathbf{y_{m,T+b-1}} & \mathbf{y_{m,T+b}} \\
\end{bmatrix}
\label{eq:B_matrix}
\end{equation}
here, the entries \( \mathbf{y}_{p,t} \) refer to time instants \( t \) when a particle \( p \) was present in the domain and are referred to herein as \textit{available data}. We address missing data within \( \mathbf{B} \) using one of the following approaches: if at least one non-null value exists for particle \( p \) within \( \mathbf{B} \), the initial guess for its value at time \( t \) is computed as the row mean of all available non-null entries for particle \( p \) in \( \mathbf{B} \); if no non-null values are available for particle \( p \) in the current \( \mathbf{B} \), then for each column \( t \), the initial guess is set to the mean of the non-null values in that column (referred to here as \textit{Mean Imputation}), imputation using the mean of subsets of the matrix defined by blocks of a given number of columns (referred to as \textit{Block-Mean Imputation}), Gappy Proper Orthogonal Decomposition (GPOD), or a combination of Mean and Block-Mean Imputation. The GPOD is typically used to provide approximations for missing or corrupted data, cf. exemplary applications regarding noisy PIV measurement \cite{Saini2016}, inverse design \cite{buithanhGappy}, unsteady flow estimation \cite{WILLCOX2006208}, design optimization \cite{duan2012gappy}, or flows with deforming meshes \cite{FRENO2014145}. However, since the missing values in SPH data refer to time instants when a particle is not in the considered domain, the approximated value has no physical meaning and only completes the matrix $B$ to compute the SVD. The only desired property of the imputed values is that they should not raise the rank of the matrix so that a sound reduction can be achieved. The basic idea behind the GPOD is to use the SVD to improve the guessed data for the missing values, initially applied by \cite{Everson95} and later modified by \cite{Gunes2006, Raben2012}.
According to Gunes algorithm for GPOD \cite{Gunes2006, Raben2012, Saini2016}:
\begin{enumerate}
    \item The temporal mean of the available data within \( \mathbf{B} \) is used as an initial guess to replace the gaps.
    \item The truncated SVD of this new full bunch matrix is computed.
    \item The gaps within the original \( \mathbf{B} \) are filled using values from reconstructing the truncated SVD of \( \mathbf{B} \) ; the original data remains unchanged.
    \item Repeat these steps until convergence occurs, e.g., the difference between the reconstructed values of two iterations is within a desired tolerance, the rank of the modified matrix remains unchanged, or a maximum number of iterations is met.
\end{enumerate}

Particles injected past the last update will also miss left-singular vectors. One approach could be to divide the bunch-matrix into $B \in \mathbb{R}^{m_1 \times b}$ to add the new columns and $C \in \mathbb{R}^{m_2 \times (T+b)}$ to add the new rows, which is compatible with the algorithm in Sec. \ref{sec:iSVD}. Adding rows, which corresponds to adding $C$ to the SVD of $Y$, is analogous to adding the columns of $C^\top$ to $Y^\top = VSU^\top$. However, in addition to computing the SVD of $K_B \in \mathbb{R}^{(q + b) \times (q + b)}$, this requires a second SVD on $K_C \in \mathbb{R}^{(q + m_2) \times (q + m_2)}$. The number of particles injected between updates is often immense. Considering that SVD's need to be computed for each $b$ time steps and their computational effort usually scales with $O(mn \min(m, n))$, the additional efforts of $K_C$ can become particularly costly.
We decided to substitute the missing \( U \)-rows for the $m_2$ new particles with existing \( U \)-rows. This does not affect the singular values as well as the singular vectors of the existing SVD. One can use any row of $U$. For simplicity, when $m_2 < m_1$ --which is the most common case--, the bottom $m_2$ rows of $U$ are utilized:
\begin{equation}
\left[
\begin{matrix}
u_{1,1} & \cdots & u_{1,q} \\
\vdots & \ddots & \vdots \\
\color{blue}\mathbf{u_{m_1-m_2,1}} & \color{blue}\mathbf{\cdots} & \color{blue}\mathbf{u_{m_1-m_2,q}} \\
\color{blue}\vdots & \color{blue}\ddots & \color{blue}\vdots \\
\color{blue}\mathbf{u_{m_1,1}} & \color{blue}\mathbf{\cdots} & \color{blue}\mathbf{u_{m_1,q}} \\
\end{matrix}
\right]
\quad
\overset{\text{\huge $\longrightarrow$}}{}
\quad
\left[
\begin{matrix}
u_{1,1} & \cdots & u_{1,q} \\
\vdots & \ddots & \vdots \\
\color{blue}\mathbf{u_{m_1-m_2,1}} & \color{blue}\mathbf{\cdots} & \color{blue}\mathbf{u_{m_1-m_2,q}} \\
\color{blue}\vdots & \color{blue}\ddots & \color{blue}\vdots \\
\color{blue}\mathbf{u_{m_1,1}} & \color{blue}\mathbf{\cdots} & \color{blue}\mathbf{u_{m_1,q}} \\
\color{blue}\mathbf{u_{m_1-m_2,1}} & \color{blue}\mathbf{\cdots} & \color{blue}\mathbf{u_{m_1-m_2,q}} \\
\color{blue}\vdots & \color{blue}\ddots & \color{blue}\vdots \\
\color{blue}\mathbf{u_{m_1,1}} & \color{blue}\mathbf{\cdots} & \color{blue}\mathbf{u_{m_1,q}} \\
\end{matrix}
\right]
\label{eq:matrix_transition}
\end{equation}
Mind that adding rows to $\mathbf{U}$ is analogous to adding rows to $\mathbf{Y}$. Opting to replace the missing rows with existing rows, the data prior to a particle's injection are copies of data related to particles that were already active at that time instant of the simulation. Mind that these values are not physically meaningful and are discarded during post-processing since the new particle was not active yet.
If the data matrix $Y$ contains more than one field, the rows to be used depend on how the data has been organized when the snapshots were collected. However, for the remainder of this work, each field undergoes its own SVD and all fields are treated independently.

\subsection{Parallelization}
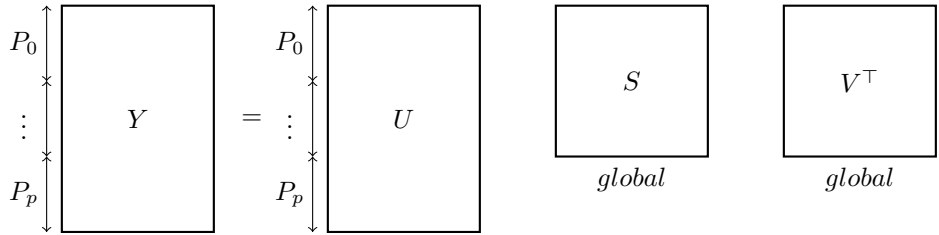
\begin{figure}[ht!]
        \centering
    \iftoggle{tikzExternal}{
        \tikzsetnextfilename{svd-par-sketch}
        \begin{tikzpicture}
              \draw[thick] (0, 0) rectangle (2, 3);
              \node at (1, 1.5) {\(Y\)};
              \draw[<->] (-0.2, 3) -- (-0.2, 2);
              \draw[<->] (-0.2, 2) -- (-0.2, 1);
              \draw[<->] (-0.2, 1) -- (-0.2, 0);
              \node at (-.5, 2.5) {\(P_0\)};
              \node at (-.5, 1.5) {\(\vdots\)};
              \node at (-.5, .5) {\(P_p\)};
              
              \node at (2.5, 1.5) {\(=\)};
        
              \draw[thick] (3.5, 0) rectangle (5.5, 3);
              \node at (4.5, 1.5) {\(U\)};
              \draw[<->] (3.3, 3) -- (3.3, 2);
              \draw[<->] (3.3, 2) -- (3.3, 1);
              \draw[<->] (3.3, 1) -- (3.3, 0);
              \node at (3, 2.5) {\(P_0\)};
              \node at (3, 1.5) {\(\vdots\)};
              \node at (3, .5) {\(P_p\)};
              
              \draw[thick] (6.5, 1) rectangle (8.5, 3);
              \node at (7.5, 2) {\(S\)};
              \node at (7.5, .7) {\(global\)};

              \draw[thick] (9.5, 1) rectangle (11.5, 3);
              \node at (10.5, 2) {\(V^\top\)};
              \node at (10.5, .7) {\(global\)};
              
            \end{tikzpicture}
    }{
        \includegraphics{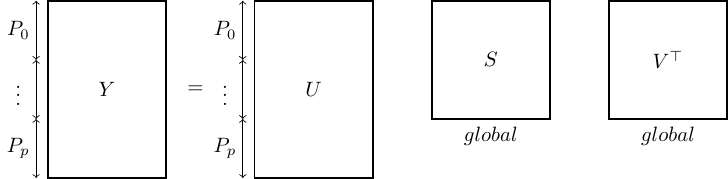} 
    }
    \caption{Matrices distribution for spatially parallel but temporally serial data in processes $P_i$ with $i \in [0,p]$.}
    \label{fig:spts_svd}
\end{figure}

When dealing with spatially parallel but temporally serial data, the data matrix $Y$ and left-singular vectors $U$ are distributed among the CPU/GPU processors. However, the singular values $S$ and the right-singular vectors $V$ coincide on all processors, cf. Fig. \ref{fig:spts_svd}.
Suppose the domain decomposition remains constant in time. In that case, all processors see data from the same amount of particles during run time, and the left-singular values, as well as the respective rows in the bunch matrix, are already placed correctly. However, if the domain decomposition changes, one must carefully handle the parallel communications so that $U$ and $B$ remain correctly aligned. Data from the employed SPH solver exacerbate this issue, as particles might travel between processes even when the domain decomposition does not change.
The two following strategies could be adopted to circumvent these issues:
\begin{enumerate}
    \item Particle-Based Row Transfer: As a particle travels, it carries its respective $U$ and $B$ rows. In this case, the changes in the particle's owner can be seen as a permutation via an elementary matrix $E$ that permutes the rows of $Y$ and $U$, i.e.,
    \begin{equation}
        EY = (EU) SV^\top \label{svd_permutation}
    \end{equation}
    
    \item Fixed Left-Singular Vectors: The left-singular vectors remain on the process used for the first update. As a particle travels, it carries its row of $B$. Before the next SVD update, the rows of $B$ that are misaligned are sent to their owner process.
\end{enumerate}
The first strategy risks having communication overhead, as a particle might travel from one process to another -- and even back before the next update. Furthermore, due to load balancing issues, new processes are activated if a minimum number of new particles are added. In this scenario, a large volume of data that scales with the number of particles, fields, modes, etc., needs to be sent to the new process. However, with the second strategy, once a particle is on the correct process, its values are already placed correctly in the bunch matrix and require no further manipulation. Additionally, large data transfers to new processes are avoided. In this work, the fixed left-singular vectors strategy is chosen. 

The proposed incremental SVD-SPH strategy is given in Alg. \ref{alg:iSVD-SPH-base}. Therein, parts related to collecting the data in $B$ while ensuring alignment between $B$ and $U$ are not included, as they depend on the specific parallelization of the flow solver. The first step is to determine if there are missing data in the bunch matrix. If this condition is met, an imputation strategy is performed as described in Sec. \ref{sec:iSVD-SPH}. Subsequently, a new SVD construction or an update is performed. Finally, the bunch-matrix is emptied.
\begin{algorithm}
\caption{General parallel incremental Singular Value Decomposition process}
\begin{algorithmic}[1]
    \State \textbf{if} missing data  
    \State \hspace{5mm} $B = \text{imputeData}(B)$ \hfill // Replace missing data, cf. Sec. \ref{sec:iSVD-SPH}
    \State \textbf{if} new SVD  
    \State \hspace{5mm} $U, S, V = \text{initialSVD}(U, S, V, B)$ \hfill // Compute the initial SVD, cf. Alg. \ref{alg:iSVD-SPH-initialSVD}
    \State \textbf{else} 
    \State \hspace{5mm} $U, S, V = \text{updateSVD}(U, S, V, B)$ \hfill // Update the existing SVD, cf. Alg. \ref{alg:iSVD-SPH-updateSVD}
    \State $\text{emptyBunchMatrix}(B)$ \hfill // Clear the bunch-matrix
\end{algorithmic}
\label{alg:iSVD-SPH-base}
\end{algorithm}
In general, serial and parallel SVD and QR decompositions can be computed using algorithms from prominent linear algebra libraries and packages, e.g. \cite{blackford1997scalapack, hernandez2005slepc}. However, due to the specific, distributed matrix structures and the parallelization as well as communication processes involved, we opt for the parallel implementation of SV and QR decompositions as described in Algs. \ref{alg:iSVD-SPH-initialSVD}-\ref{alg:iSVD-SPH-QR}.

Parallel QR decompositions follow strategies of \cite{Demmel, Benson2013DirectQF, sayadi} for tall and skinny (TS) matrices, frequently labeled TSQR. The implemented parallel TSQR decomposition is built upon the following three steps: (1) A serial QR decomposition is performed on the matrix's $B$ local blocks, resulting in local $R_1$ matrices that are subsequently gathered towards a temporary, global $\tilde{R}$ matrix, on which a second (2) serial QR decomposition is performed, offering the global $R$. Finally, local $Q$ matrices from the first serial decompositions are (3) multiplied by the local components of the second serial QR decomposition to obtain the final global $Q$. The procedure minimizes the parallel communication effort and outperforms, e.g., (modified) Gram-Schmidt procedures.
The computation of the parallel SVD involves an initial parallel TSQR on $B$, followed by a serial SVD on the resulting $R$ that provides the global $S$ and $V$. Finally, local $U$ matrices are obtained by multiplying the local $Q$ matrices from the initial TSQR decomposition with the left-singular vectors of $R$.
\begin{algorithm}
\caption{Initial Parallel Singular Value Decomposition}
\begin{algorithmic}[1]
    \State \textbf{if} $\text{parallel}$
    \State \hspace{5mm} $Q, R = \text{TSQR}(B)$ \hfill // Parallel QR decomposition (Alg. \ref{alg:iSVD-SPH-QR})
    \State \hspace{5mm} $U_R, S, V = \text{serialSVD}(R)$  \hfill // Serial SV decomposition
    \State \hspace{5mm} $U = Q U_R$ 
    \State \textbf{else}
    \State \hspace{5mm} $U, S, V = \text{serialSVD}(B)$
\end{algorithmic}
\label{alg:iSVD-SPH-initialSVD}
\end{algorithm}
\begin{algorithm}
\caption{Parallel Tall and Skinny QR Decomposition}
\begin{algorithmic}[1]
    \State \textbf{if} parallel
    \State \hspace{5mm} $Q_1, R_1 = \text{serialQR}(B)$ \hfill // Serial QR decomposition of local $B$ block
    \State \hspace{5mm} $\tilde{R} = \text{AllGather}(R_1)$ \hfill // Gather all parts of $R$
    \State \hspace{5mm} $Q_2, R = \text{serialQR}(\tilde{R})$ \hfill // Serial QR decomposition on gathered $R$
    \State \hspace{5mm} $\tilde{Q} = Q_2[\text{procId} \cdot A.\text{cols()} : (\text{procId} + 1) \cdot A.\text{cols()}, :]$ \hfill // Extract process specific data
    \State \hspace{5mm} $Q = Q_1 \tilde{Q}$
    \State \textbf{else}
    \State \hspace{5mm} $Q, R = \text{serialQR}(B)$ \hfill // Serial QR decomposition of $B$
\end{algorithmic}
\label{alg:iSVD-SPH-QR}
\end{algorithm}

The procedure to update the SVD is presented in Alg. \ref{alg:iSVD-SPH-updateSVD} and follows \cite{KUHL2024109022} with some key differences: (a) Handling missing rows in $U$, (b) performing the initial SVD on a complete bunch-matrix rather than a single column vector, (c) neglecting the specification of a truncation criterion, and (d) using the TSQR decomposition instead of a modified Gram-Schmidt procedure.
The algorithm starts by checking for a possibly empty local matrix $U$ -- which would happen if enough particles were injected since the last update to activate a new process. In that case, the local $U$ is initialized as a zero matrix instead of copying left-singular vectors from other processes, which would require additional communication effort. The remaining cases with already populated local left-singular vectors extend $U$ depending on whether the number of additional rows exceeds the number of existing rows.
The global product $U^T B$ is computed in serial mode and then made global by parallel summation, cf. \cite{KUHL2024109022}. The residual matrix $P = B - UM$ is orthogonalized using the proposed TSQR decomposition strategy from Alg \ref{alg:iSVD-SPH-QR}. An additional orthonormalization step is applied to $Q$, which accounts for a potential numerical loss of orthogonality during SVD updates \cite{Brand2006, FAREED20181942, fareed2019, FAREED2019223, bach, KUHL2024109022}. Following the enhanced itSVD algorithm outlined in \cite{KUHL2024109022}, this step precedes the construction of $K$. The global matrix $K$ is constructed and decomposed on one specific process to avoid rounding errors in the singular values. Truncated SVD results are then made global through broadcasting, and the SVD matrices are updated.
\begin{algorithm}
\caption{Update of the Incremental Singular Value Decomposition}
\begin{algorithmic}[1]
    \State \textbf{if} $U.rows() == 0$
    \State \hspace{5mm} $U = \text{Zeros}(\text{m}, r)$ \hfill // New process without existing U \\
    \hfill // m current number of rows of B assigned to this process \\
    \hfill // r current global truncation rank of U
    \State \textbf{else}
    \State \hspace{5mm} $m_1 = U.rows()$ \hfill // Number of rows of U assigned to this process at last update
    \State \hspace{5mm} $m_2 = B.rows() - U.rows()$ \hfill // Number of rows to add for this update
    \State \hspace{5mm} \textbf{if} $m_2 \leq m_1$
    \State \hspace{10mm} U = AppendBottomRows(U, m2) \hfill // Append bottom m2 rows of U to itself
    \State \hspace{5mm} \textbf{else}
    \State \hspace{10mm} $i = \text{round}\left(\frac{m_2}{m_1}\right)$ \hfill // Full copies needed
    \State \hspace{10mm} $j = m_2 - i \times m_1$ \hfill // Remaining rows needed
    \State \hspace{10mm}U = AppendMultipleTimes(U, i) \hfill //Append U to itself i times 
    \State \hspace{10mm}U = AppendBottomRows(U, j) \hfill //Append j bottom rows of U to itself

    \State $M = U^\top B$ \hfill // Compute the local matrix M
    \State \textbf{if} parallel
    \State \hspace{5mm}$M = \text{parallelSum}(M)$ \hfill // Sum M from all processes to make it global
    \State $P = B - UM$ \hfill  
    \State $Q_p, R_p = \text{TSQR}(P)$ \hfill 
    \State $Q = \begin{bmatrix} U & Q_p \end{bmatrix}$ \hfill
    \State $Q_q, R_q = \text{TSQR}(Q)$ 
    \State \textbf{if} $procId == 0$
    \State \hspace{5mm} $K = R_q \begin{bmatrix} S & U^T B \\ 0 & R_p \end{bmatrix}$  
    \State \hspace{5mm} $U', S', V' = \text{serialSVD}(K)$ \hfill  
    \State \hspace{5mm} \textbf{if} adaptive truncation
    \State \hspace{10mm} q = computeTruncationRank($S$) \hfill //Criteria-dependent truncation 
    \State \hspace{5mm} \textbf{else} 
    \State \hspace{10mm} q = constant \hfill // Truncation by fixed value
    \State \hspace{5mm} $U' = U'(:,1:q)$   
    \State \hspace{5mm} $S = S'(1:q)$  
    \State \hspace{5mm} $V' = V'(:,1:q)$ 
    \State $Bcast(U', S, V')$ \hfill // Broadcast truncated results to all processes
    \State $R = \begin{bmatrix} V & 0 \\ 0 & I \end{bmatrix}$  
    \State $U = Q_q U'$  
    \State $V = R  V'$ 

\end{algorithmic}
\label{alg:iSVD-SPH-updateSVD}
\end{algorithm}

All multi-CPU operations are performed with the Message Passing Interface (MPI) protocol, while the GPU operations are performed with the Compute Unified Device Architecture (CUDA) along with the \textit{cuSOLVER} and \textit{cuBLAS} libraries. Serial matrix operations and decompositions utilize the \emph{Eigen} library \cite{eigenweb}.

\section{Validation}
The subsequent validation displays different aspects of the reconstruction. First, the quality of the reconstruction for different ranks is discussed. To this end, we use the relative reconstruction error
\begin{equation}
    \varepsilon(\phi (\tau)) = 100 \cdot \max \left( \|\frac{\phi _{\text{FOM}}(\tau) - \phi _{\text{ROM}}(\tau)}{\phi _{\text{ref}}}\| \right) \, , \label{eq: max reconstruction error}
\end{equation}
which observes the maximum difference between \textit{Full Order Model} (SPH) and \textit{Reduced Order Model} (SVD) for a field $\phi$ at time $\tau$, with respect to the maximum absolute value during the complete simulation $\phi _{\text{ref}}$, where $\| \cdot \|$ represents the absolute value. The SVD is either adaptively truncated by neglecting all singular values that are five orders of magnitude smaller than the first singular value or always truncated at a fixed rank.
Secondly, the overhead of the procedures is assessed, which involves both the computational time and the associated storage/memory efforts. For short-term dynamics with low periodicity and limited temporal correlations, a segmentation of the time window under consideration for the SVD could be advantageous. The latter is associated with separate ROMs for each temporal segment combined with adaptive truncation criteria. Related computational advantages are, therefore, not guaranteed and will be investigated in the first case of this study. In this context, we will also examine the changes in the reconstruction error when moving from one segment to another.

All simulations were conducted on an AMD EPYC 7573X 32-Core Processor (2 sockets), 512GiB of RAM, and an NVIDIA RTX A6000 GPU (48GiB memory) for the 3D application.

\subsection{2D Dam Break}
\begin{figure}
\centering
\includegraphics[width=0.75\columnwidth,trim=1cm 1cm 1cm 1cm,clip]{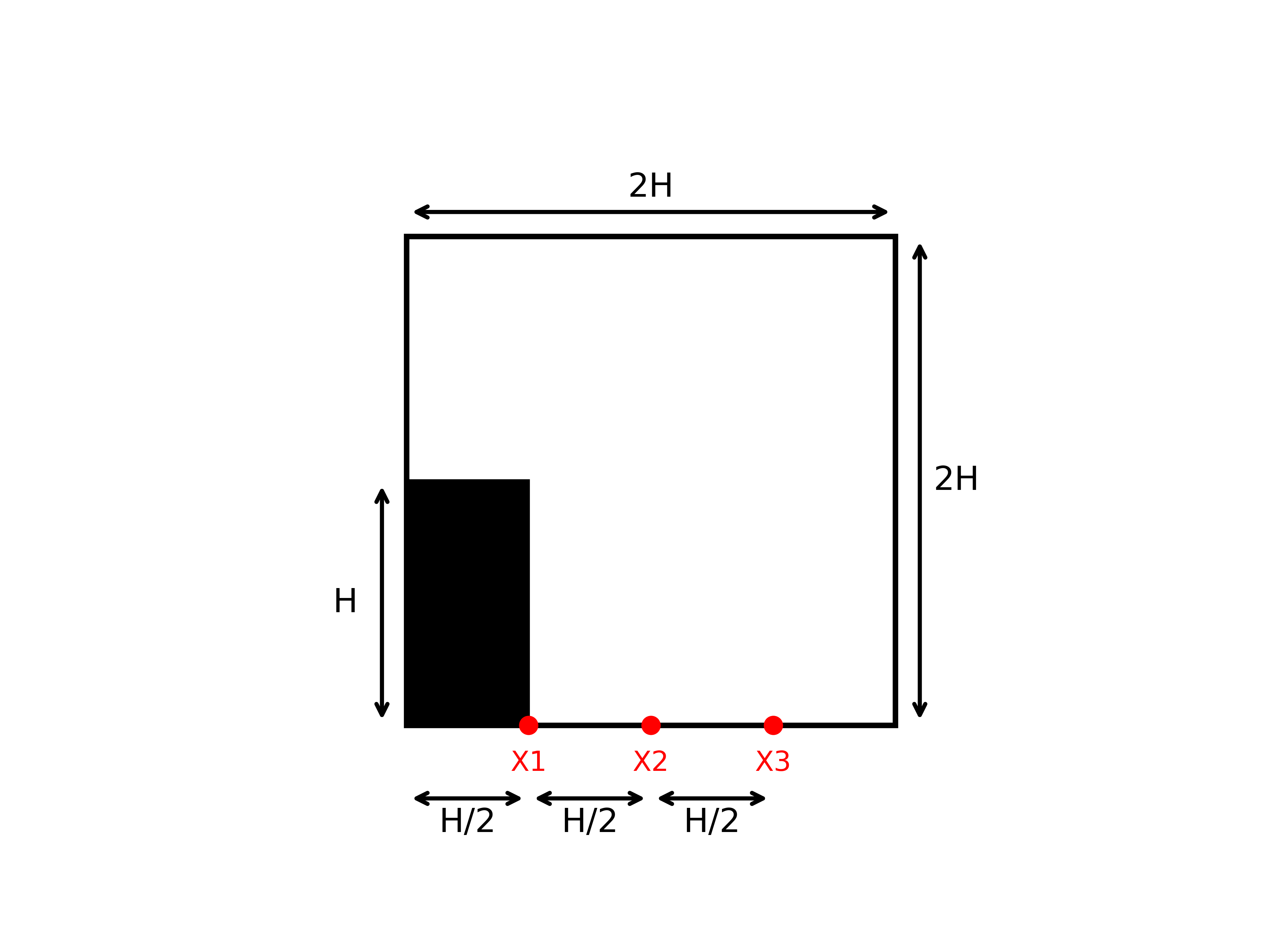}
\caption{2D Dam Break: Illustration of the initial configuration. Three height measurement probes (X1, X2, X3) are highlighted in red.}
\label{fig:dambreakSketch}
\end{figure}

The 2D dam break simulation models the temporal evolution of an initially rectangular column of water ($\rho_0=1000$ kg/m$^3$) under the influence of pressure and gravity. The complex sloshing motion involves different wave patterns and splashing as the water interacts with the walls. The test case is, therefore, frequently used to scrutinize SPH methods, see \cite{ANTUONO2021104806, YOO2024112930}.
The considered domain refers to a closed box, cf. Fig. \ref{fig:dambreakSketch}, i.e., no particle can enter and leave the domain. This allows for an assessment of the incremental SVD, particularly its ability to capture the particle positions without entering missing data.

 The square domain spans $2H \times 2H$, where $H=2m$ denotes the initial water column height. The initial width of the water column refers to $H/2$, and the initial particle spacing is assigned to $\Delta x = 0.02m$, which amounts to 5000 fluid particles. The simulation is performed over 20 000 time steps and terminates before the flow comes to rest, i.e., after the waterfront hits the downstream wall, goes up, and then comes back down. The Courant number criterion (\ref{eq:CFL}) controls the time step. To preserve a small Mach number $M\le0.1$, the numerical speed of sound is assigned to $c_0 = 100 \, \text{m/s}$ which is well above an estimated Torricelli speed of $\sqrt{2 \, g \, H} \approx 6.3 m/s$, with $\mathbf{g} = -g \mathbf{e}_z $ and $ g= 9.81 m/s^2$. Mind that the bunch size used for the SVD updates is assigned to $b=400$ in this case.

\begin{figure}
\centering
\captionsetup[subfloat]{margin=10pt, format=hang, singlelinecheck=false, justification=centering}
\subfloat[\; $q=4$ modes]{%
    \includegraphics[width=0.5\textwidth]{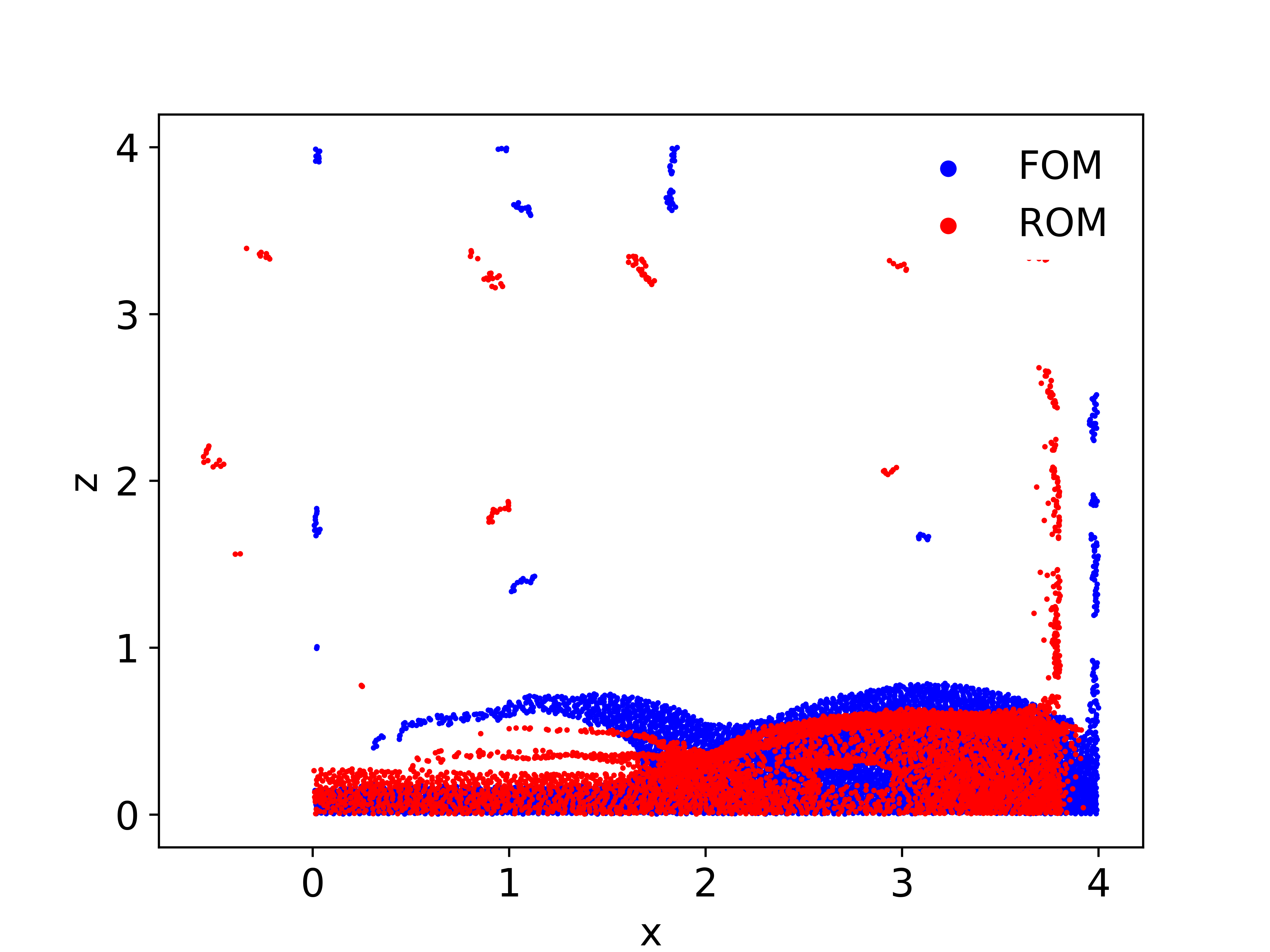}%
    \label{fig:damBreakPosition4mode}%
}
\subfloat[\; $q=8$ modes]{%
    \includegraphics[width=0.5\textwidth]{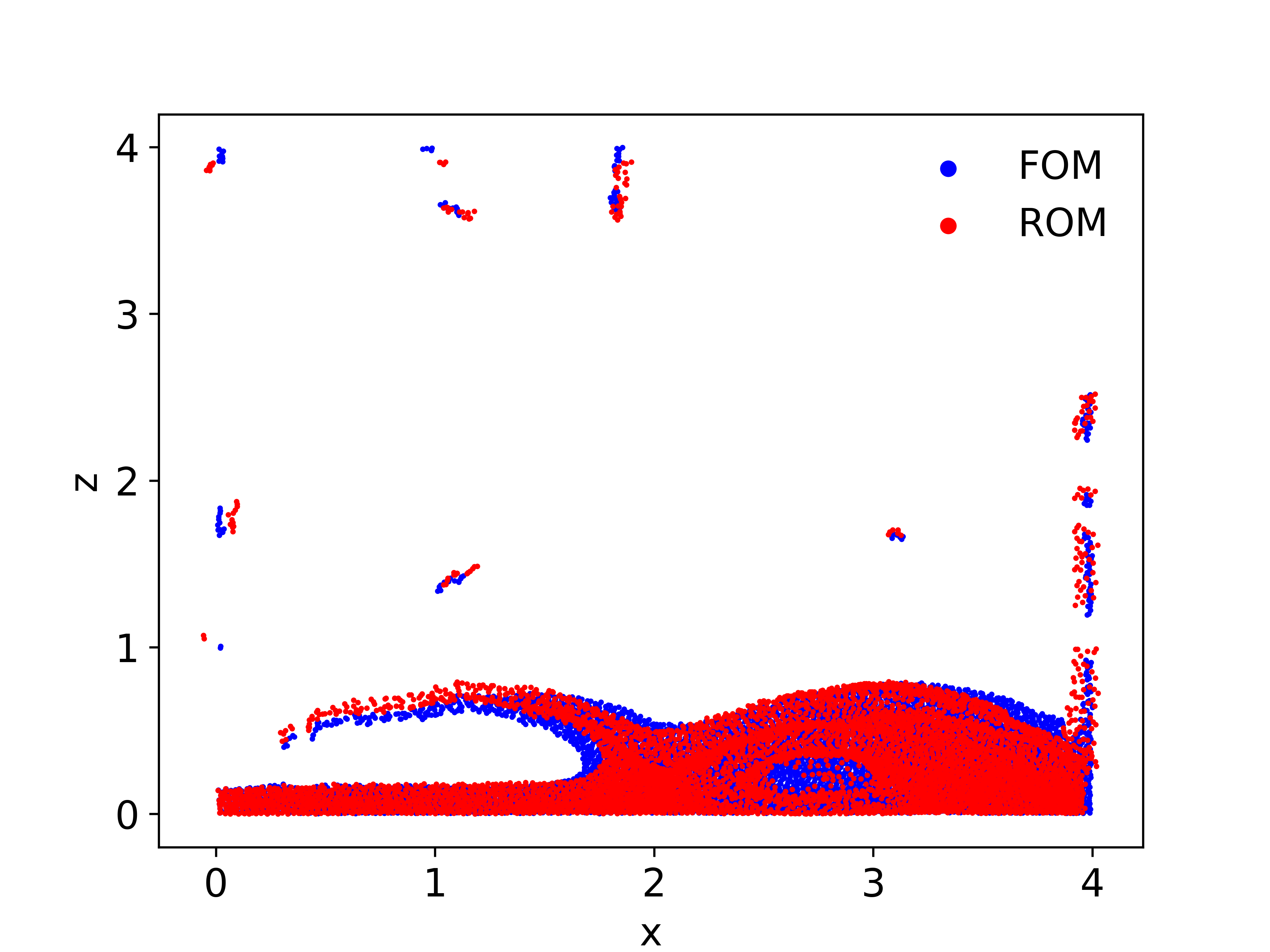}%
    \label{fig:damBreakPosition8mode}%
}
\hfill
\subfloat[\; $q=16$ modes]{%
    \includegraphics[width=0.5\textwidth]{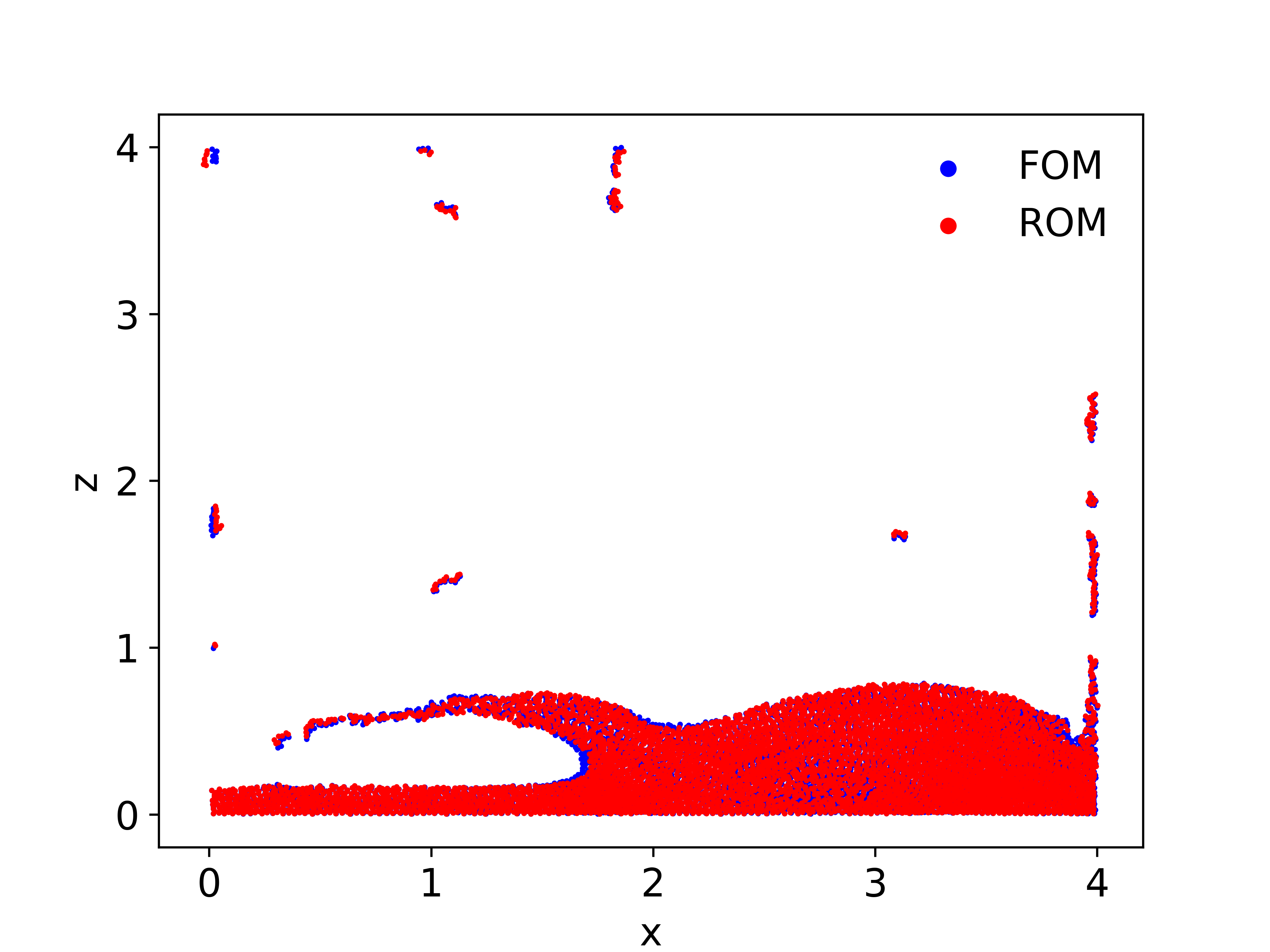}%
    \label{fig:damBreakPosition16mode}%
}
\subfloat[\; $q=32$ modes]{%
    \includegraphics[width=0.5\textwidth]{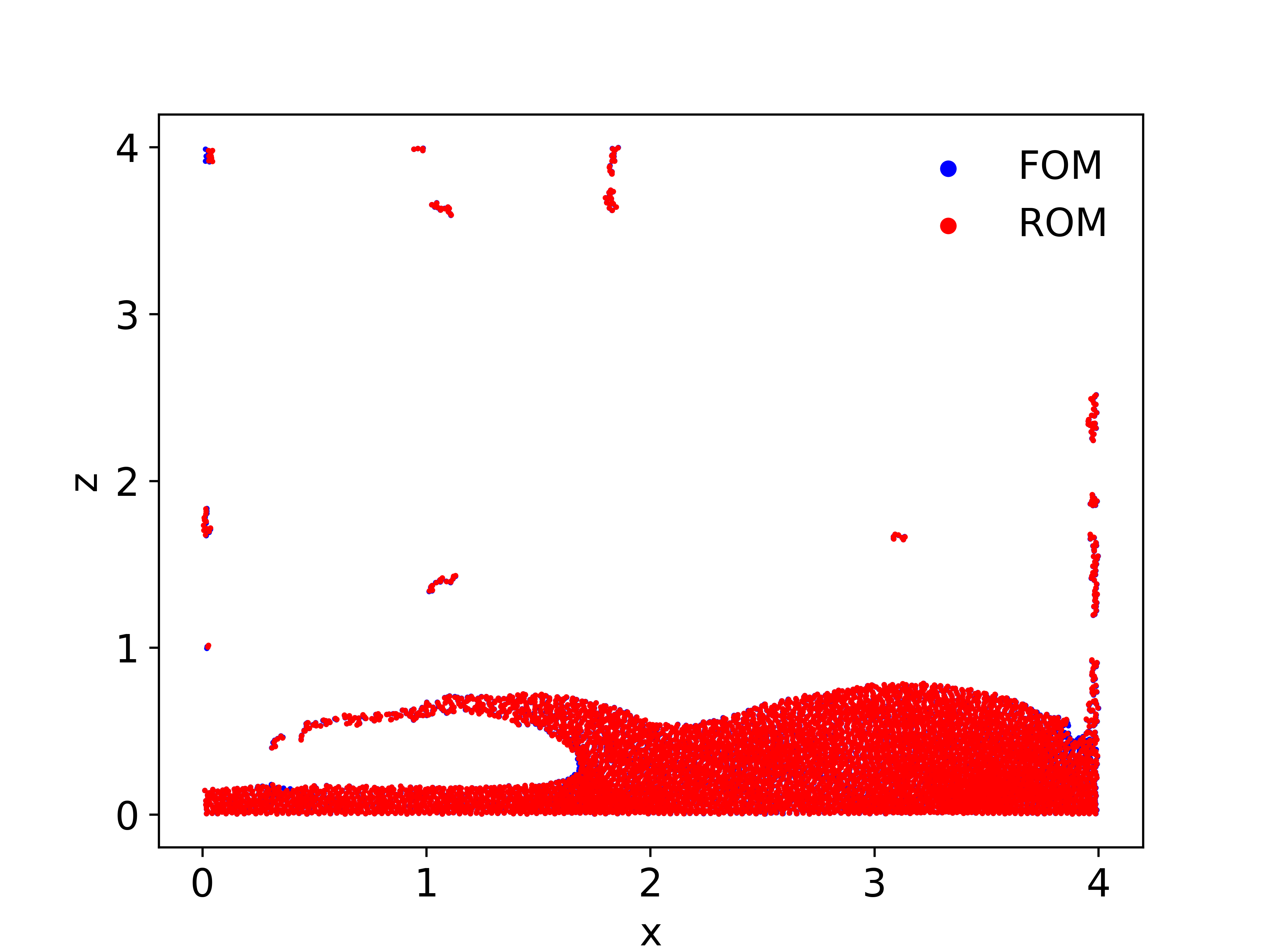}%
    \label{fig:damBreakPosition32mode}%
}
\hfill
\subfloat[\; $q=50$ modes]{%
    \includegraphics[width=0.5\textwidth]{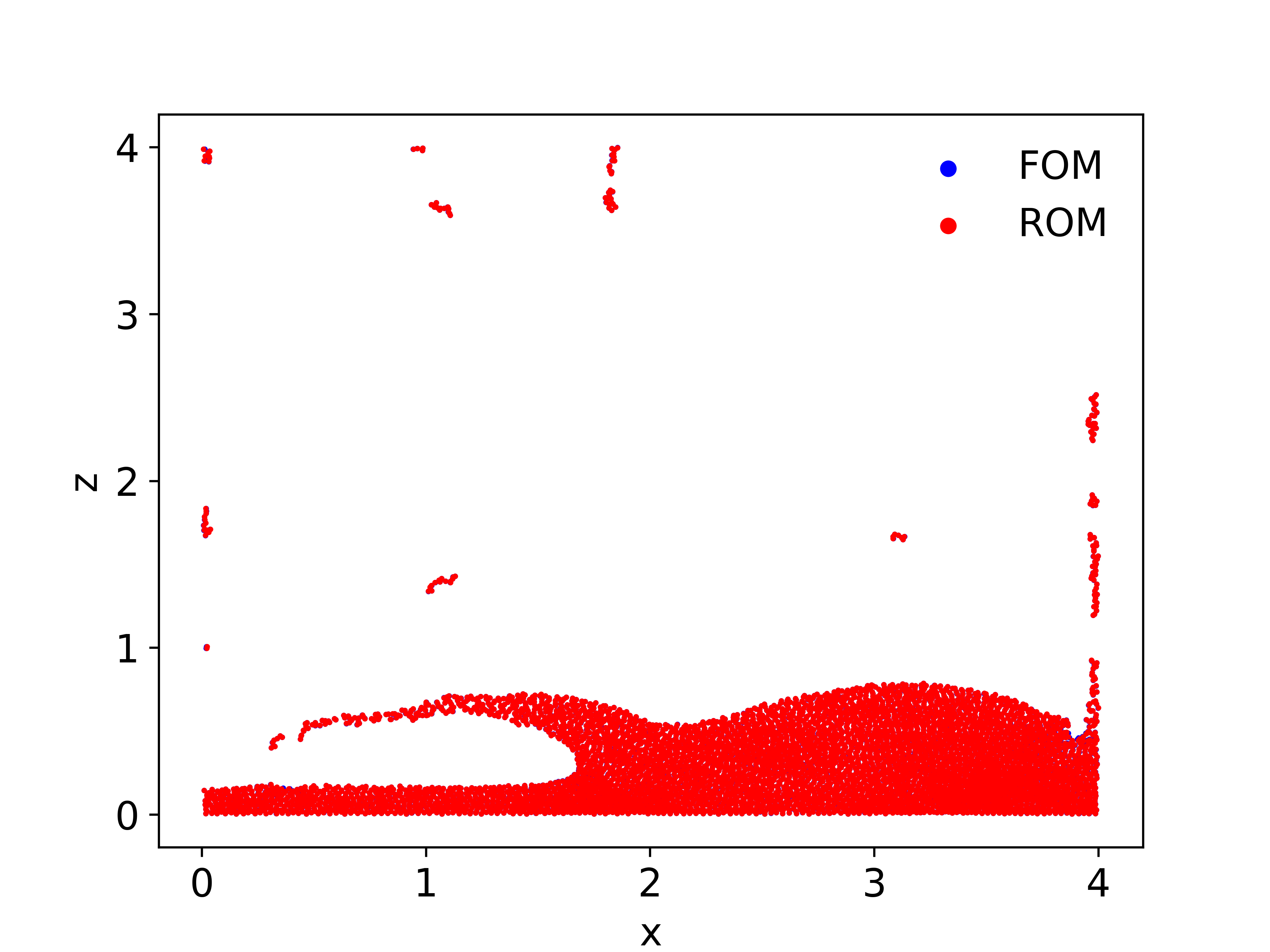}%
    \label{fig:damBreakPosition50mode}%
}
\subfloat[\; $q=100$ modes]{%
    \includegraphics[width=0.5\textwidth]{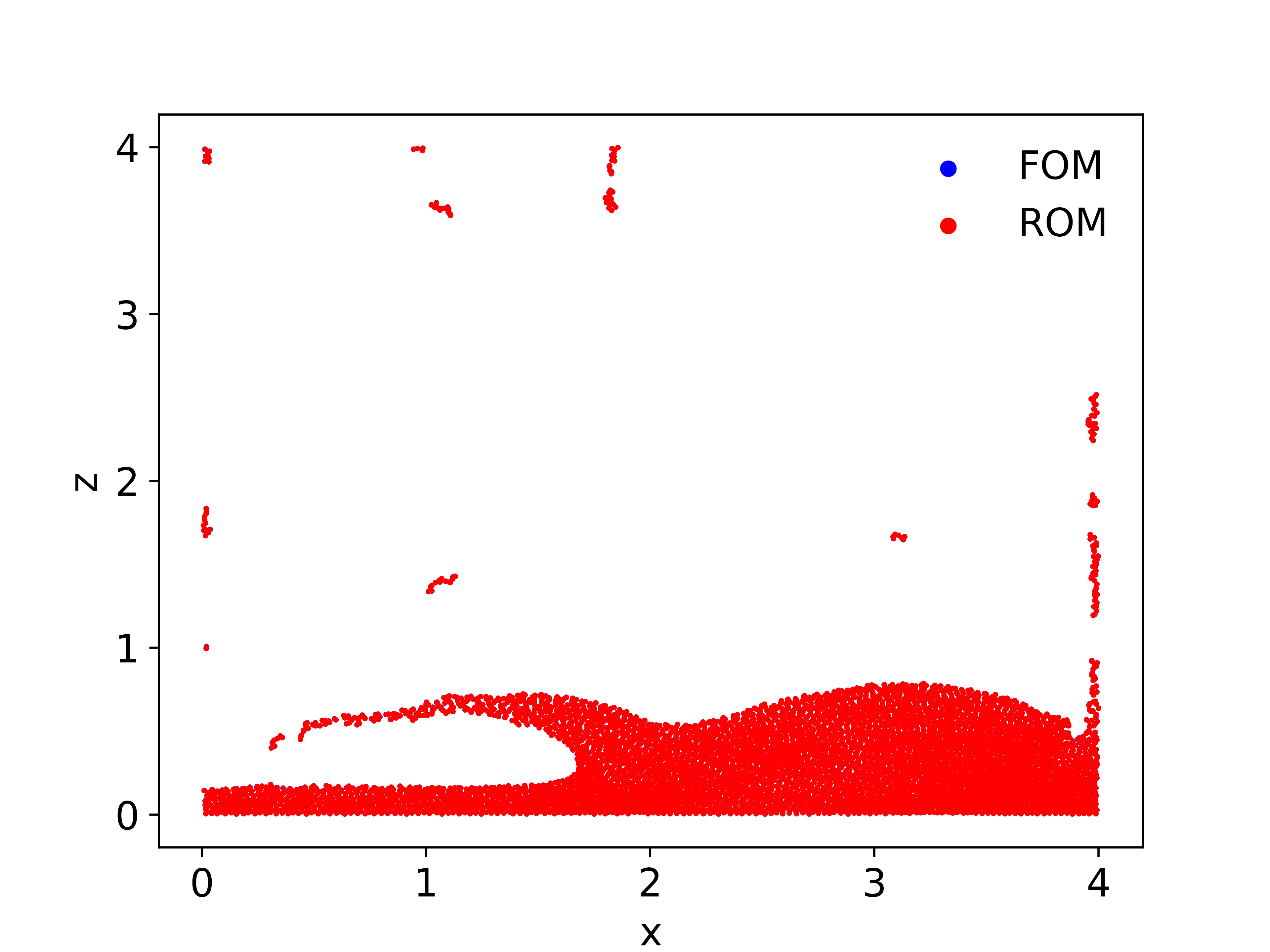}%
    \label{fig:damBreakPosition100mode}%
}
\caption{2D Dam Break: Snap-shot of the particle positions at the last simulated time step ($t = 2.26 \text{s}$) obtained from the original FOM (blue; SPH) and the SVD-based reconstruction (red) for six different numbers of considered modes.}
\label{fig:dambreakposition}
\end{figure}

Figure \ref{fig:dambreakposition} displays the water body at the final time of the simulation. A FOM solution (blue) and a ROM solution (red) with varied accuracy are compared in all six graphs. As indicated by the figure, an insufficient number of considered SVD modes yields poor reconstructions of the particle positions and the shape of the water body, and reconstructed particles might even leave the domain (cf. Fig. \ref{fig:damBreakPosition4mode}). Conversely, using enough modes yields a reconstruction that closely matches the original data. However, results indicate that a sufficient reconstruction of the body of water can be obtained with approximately 50 modes, which refers to 0.25\% of the theoretically available modes for the non-segmented, single-window approach. 

\begin{figure}
\centering
\includegraphics[width=0.5\textwidth]{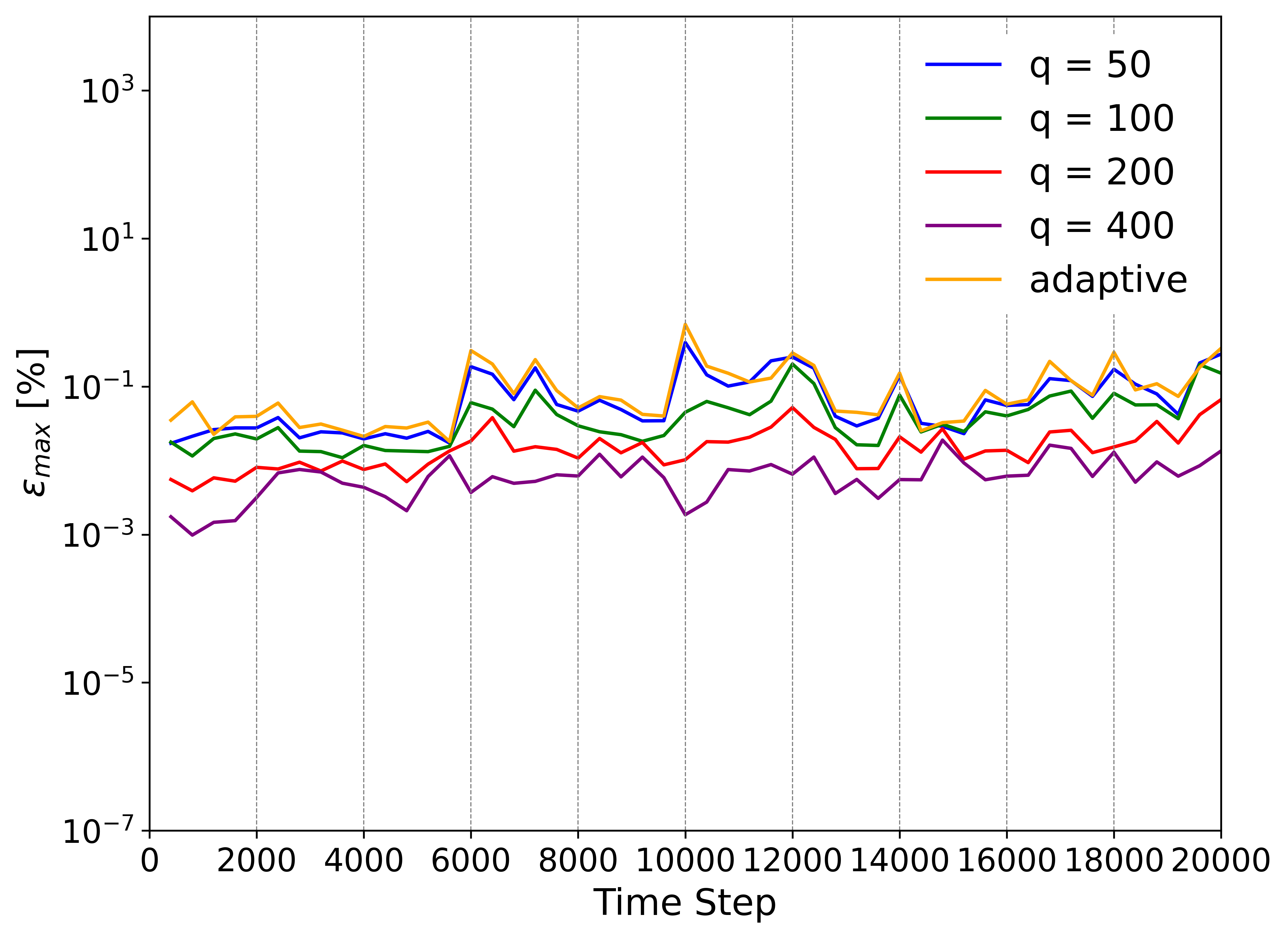}%
\hfill
\includegraphics[width=0.5\textwidth]{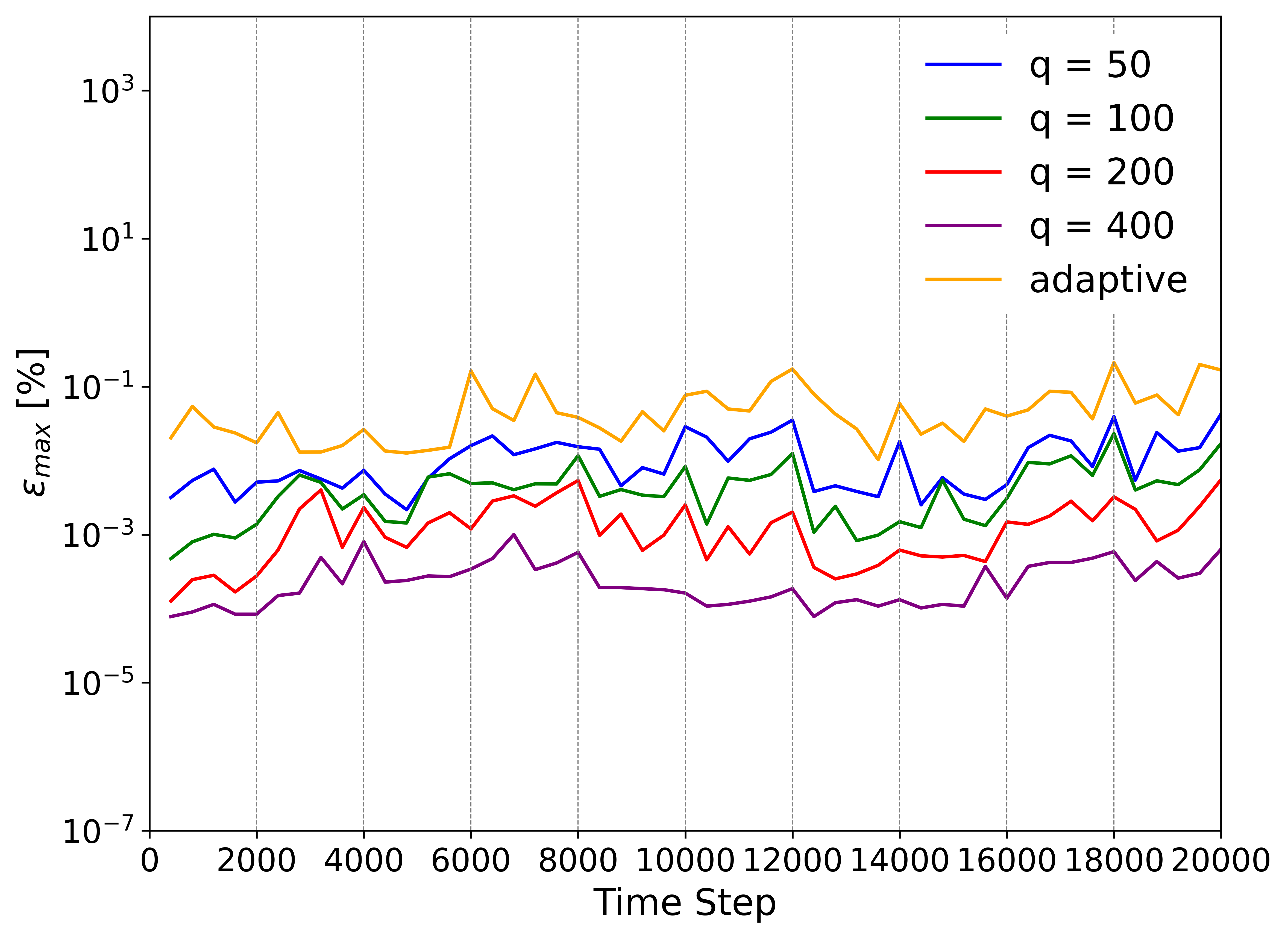}
\caption{2D Dam Break: Temporal evolution of the maximum density reconstruction error in percent obtained with different truncation ranks $q$. Left graph shows single window/SVD results and right graph displays results of segmentation into 10 (equally sized) windows/SVDs, both in combination with a bunch matrix width $b=400$.}
\label{fig:dambreakerrorr}
\end{figure}

\begin{figure}
\centering
\includegraphics[width=0.5\textwidth]{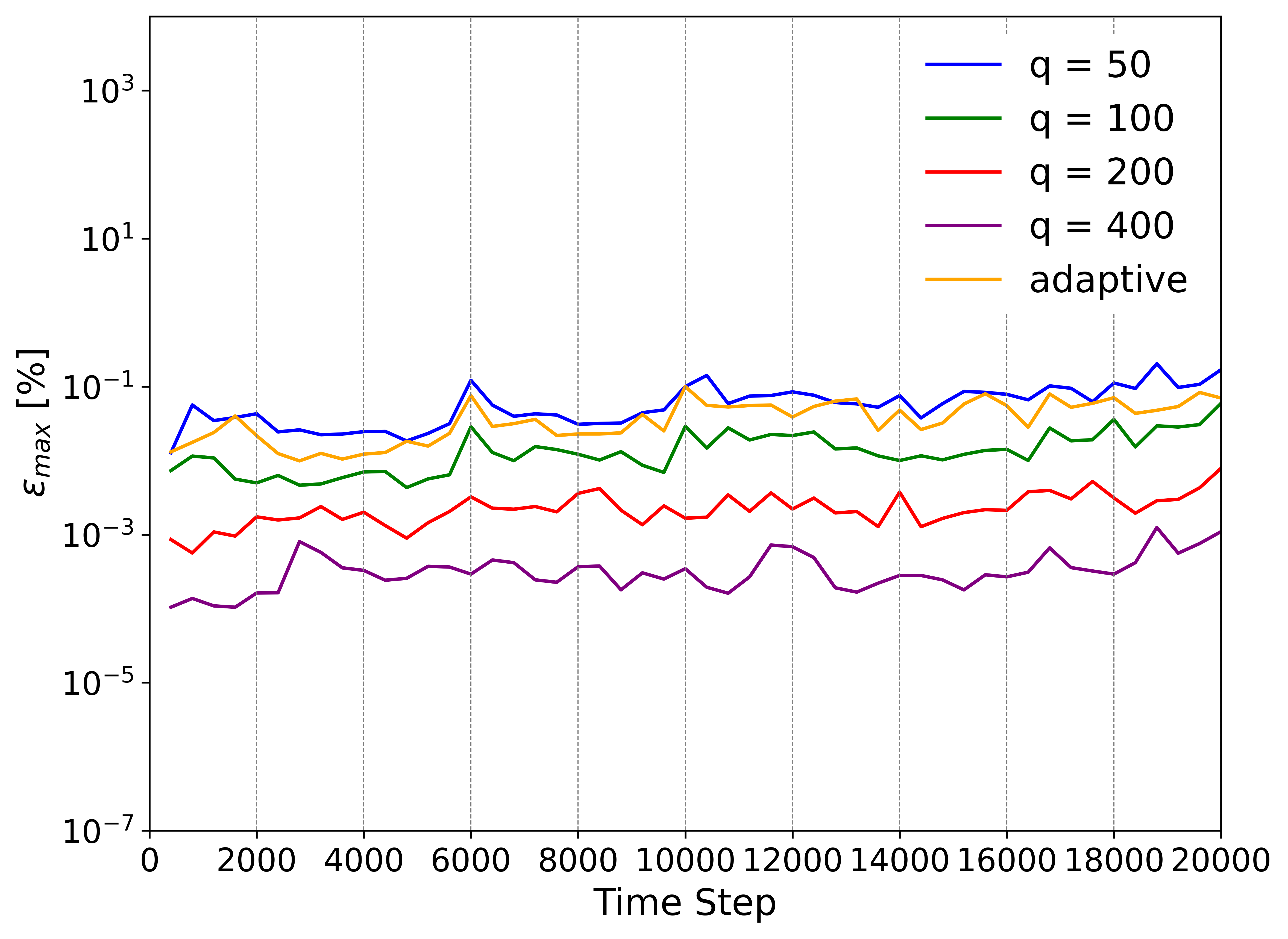}%
\hfill
\includegraphics[width=0.5\textwidth]{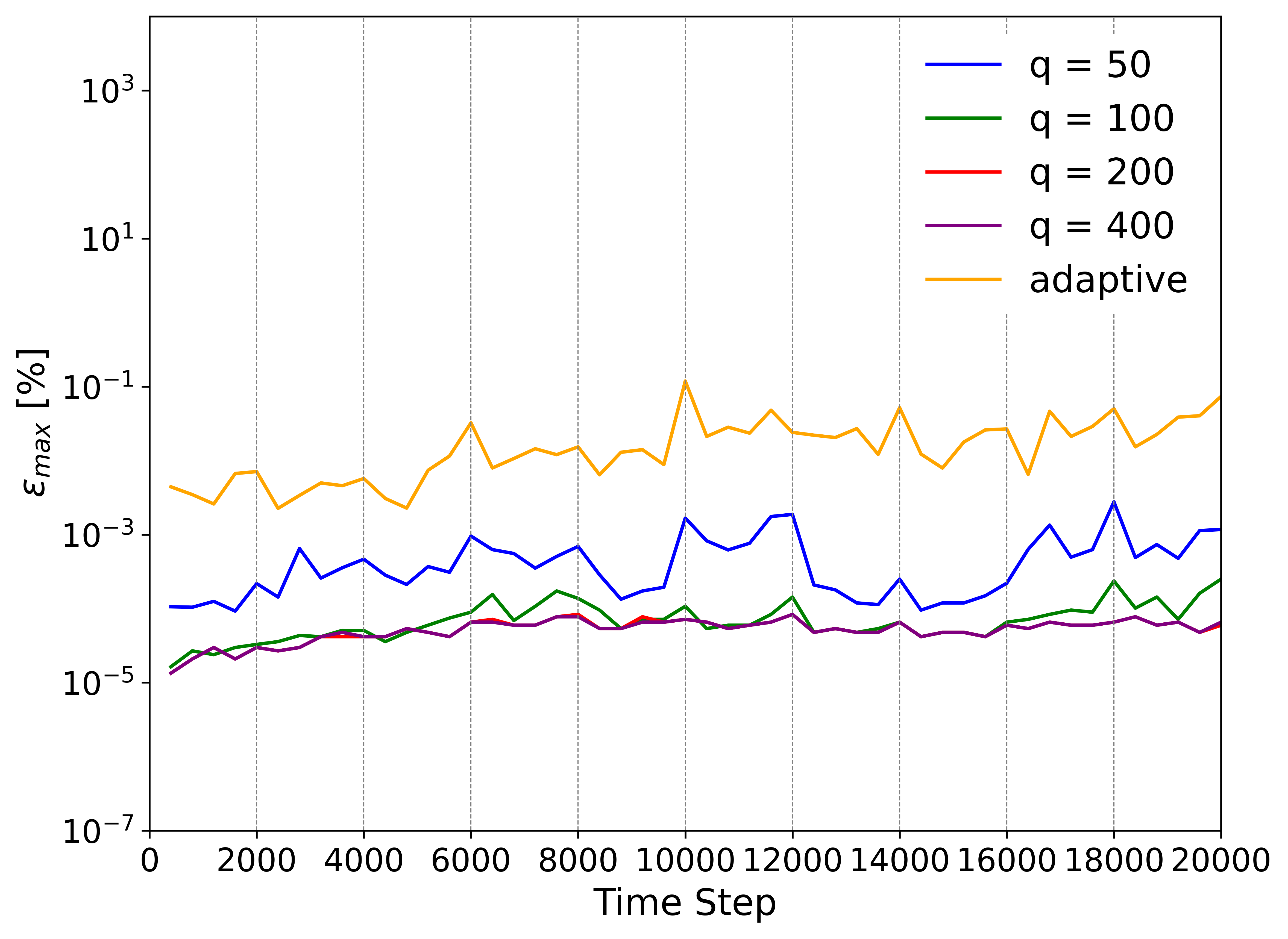}
\caption{2D Dam Break: Temporal evolution of the maximum horizontal position reconstruction error in percent obtained with different truncation ranks $q$. Left graph shows single window/SVD results and right graph displays results of segmentation into 10 (equally sized) windows/SVDs, both in combination with a bunch matrix width $b=400$.}
\label{fig:dambreakerrorx}
\end{figure}

\begin{figure}
\centering
\includegraphics[width=0.5\textwidth]{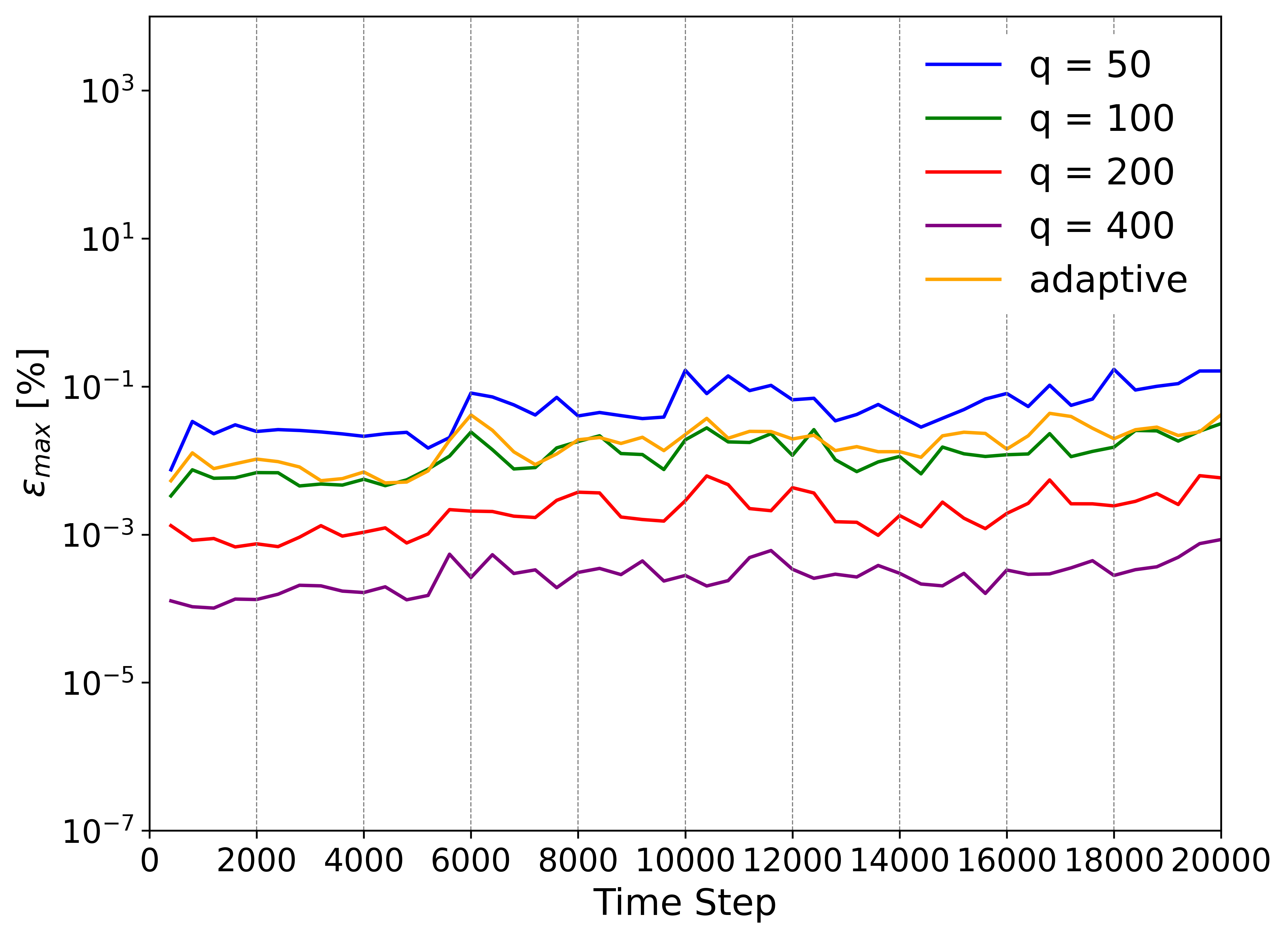}%
\hfill
\includegraphics[width=0.5\textwidth]{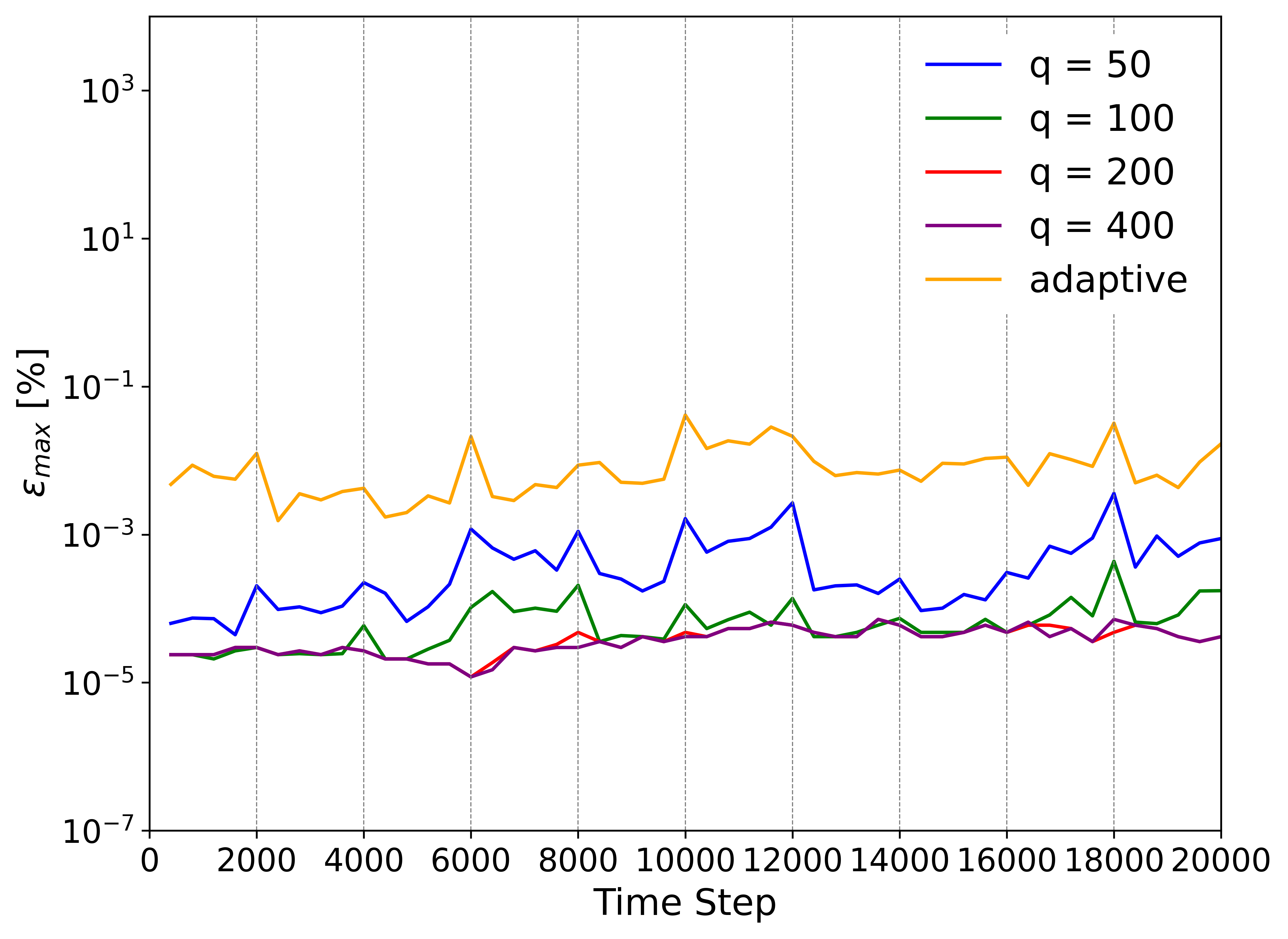}
\caption{2D Dam Break: Temporal evolution of the maximum vertical position reconstruction error in percent obtained with different truncation ranks $q$. Left graph shows single window/SVD results and right graph displays results of segmentation into 10 (equally sized) windows/SVDs, both in combination with a bunch matrix width $b=400$.}
\label{fig:dambreakerrorz}
\end{figure}

\begin{figure}
\centering
\includegraphics[width=0.5\textwidth]{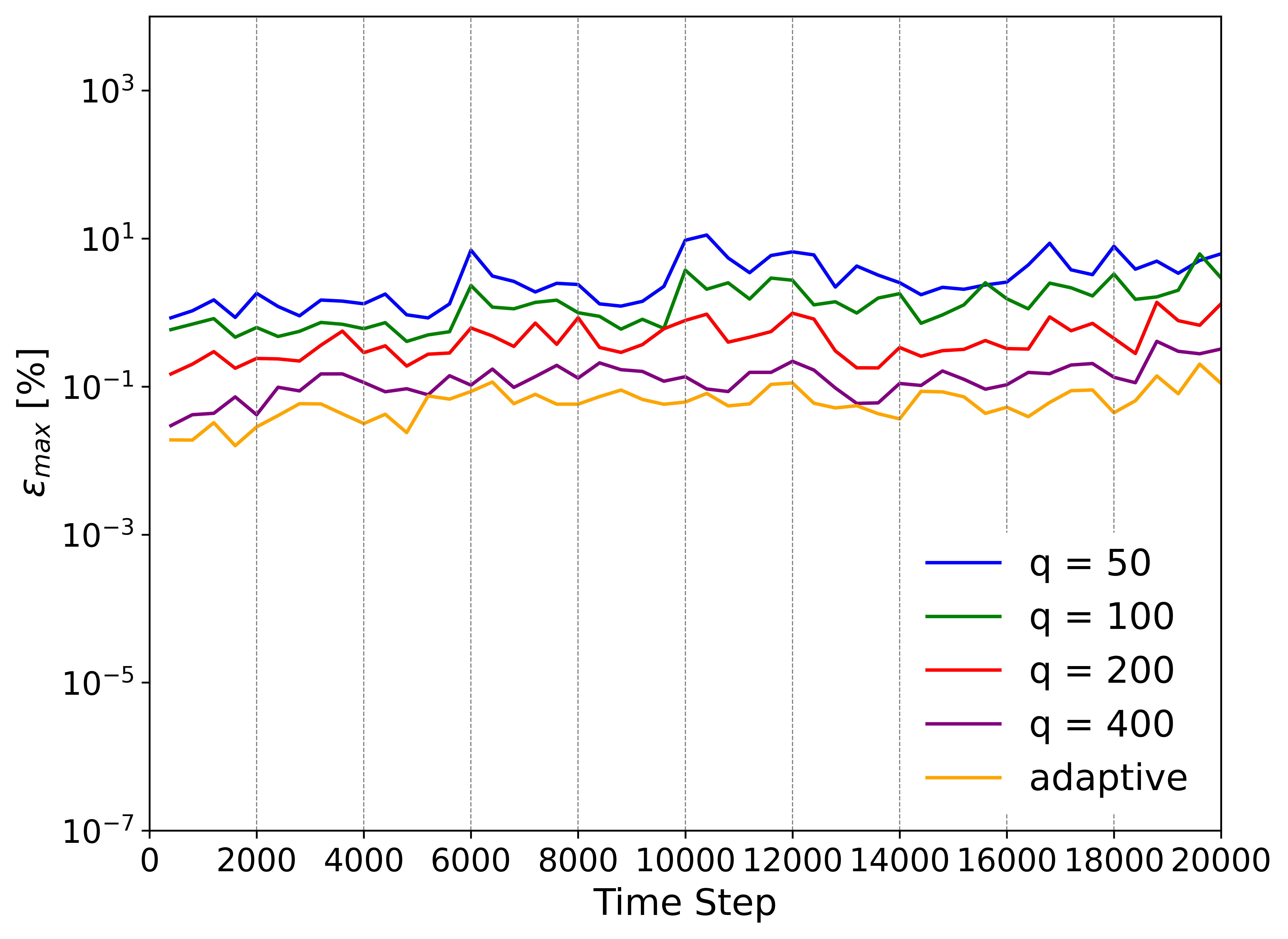}%
\hfill
\includegraphics[width=0.5\textwidth]{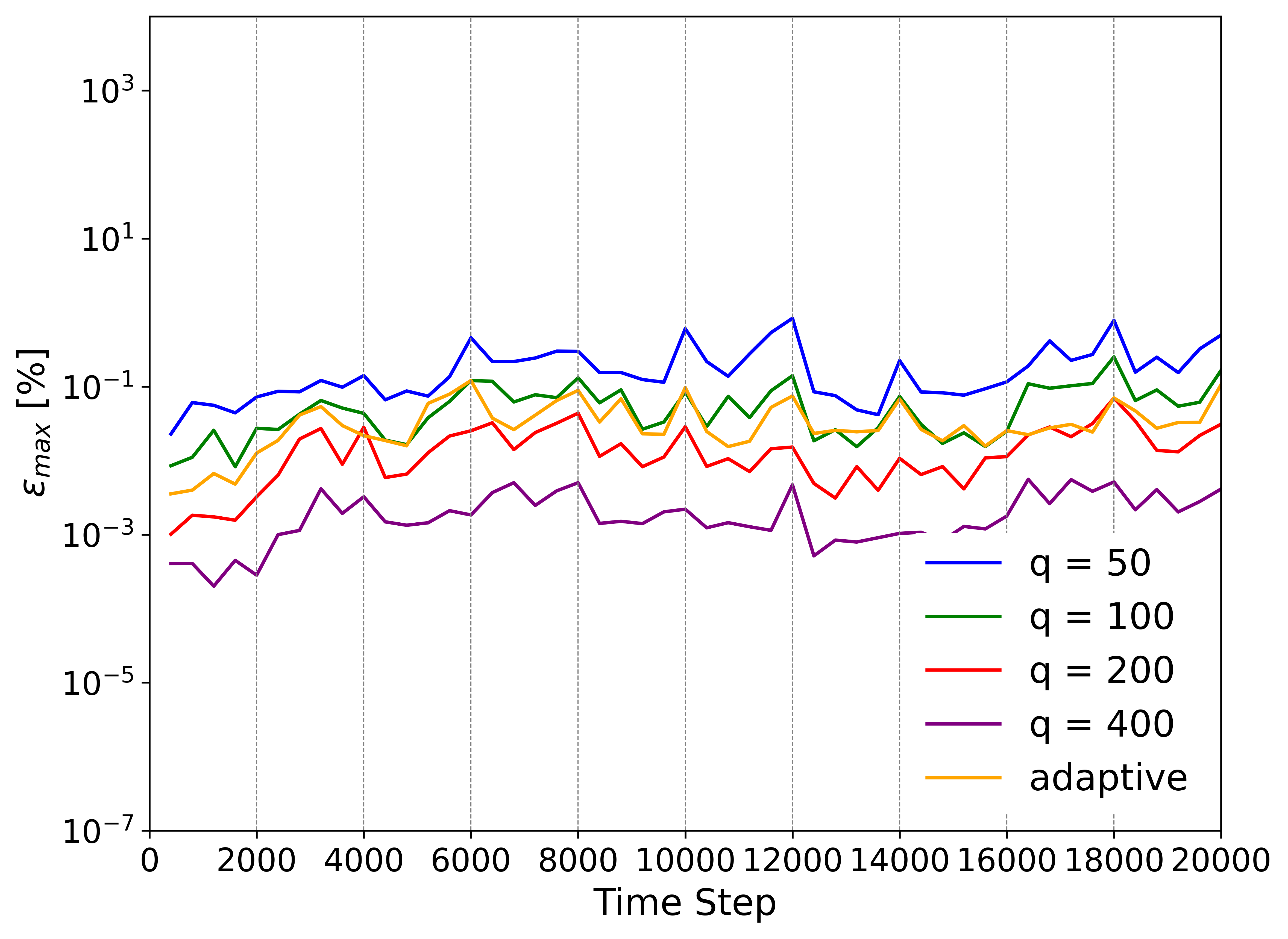}
\caption{2D Dam Break: Temporal evolution of the maximum horizontal velocity reconstruction error in percent obtained with different truncation ranks $q$. Left graph shows single window/SVD results and right graph displays results of segmentation into 10 (equally sized) windows/SVDs, both in combination with a bunch matrix width $b=400$.}
\label{fig:dambreakerroru}
\end{figure}

\begin{figure}
\centering
\includegraphics[width=0.5\textwidth]{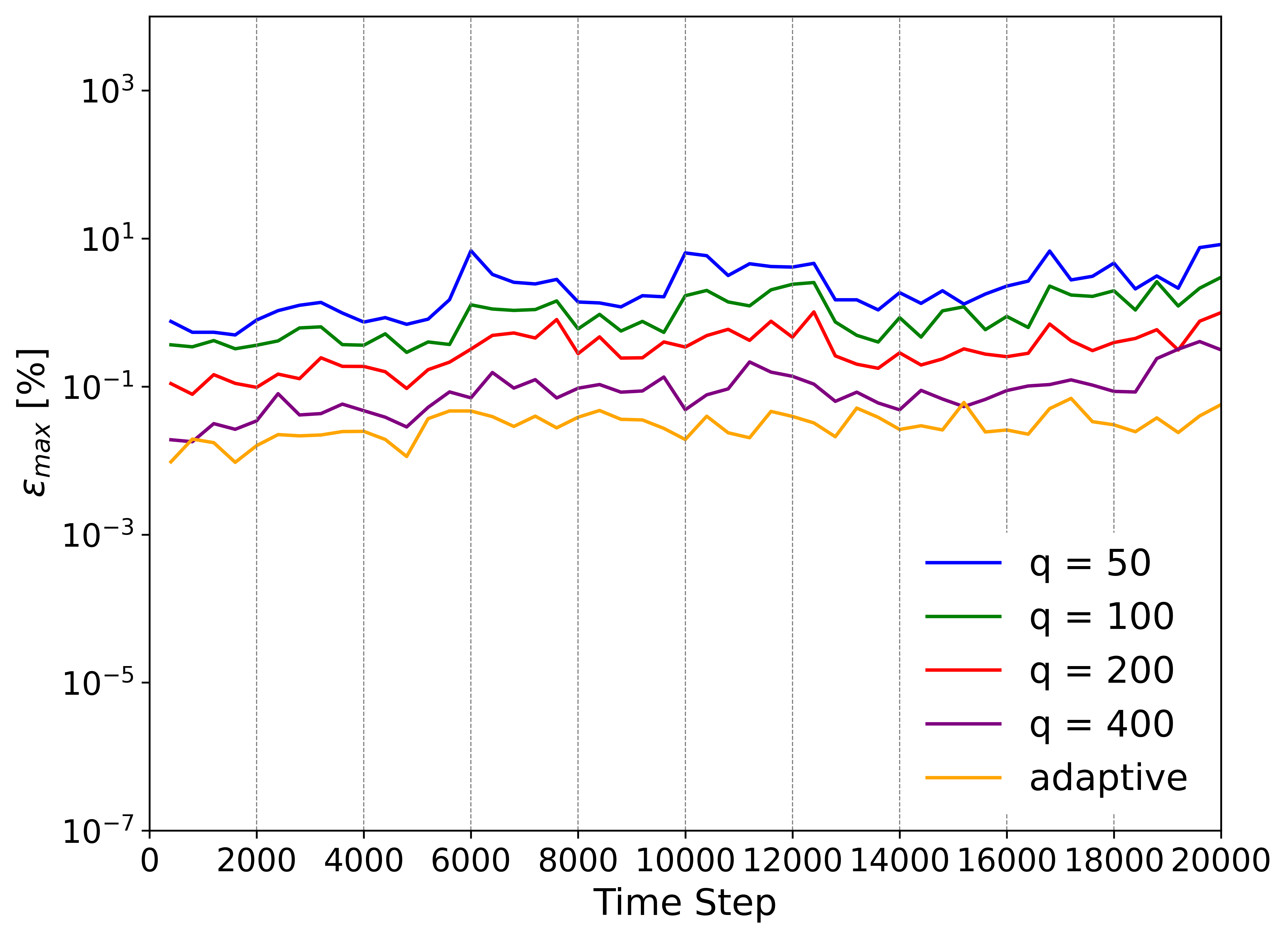}%
\hfill
\includegraphics[width=0.5\textwidth]{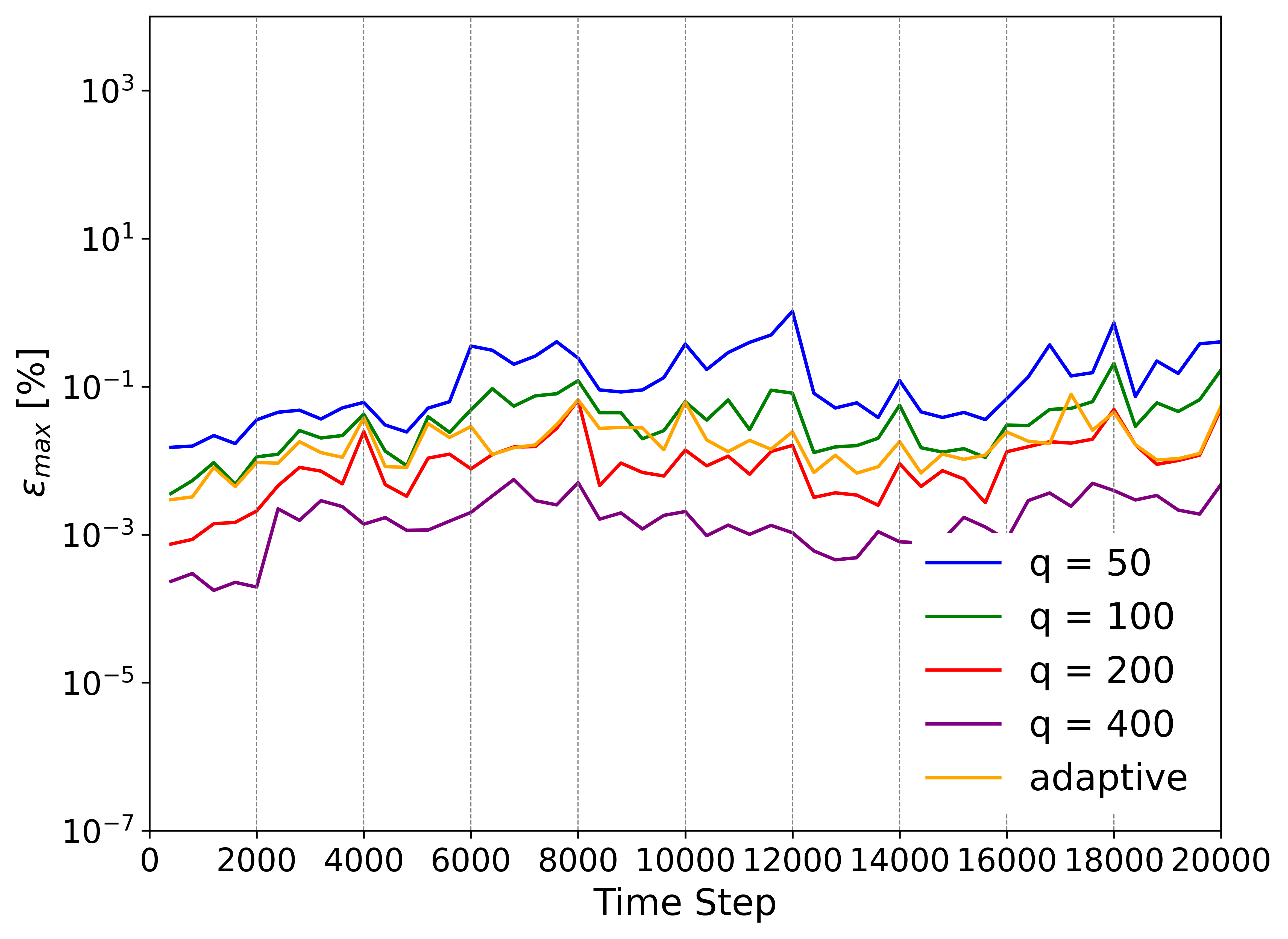}
\caption{2D Dam Break: Temporal evolution of the maximum vertical velocity reconstruction error in percent obtained with different truncation ranks $q$. Left graph shows single window/SVD results and right graph displays results of segmentation into 10 (equally sized) windows/SVDs, both in combination with a bunch matrix width $b=400$.}
\label{fig:dambreakerrorw}
\end{figure}

A more quantitative assessment of the reconstruction by the reduced order model is shown in Figures \ref{fig:dambreakerrorr}-\ref{fig:dambreakerrorw}. The figures compare the temporal evolution of the reconstruction error for the adaptive rank truncation (orange line) with results obtained in combination with four different fixed truncation ranks $q \in [50,100,200,400]$, that would be considered sufficient based on the visual inspection of the reconstructed water body. Attention is confined to the error obtained for the density, velocity, and position reconstruction. In each figure, two different approaches are presented. The respective left panels depict the results for a single window SVD, and the right panels refer to a subdivision into ten distinct, consecutive time windows that cover an equal amount of time steps with ten respective SVDs.

\begin{figure}
\centering
\includegraphics[width=0.5\textwidth]{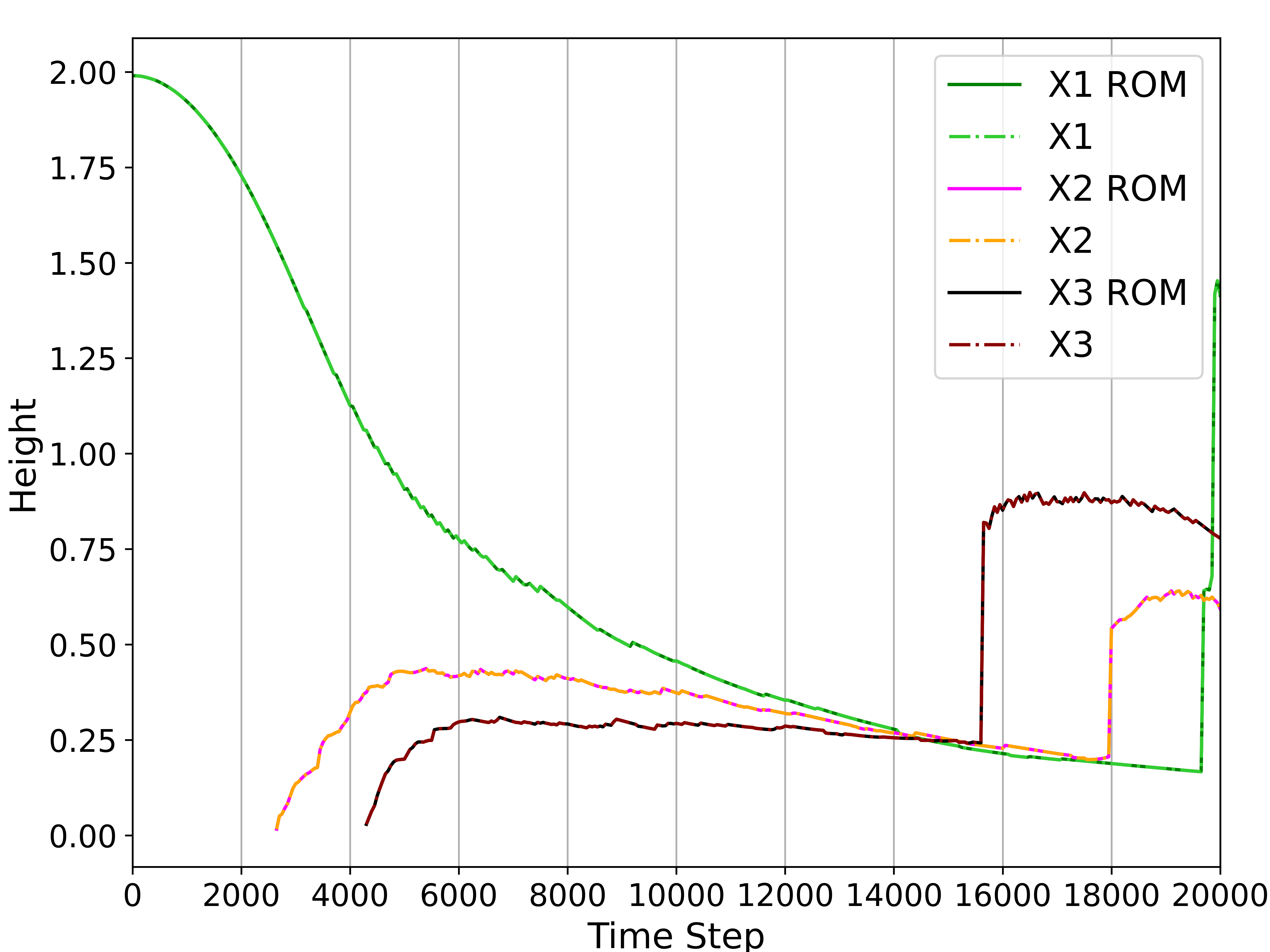}%
\hfill
\includegraphics[width=0.5\textwidth]{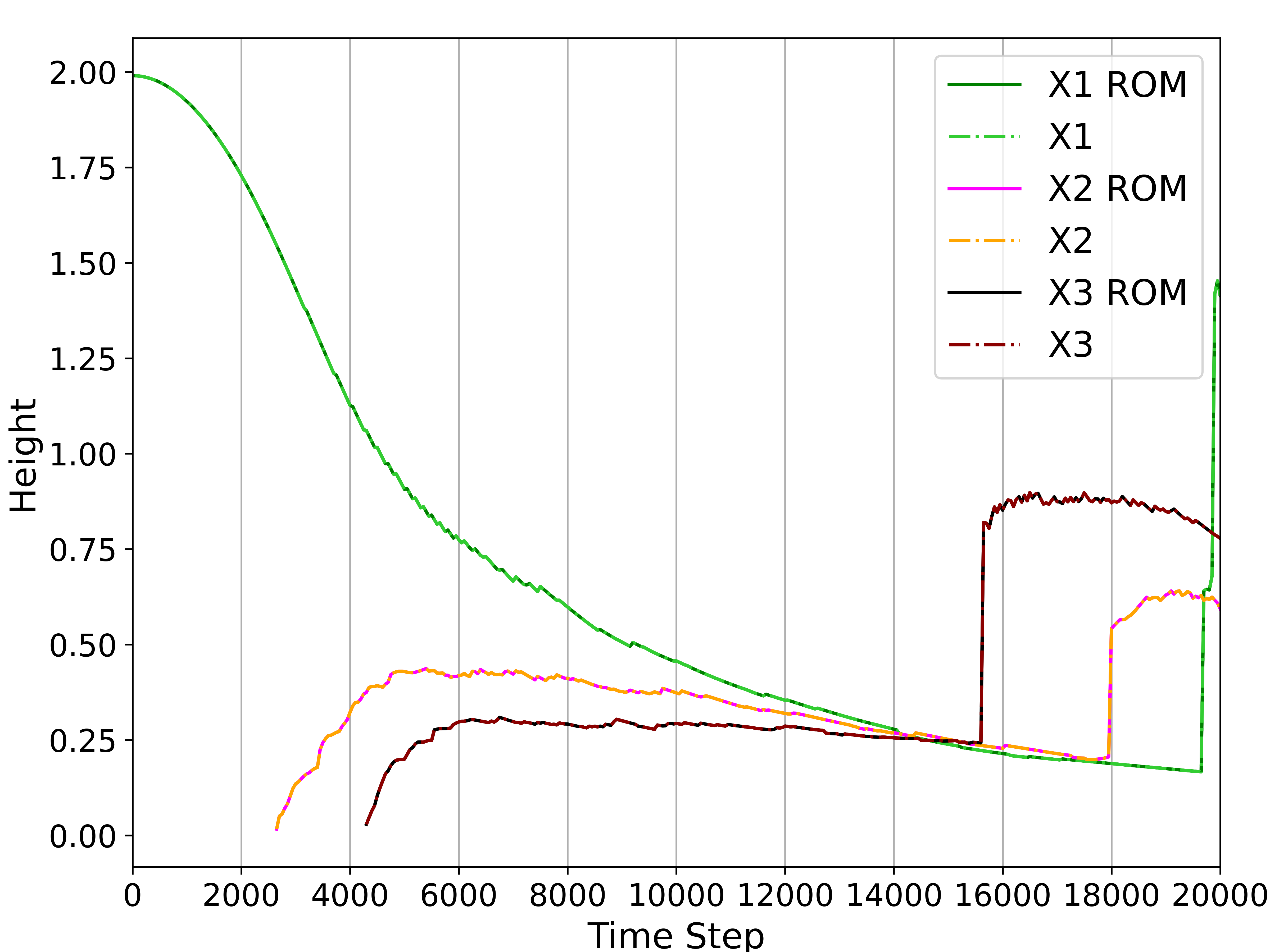}
\caption{2D Dam Break: Temporal evolution of the water level (height) at the horizontal locations X1, X2, and X3, obtained from the ROM with adaptive truncation. Left graph shows single window/SVD results and right graph displays results of a segmentation into 10 (equally sized) windows/SVDs, both in combination with a bunch matrix width $b=400$.}
\label{fig:dambreakWaterLevel}
\end{figure}

In order to ensure efficiency, the adaptive truncation variant is preferred. When attention is directed to the reconstruction of the density and positions, cf. Figs. \ref{fig:dambreakerrorr}-\ref{fig:dambreakerrorz}, it appears that the adaptive truncation essentially corresponds to a truncation at $q=50$,  essentially confirming the results of Fig. \ref{fig:dambreakposition}. Except for very few outliers, maximum relative error magnitudes below one per cent are achievable with the adaptive scheme. The picture changes for the reconstruction of the velocities, displayed in Figs. \ref{fig:dambreakerroru}-\ref{fig:dambreakerrorw}, where the adaptive truncation agrees with the fixed-rank truncation at $q=200$ or $q=400$ for a similar error level, which reveals more complex dynamics of the velocities and a smaller degree of compression. The large discrepancy of the required rank recommends separate model reductions for each field considered to increase the efficiency. Fig. \ref{fig:dambreakWaterLevel} depicts the temporal evolution of the computed water level (height) at the locations X1, X2, and X3, cf. Fig. \ref{fig:dambreakSketch}. The figure shows excellent agreement between the FOM and the ROM results. Note that isolated splashes and droplets were not taken into account when measuring the height.

\begin{figure}
\centering
\captionsetup[subfloat]{margin=10pt, format=hang, singlelinecheck=false, justification=centering}
\subfloat[1 window]{%
    \includegraphics[width=0.5\textwidth]{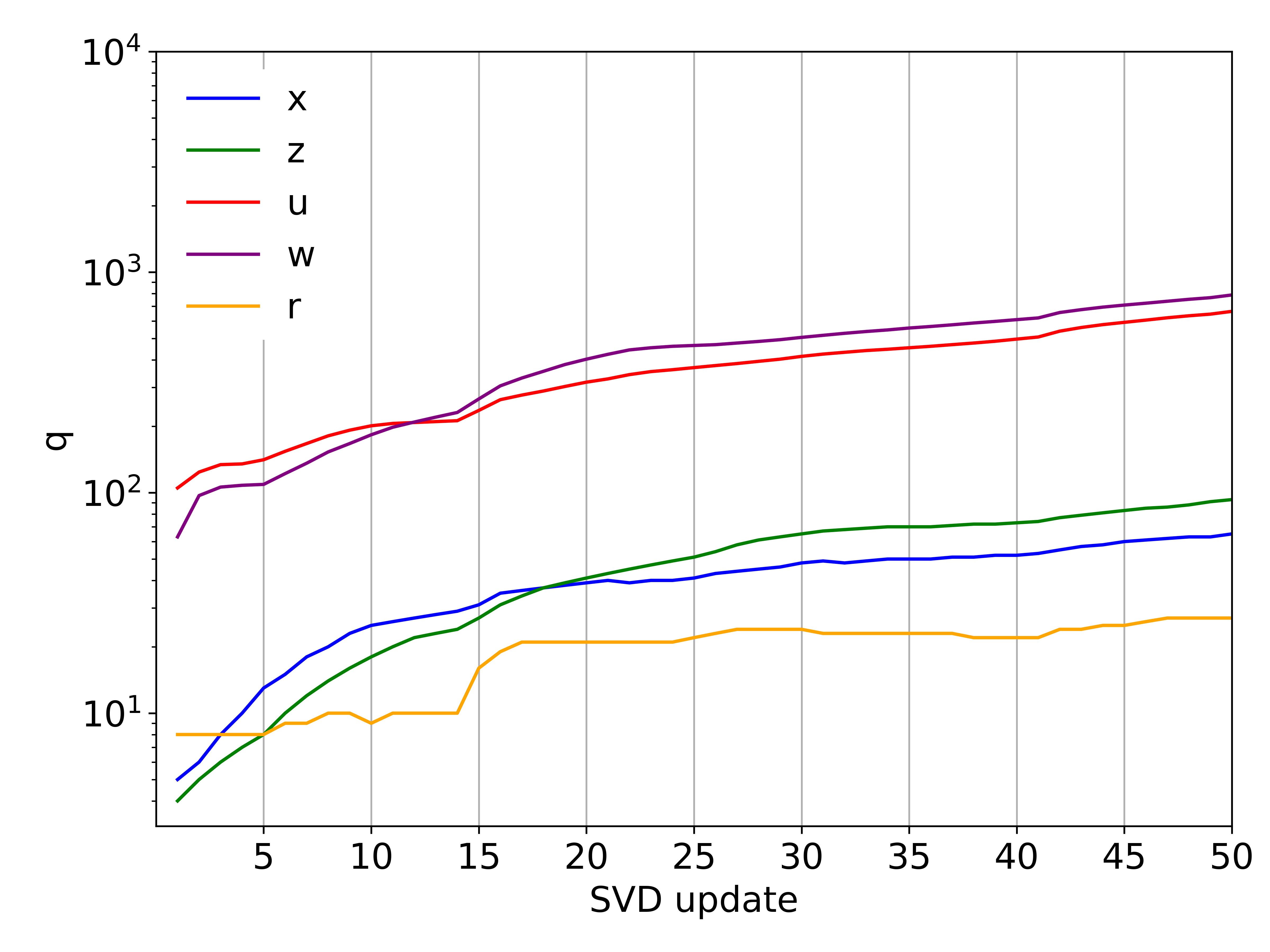}%
    \label{fig:dambreakrank1}%
}
\subfloat[10 windows]{%
    \includegraphics[width=0.5\textwidth]{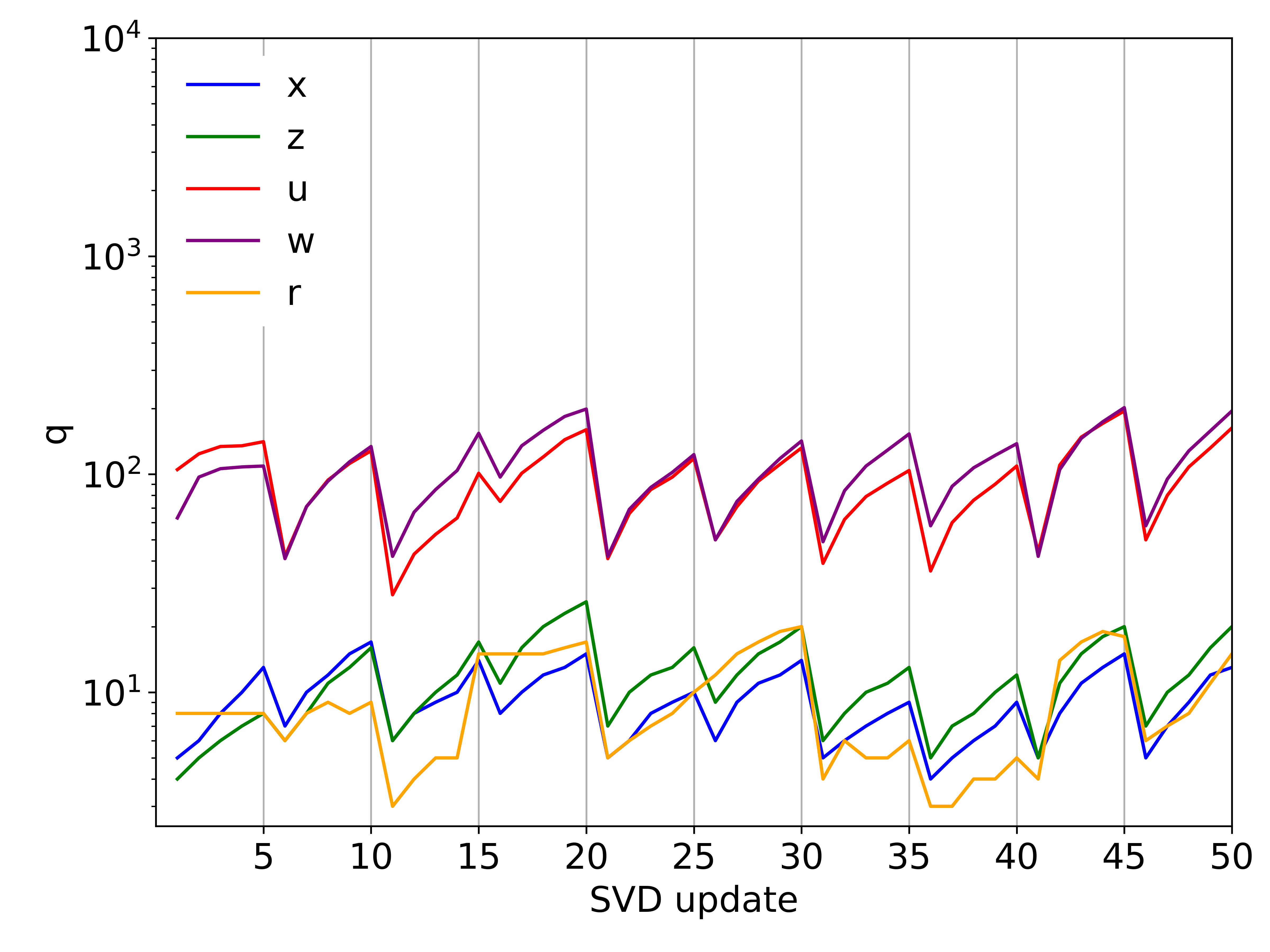}%
    \label{fig:dambreakrank10}%
}
\caption{2D Dam Break: Evolution of the truncation rank over SVD update for (a) one and (b) ten windows.}
\label{fig:dambreakrank}
\end{figure}

Due to the non-periodic behavior of the dam break flow, more information is continuously added to the SVD as the simulation progresses, and the truncation ranks for each field keep growing, cf. Fig. \ref{fig:dambreakrank1}. Having multiple SVDs that handle separate time windows might help mitigate this problem. Let us consider two exemplary setups: one option in which a single SVD covers all 20,000 time steps, and a second option in which 10 separate SVDs each cover 2,000 non-overlapping time steps. Both setups feature the same bunch width, i.e. $b=400$. Although the segmented SVD approach, in combination with adaptive truncation, always shows a better agreement between FOM and ROM, there may be doubts about the associated data effort. For the example employed above, i.e., ten vs. one window, Fig. \ref{fig:dambreakrank10} reveals an average maximum rank for the vertical velocity $w$ above 100. When multiplied by ten windows, this is more than the maximum rank of the single-window setup indicated in Fig. \ref{fig:dambreakrank1}, which is slightly below 800 and would give the impression that the 10-window setup is less efficient than the single-window setup. However, the associated data effort for multiple windows is estimated by more than simply adding up the maximum rank for all windows. Considering that the size of the left singular vector is determined by the product of the rank and the number of particles, the size of the singular values corresponds to the rank. The product of the number of time steps and the rank determines the size of the right singular vector. The following equation therefore gives the compression ratio for the windowing strategy \( CR_w \):
\begin{equation}
CR_w = \frac{\sum_{\textrm{Fields}}q^1}{\sum_{i=1}^{N_w} (\sum_{\textrm{Fields}}q^{N_w}(i))} \; \left(  \frac{N_p + 1 + N_s}{ N_p + 1 + \frac{N_s}{N_w}}\right) \, .
\label{eq:crw}
\end{equation}

In Eqn. (\ref{eq:crw}), the superscript indicates the total number of windows: \( q^1 \) represents the final rank of a particular field for a single window, while \( q^{N_w}(i) \) denotes the corresponding final rank for the \(i\)-th window. \( N_s \) is the total number of time steps, \( N_p \) is the number of particles, \( N_w \) is the number of windows. For huge particle numbers that significantly exceed the number of time steps, i.e., $N_s/N_p << 1$, the second ratio in (\ref{eq:crw}) is close to unity. Moreover, the  factor in brackets in equation (\ref{eq:crw}) tends to a $(N_s/N_p) \to 1$ for large numbers of windows $N_W$, while the total number of involved ranks in the denominator of the first factor increases to much larger values. Note that in cases featuring open boundaries, i.e. inlets and outlets, $N_p$ depends on the respective window and the definition is altered towards
\begin{equation}
CR_w = \frac{[\sum_{\textrm{Fields}} q^1] \;  (N_p^1 + 1 + N_s)}{\sum_{i=1}^{N_w} [(\sum_{\textrm{Fields}} q^{N_w}(i)) (N_p^{N_w}(i) + 1 + \frac{N_s}{N_w})]}.
\label{eq:crw_Np(i)}
\end{equation}

Table \ref{table:CR_w} displays the resulting compression ratio for different numbers of windows $N_w=2-50$ and indicates that the segmented, piece-wise approach indeed achieves the best degree of compression due to the small amount of particles and the more significant number of time steps for $N_W=10$. Mind that $CR_w=1$ refers to the single window approach. Table \ref{table:DB-CPU} depicts the computational surplus (wall-clock time) of the ROM, normalized with the respective SPH simulation time. The table shows the development of the relative computing effort as a function of the number of windows $N_W$ and the truncation rank $q$. It is seen that the effort scales almost linearly with the truncation rank for a single window. Furthermore, a strong positive influence of increasing window numbers for high truncation ranks is noticeable. While each new SVD has a constant computational cost relative to the truncation, the cost of an update does not scale linearly with the number of retained modes, making the influence of the number of windows dependent on the rank. As the number of windows increases, the number of new SVD computations grows, which causes the overall cost of the algorithm to converge towards a limit where only new SVDs are computed, rather than updates being applied to existing ones. In the limit case, i.e., for $N_w=N_s/b=$ 50  windows for the considered case, only new SVDs are computed, leading to uniform computational cost across all truncations in the last column of Table \ref{table:DB-CPU}. Note that the computational overhead with adaptive tuncation for $N_w=10$ windows is less than the overhead with $q=50$ for a single window.

\begin{table}[ht!]
\centering
\begin{tabular*}{\textwidth}{@{\extracolsep{\fill}} c c c c c c}
\hline 
$N_w$ & 2 & 5 & 10 & 25 & 50 \\ \hline 
$CR_w$ & 1.31 & 1.69 & 1.76 & 1.56 & 1.30 \\  
$\sum_i (\sum_{\textrm{Fields}} q^{N_w}(i))$ & 2082 & 2697 & 3320 & 4526 & 5824 \\ \hline
\end{tabular*}
\caption{2D Dam Break: Compression ratio for the windowing strategy $CR_w$ obtained with different number of windows $N_w$ using 5000 particles and 20 000 time steps ($\sum_{\textrm{Fields}} q^1=1637$).
}
\label{table:CR_w}
\end{table}

\begin{table}[ht!]
\centering
\begin{tabular*}{\textwidth}{@{\extracolsep{\fill}} c c c c c c c}
\hline
 ${t^{SVD}}/{t^{SPH}} [\%]$ & $N_w=1$ & $N_w=2$ & $N_w=5$ & $N_w=10$ & $N_w=25$ & $N_w=50$ \\  \hline
$q=50 $ & 14.92 & 14.27 & 13.7 & 13.14 & 11.72 & 9.31 \\  
$q=100 $ & 17.53 & 16.83 & 16.14 & 15.21 & 12.97 & 9.31 \\  
$q=200 $ & 24.90 & 23.13 & 21.70 & 20.14 & 16.08 & 9.31 \\  
$q=400 $ & 43.95 & 39.85 & 36.45 & 33.08 & 24.03 & 9.31 \\  
$\textrm{adaptive} \, q$ & 27.84 & 19.05 & 14.54 & 12.85 & 11.07 & 9.31 \\ \hline
\end{tabular*}
\caption{2D Dam Break: Computational overhead $\frac{t^{SVD}}{t^{SPH}}$ in percent for the windowing strategy obtained with different number of windows $N_w$ and truncation ranks $q$ using 5000 particles and 20 000 time steps. 
}
\label{table:DB-CPU}
\end{table}

The incremental SVD accurately reproduces meshless particle flow data in  agreement with the FOM. While segmented, piece-wise SVD methods reduce the  computational overhead, it only saves memory in combination with an adaptive truncation. In the remainder of this paper, truncation is assumed to be adaptive, and the application of the windowing strategy is always specified.

\subsection{2D Impinging Jet}
\label{sec:jet}
\begin{figure}[h]
\centering
\captionsetup[subfloat]{margin=10pt, format=hang, singlelinecheck=false, justification=centering}
\subfloat[Initial state]{%
    \includegraphics[width=0.5\textwidth]{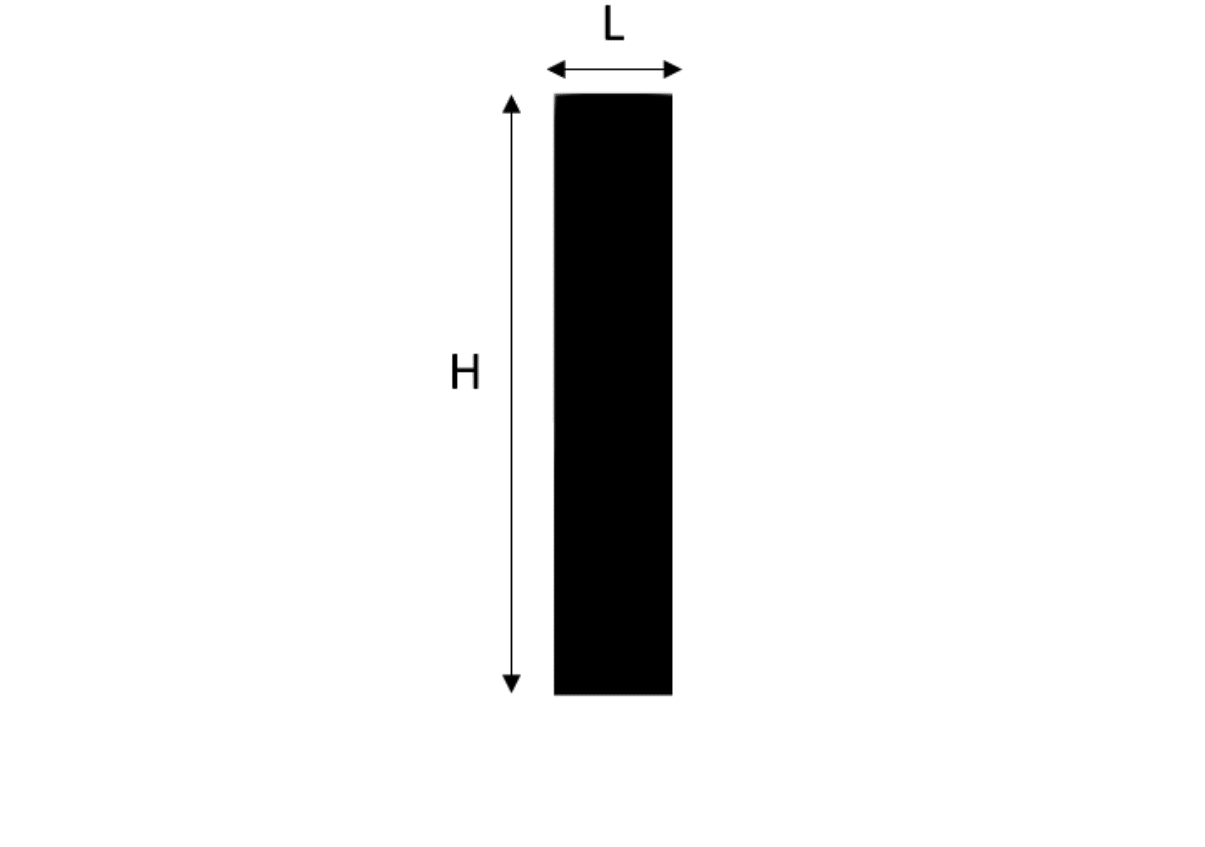}%
    \label{fig:jetinitial}%
}
\subfloat[Fully developed jet]{%
    \includegraphics[width=0.5\textwidth]{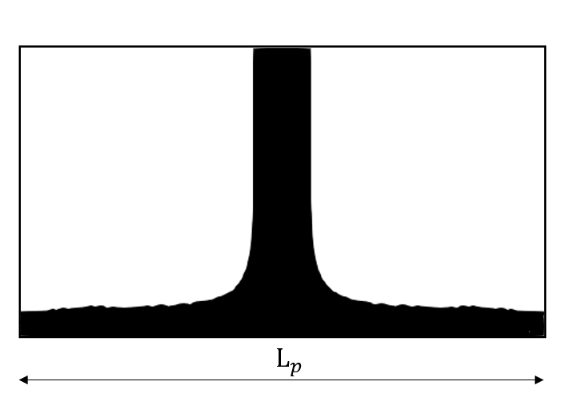}%
    \label{fig:jetdeveloped}%
}
\caption{2D Impinging Jet: Illustration of (a) the initial configuration/domain and (b) the fully developed state for the investigated.} 
\end{figure}

The second test case concerns the simulation of an impinging jet. The rectangular domain has a horizontal bottom boundary corresponding to a flat plate with a length $L_p = 0.18$m, and the height of the domain reads  $H=0.1$m. Particles are injected vertically through a confined, centered area of width $L=0.02$m from the top with a velocity $w_{in} = -20 \, \text{m/s}$ and flow towards the plate along its normal. Figure \ref{fig:jetinitial} illustrates the initial particle configuration that  employs a regular Cartesian lattice distance $\Delta x=0.0005$ and amounts to 8000 particles. The speed of sound is assigned to $c_0 = 200 \, \text{m/s}$, i.e. $w_{in}/c_0 = 0.1$, and the influence of gravity is neglected.

The simulation is performed over $N_s =$ 5 000 time steps and terminates in a fully developed state illustrated in  Fig. \ref{fig:jetdeveloped}. 
The total number of active particles in the fully-developed  (pseudo-steady) state oscillates around $N_p=$ \SI{15500}{}.

Following the impact, 
the fluid is horizontally displaced, eventually reaching the domain's vertical borders and leaving it.
Any particle outside the rectangular computational domain $(L_p \times H)$ is removed from the SPH calculation.  The presence of  inlet and outlet sections distinguishes this case from the dam break case investigate in the previous section. At the same time, it makes this case interesting for assessing how imputing/assigning missing values for the empty positions of the data matrices affects the accuracy of the reconstruction, the achievable compression and the computational effort. The section therefore focuses on discussions on the influence of (a) the imputation as well as (b) the windowing strategy on the accuracy, the compression rate and the related CPU effort.

All four imputation strategies described in Sec. \ref{sec:iSVD-SPH},
i.e. mean (MI), block-mean (BMI), hybrid MI-BMI and gappy POD (GPOD) are assessed here. 
The hybrid mean and block-mean imputation, denoted as $h_{MI-BMI}$, refers to a situation, where a bunch matrix with a standard deviation below unity after the MI is replaced by BMI.
Furthermore, the GPOD approach is assessed  for three different (fixed) iteration numbers, i.e.,  GPOD(i), where $i \in [1,5,10]$.  

\begin{figure}
\centering
\captionsetup[subfloat]{margin=10pt, format=hang, singlelinecheck=false, justification=centering}
\subfloat[\; FOM time step = 250]{%
    \includegraphics[width=0.5\textwidth]{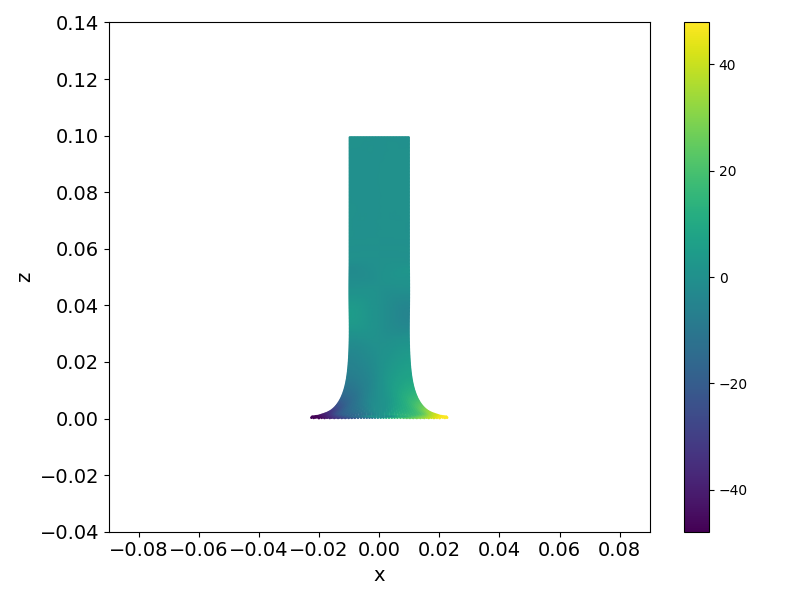}%
}
\subfloat[\; ROM time step = 250]{%
    \includegraphics[width=0.5\textwidth]{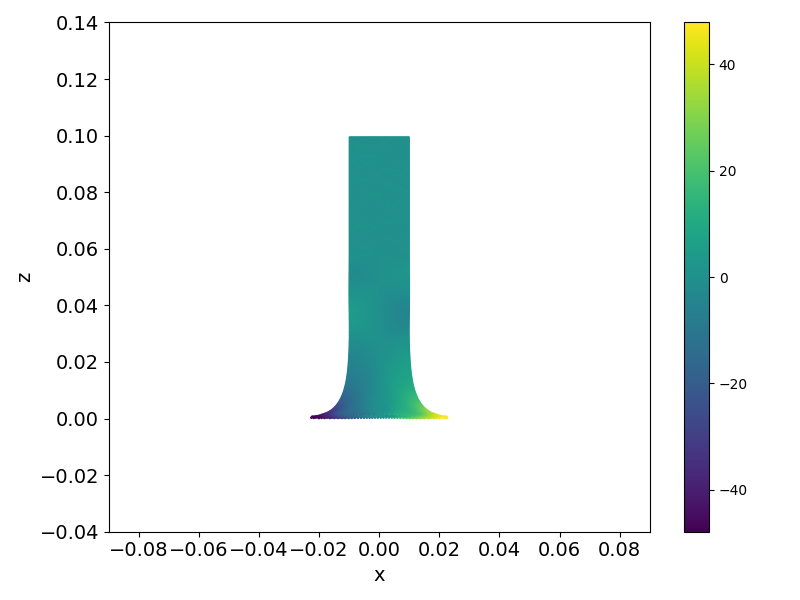}%
}
\hfill
\subfloat[\; FOM time step = 2500]{%
    \includegraphics[width=0.5\textwidth]{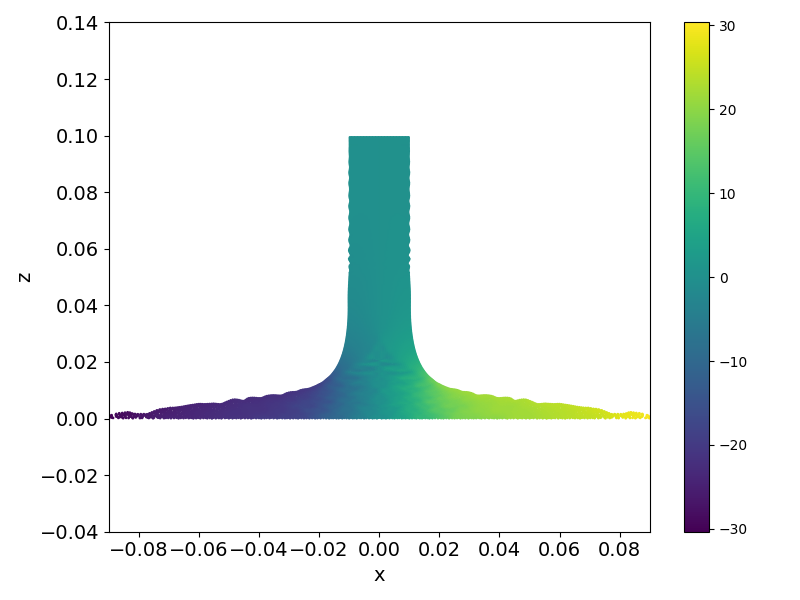}%
}
\subfloat[\; ROM time step = 2500]{%
    \includegraphics[width=0.5\textwidth]{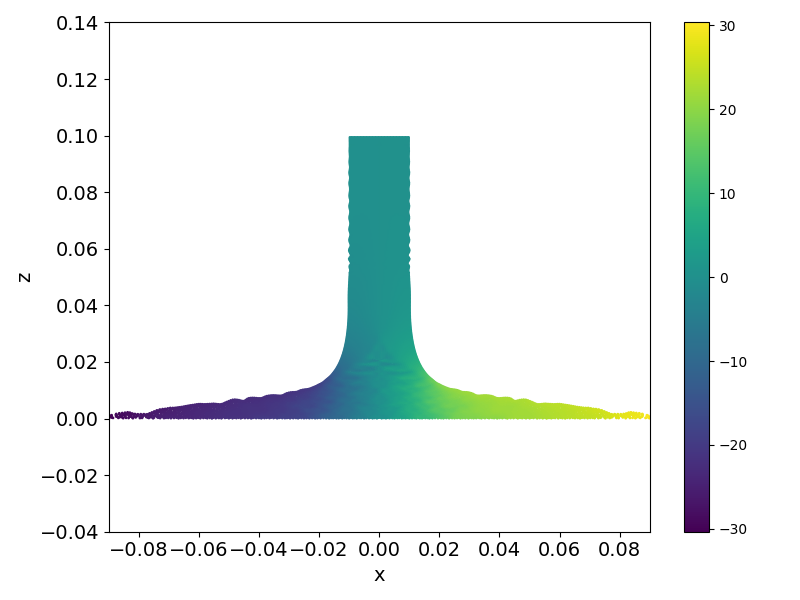}%
}
\hfill
\subfloat[\; FOM time step = 5000]{%
    \includegraphics[width=0.5\textwidth]{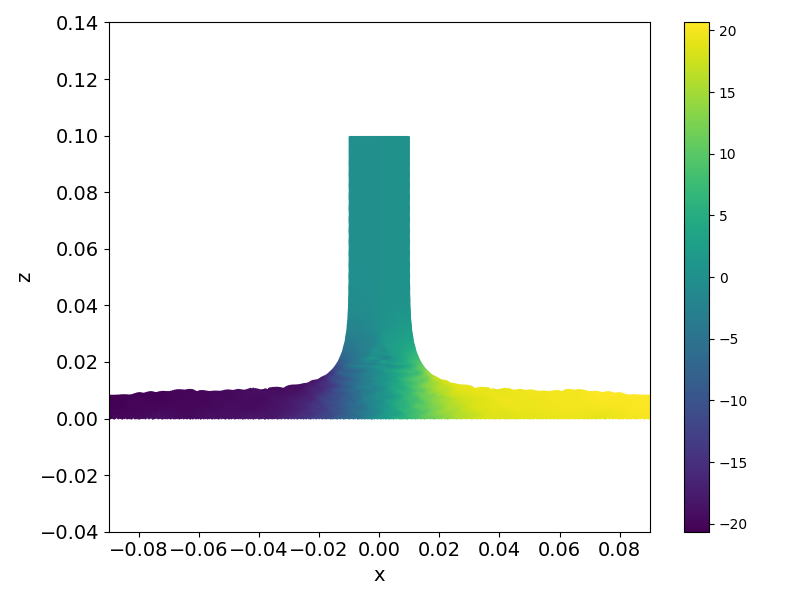}%
}
\subfloat[\; ROM time step = 5000]{%
    \includegraphics[width=0.5\textwidth]{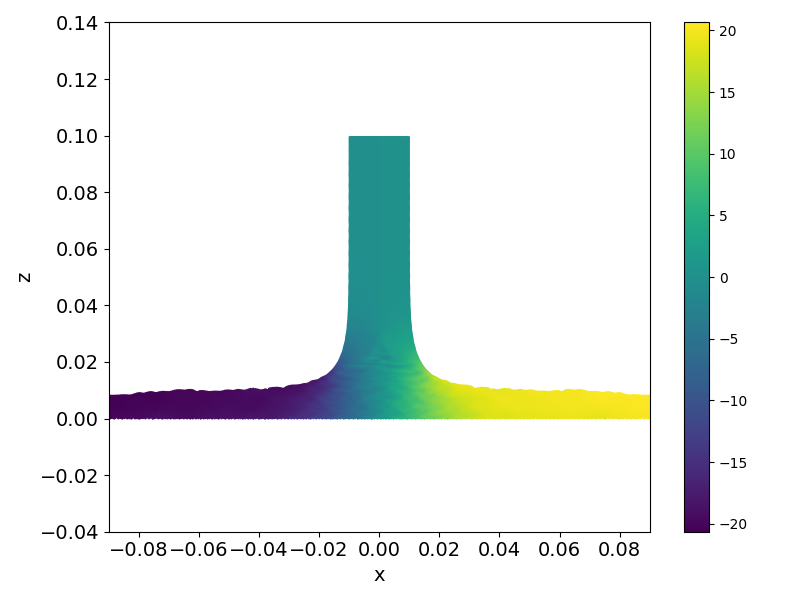}%
}
\caption{2D Impinging Jet: Snap-shot of the particle positions colored by u-velocity field obtained from the original FOM (left; SPH) and the SVD-based reconstruction (right) at three different time instants.}
\label{fig:jetqualitative}
\end{figure}

\begin{figure}
\centering
\captionsetup[subfloat]{margin=10pt, format=hang, singlelinecheck=false, justification=centering}
\subfloat[\; FOM time step = 250]{%
    \includegraphics[width=0.5\textwidth]{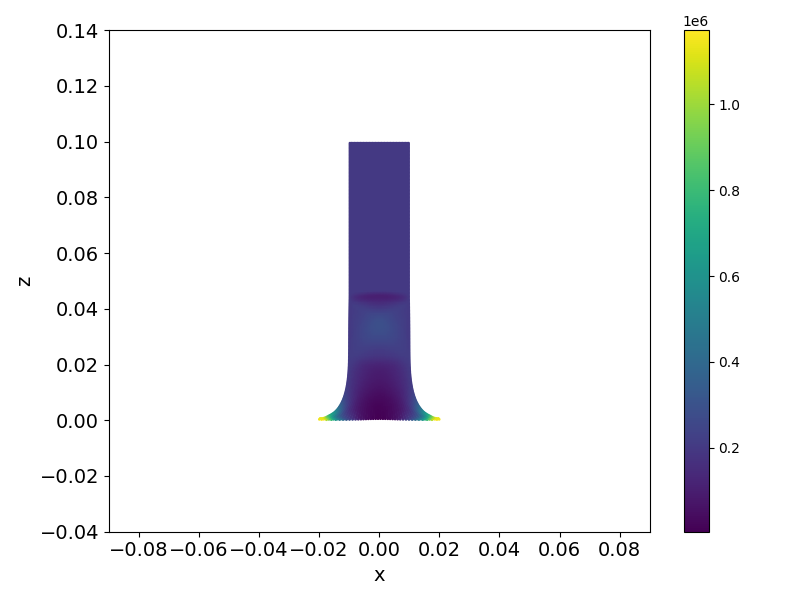}%
}
\subfloat[\; ROM time step = 250]{%
    \includegraphics[width=0.5\textwidth]{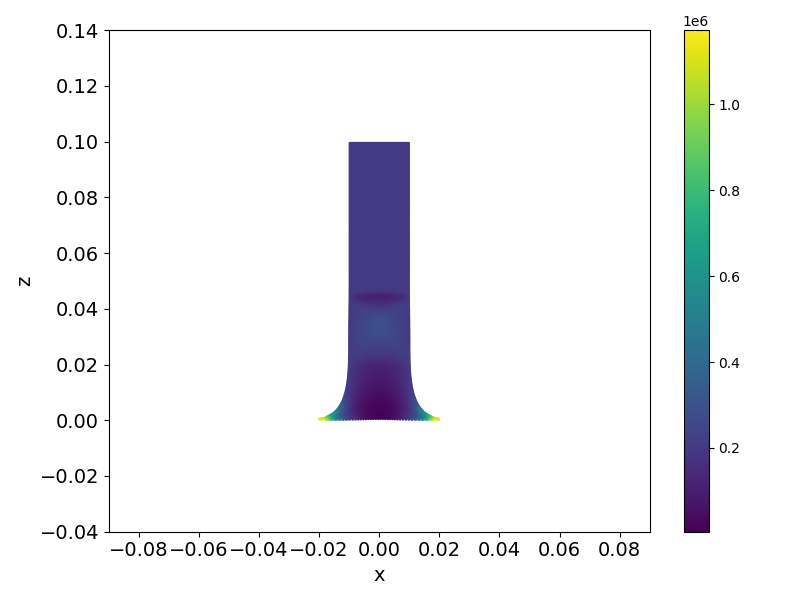}%
}
\hfill
\subfloat[\; FOM time step = 2500]{%
    \includegraphics[width=0.5\textwidth]{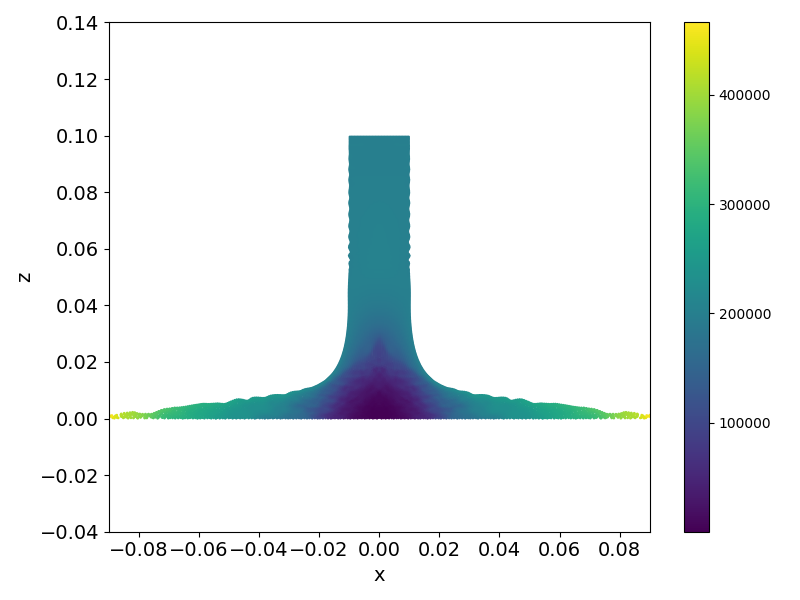}%
}
\subfloat[\; ROM time step = 2500]{%
    \includegraphics[width=0.5\textwidth]{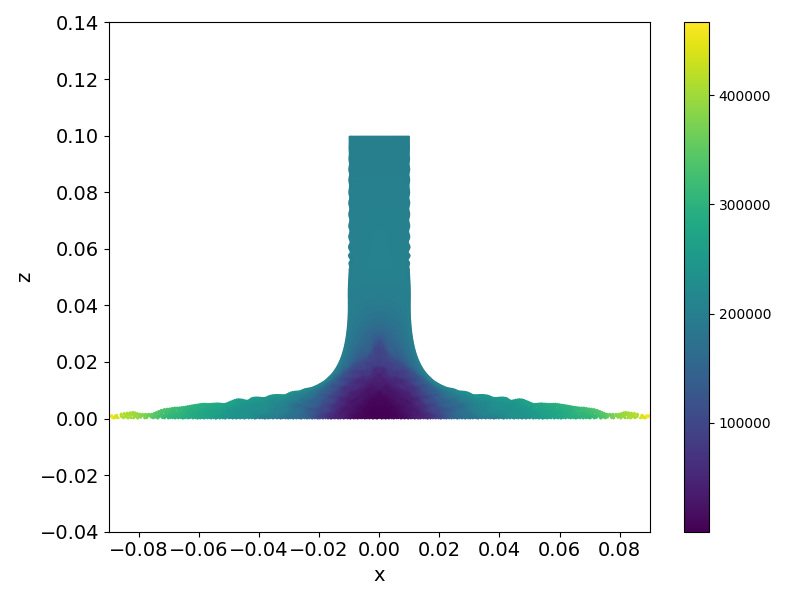}%
}
\hfill
\subfloat[\; FOM time step = 5000]{%
    \includegraphics[width=0.5\textwidth]{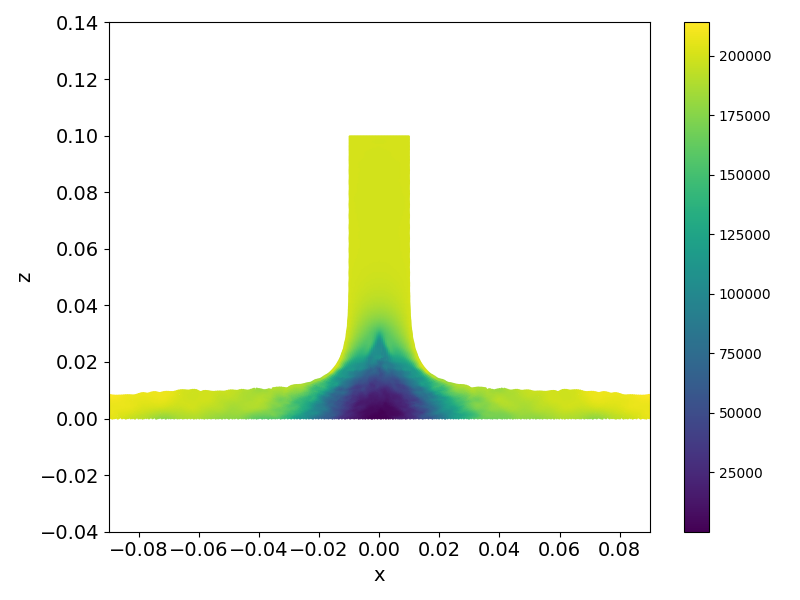}%
}
\subfloat[\; ROM time step = 5000]{%
    \includegraphics[width=0.5\textwidth]{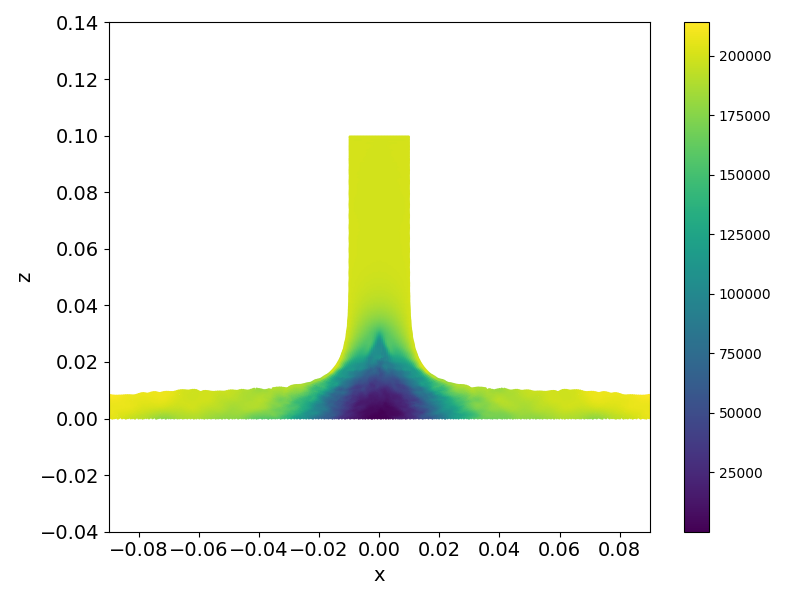}%
}
\caption{2D Impinging Jet: Snap-shot of the particle positions colored by dynamic pressure field obtained from the original FOM (left; SPH) and the SVD-based reconstruction (right) at three different time instants.}
\label{fig:jetqualitative_dynamicPressure}
\end{figure}

Figure \ref{fig:jetqualitative} and \ref{fig:jetqualitative_dynamicPressure} display two qualitative comparison of the reconstructed jet and the full-order SPH data extracted at three different time instants, with $b=250$, for the u-velocity field and dynamic pressure, respectively.

In line with Eqn. \eqref{eq: max reconstruction error}, we build an error matrix $E \in \mathbb{R}^{5\times5000}$, where each row is assigned to one of the five physical fields stored — here in particular the two components of position, the two components of velocity, and the density — and each column corresponds to a particular time step. This matrix's mean and maximum values are indicated by $\varepsilon_{mean}$ and $\varepsilon_{max}$, respectively. Table \ref{table:IJ-ImpStrat} shows the final rank $q_\phi$ of a field $\phi$, the sum of all field ranks $\sum q$, the error measures $\varepsilon_{mean}$ and $\varepsilon_{max}$, as well as the relative computational overhead $t^{SVD}/t^{SPH} \cdot 100\%$ for all utilized imputation strategies. Therein, $t^{SVD}$ and $t^{SPH}$ refer to the measured wall clock time of the ROM construction and the flow solver integration, respectively. The simulations are conducted thrice to underline the credibility of the measured wall clock time, and their mean value is reported. Investigations of this first step are performed for a fixed bunch width of $b=250$ and no windowing was employed. While the different imputation strategies do not significantly affect the accuracy, the associated number of modes and computational effort vary significantly. Considering that the imputation strategies were tested on the same simulation, the number of retained modes directly relates to the compression rate. As expected, the GPOD with 10 iterations achieves the best compression and requires the least number of modes. On the other hand, this example has a CPU overhead that is almost as expensive as the SPH simulation. 
The BMI method is algorithmically very simple and requires only a moderate computational effort in the range of 30\%. However, it still delivers competitive compression rates. The strategy can be further optimized slightly in combination with the hybrid $h_{MI-BMI}$.
Hence, the strategy we employ for the remaining investigations is the hybrid $h_{MI-BMI}$ approach.

\begin{table}[ht!]
\centering
\begin{tabular*}{\textwidth}{@{\extracolsep{\fill}} c c c c c c c c c c}
\hline
    & $q_x$ & $q_z$ & $q_u$ & $q_w$ & $q_\rho$ & $\sum q_\phi$& $\varepsilon_{mean}[\%]$ & $\varepsilon_{max}[\%]$ & $t^{SVD}/t^{SPH} \cdot 100\%$ \\
\hline
MI      & 856 & 254 & 523 & 480 & 160 &2391& 0.071 & 0.377 & 31.34  \\

BMI & 301 & 236 & 593 & 598 & 176 &1904&0.070& 0.235 & 29.49  \\

GPOD(1)  & 663 & 253 & 515 & 472 & 162 &2191&0.081 & 0.373 & 37.64  \\

GPOD(5)  & 472 & 204 & 511 & 466 & 163 &1948&0.077 & 0.376 & 62.89  \\

GPOD(10) & 364 & 179 & 509 & 463 & 163 & 1813& 0.075 & 0.374 & 95.60  \\

$h_{MI-BMI}$ & 301& 236 & 523 & 480 & 160 & 1818& 0.071 & 0.377 & 27.37 \\
\hline

\end{tabular*}
\caption{Reconstruction of the 2D impinging jet results for 5000 time steps using 6 different imputation strategies without windowing: Comparison of mode counts $q_\phi$, mean and maximum reconstruction error $\varepsilon$, as well as relative computational overhead $t^{SVD}/t^{SPH}$ for the mean imputation (MI), block-mean imputation (BMI), three Gappy POD methods and the hybrid mean/block-mean $h_{MI-BMI}$ imputation strategy, from top to bottom, respectively.}
\label{table:IJ-ImpStrat}
\end{table}

In a second step, the windowing strategy is assessed for the $h_{MI-BMI}$. The number of particles included in a given SVD and the number of time steps assigned to the SVD are influencing parameters 
on the compression ratio (cf. Eqn. \ref{eq:crw_Np(i)}) and the CPU effort.
However, since we also store values for inactive particles, the compression ratio (CR) is measured against a Full-Order Full-Storage (FOFS) approach. For each time step ($n_t$), one needs to store one numerical value per field ($n_f$) per active particle ($n_p$), allowing for a hypothetical FOFS approach's storing effort, proportional to approximately $ c \,  \mathrm{n_f} \, \sum_{i = 1}^{i = n_t} \, \mathrm{n_p(i)}$, where $c$ denotes the required size for storing a number, that depends, e.g., on the underlying single- or double precision metric. Table \ref{table:IJ-Wind-CR} presents the results for three different windowing configurations and four different bunch-widths. The compression ratio moderately varies between $7.72 \leq CR \leq 9.03$ and reveals minor benefits for the single window
approach.
The  relative computational overhead $t^{SVD}/t^{SPH} \cdot 100 \%$ is provided in Table \ref{table:IJ-Wind-Time}. In contrast to the compression ratio, the computational overhead varies substantially, between $7.6 \leq t^{SVD}/t^{SPH} \cdot 100 \% \leq 50.0$. The overhead lowers as the number of windows increases for all considered bunch-widths. Additionally, a larger number of windows seems to favor smaller bunch widths, as minimal efforts
are continuously shifted towards smaller bunch-widths when the amount of windows increases. 

\begin{table}[ht!]
\centering
\begin{tabular*}{\textwidth}{@{\extracolsep{\fill}} c c c c c}
\hline
CR & $b=50$ & $b=125$ & $b=250$ & $b=625$ \\
\hline
$N_w=1$ & 7.72 & 7.74 & 7.80 & 7.79\\
$N_w=2$ & 8.78 & 8.88 & 8.96 & 9.03\\
$N_w=4$ & 8.26 & 8.38 & 8.46 & 8.52\\
\hline  
\end{tabular*}
\caption{Reconstruction of the 2D impinging jet results for 5000 time steps using the  $h_{MI-BMI}$ imputation strategy: Compression Ratio (CR) of the Reduced-Order Models measured against a Full-Order Full-Storage approach for different windows $N_w$ and bunch-widths $b$ sizes.}
\label{table:IJ-Wind-CR}
\end{table}

\begin{table}[ht!]
\centering
\begin{tabular*}{\textwidth}{@{\extracolsep{\fill}} c c c c c}
\hline
$t^{SVD}/t^{SPH} \cdot 100 \%$ & $b=50$ & $b=125$ & $b=250$ & $b=625$ \\
\hline
$N_w=1$ & 50.0 & 30.4 & 27.37 & 37.6\\
$N_w=2$ & 17.6 & 13.1 & 14.8 & 26.9\\
$N_w=4$ & 7.60 & 7.60 & 10.0 & 22.1\\
\hline  
\end{tabular*}
\caption{Reconstruction of the 2D impinging jet results for 5000 time steps using the  $h_{MI-BMI}$ imputation strategy: Relative computational overhead $t^{SVD}/t^{SPH} \cdot 100\%$ of the Singular Value Decomposition against the flow solver's simulation time for different windows $N_w$ and bunch-widths $b$ sizes.}
\label{table:IJ-Wind-Time}
\end{table}

\subsection{3D Pelton Runner}
\label{subsec:3d_pelton}
\begin{figure}
\centering
\captionsetup[subfloat]{margin=10pt, format=hang, singlelinecheck=false, justification=centering}
\subfloat[]{%
    \iftoggle{tikzExternal}{
        \tikzsetnextfilename{pelton-sketch-left}
        \begin{tikzpicture}
            \node[anchor=south west,inner sep=0] (image) at (0,0) {\includegraphics[width=0.5\textwidth]{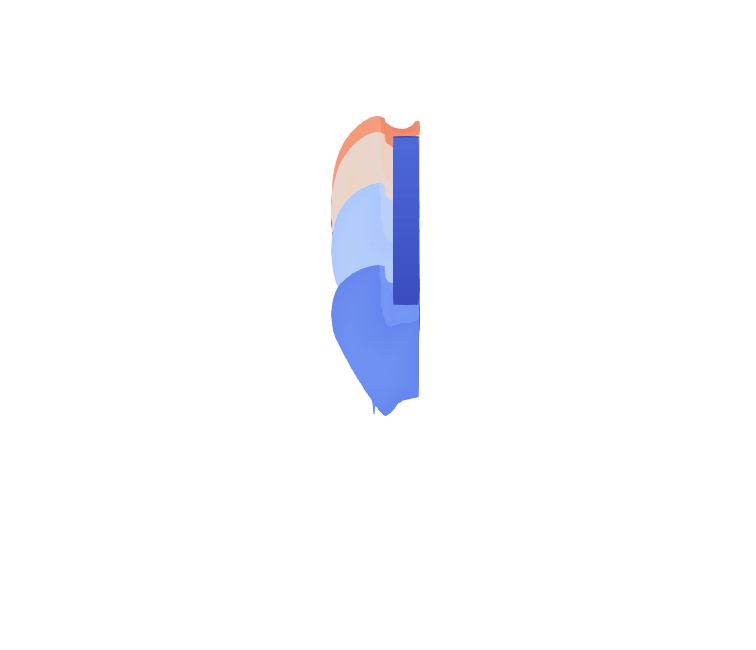}};
            \begin{scope}[x={(image.south east)},y={(image.north west)}]
                \draw[dashed,thick] (0.555,0.2) -- (0.555,0.3);
                \draw[thick] (0.555,0.3) -- (0.555,0.9);      
                \draw[dashed,thick] (0.555,0.9) -- (0.555,1); 
                \draw[thick, ->] (0.1, 0.1) -- (0.1, 0.2) node[above] {$y$};
                \draw[thick, ->] (0.1, 0.1) -- (0.0, 0.1) node[left] {$z$};
            \end{scope}
            \end{tikzpicture}%
        }{
        \includegraphics{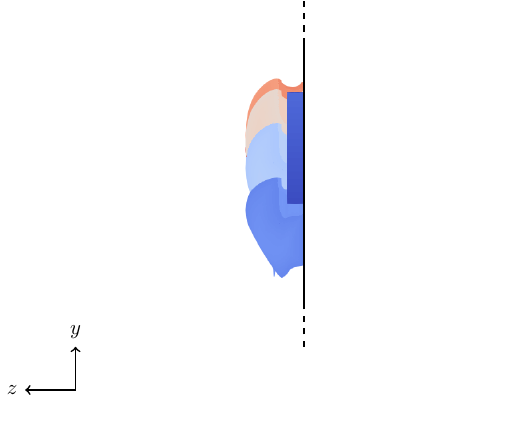} 
    }
}
\subfloat[]{%
    \iftoggle{tikzExternal}{
        \tikzsetnextfilename{pelton-sketch-right}
        \begin{tikzpicture}
            \node[anchor=south west,inner sep=0] (image) at (0,0) {\includegraphics[width=0.5\textwidth]{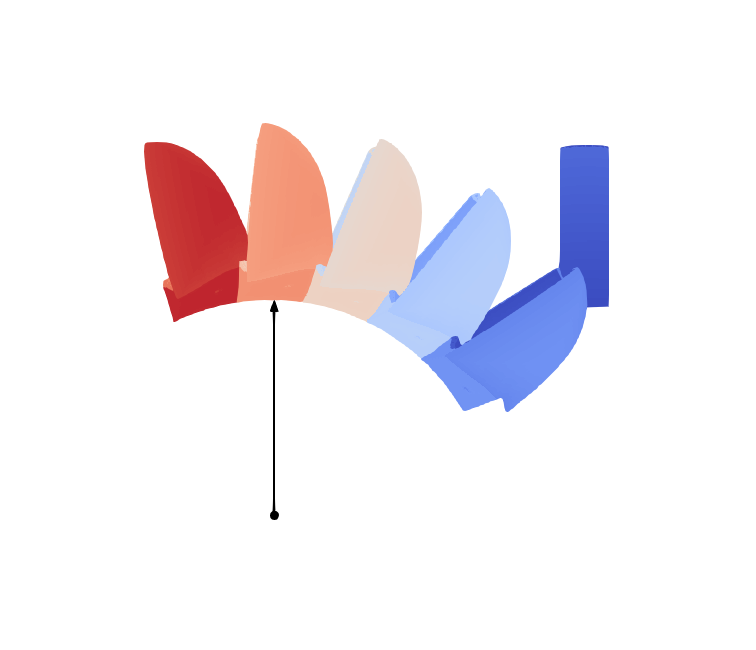}};
            \begin{scope}[x={(image.south east)},y={(image.north west)}]
                \node[black] at (0.51,0.3) {0.128 mm};
                \draw[thick, ->] (0.775, 0.9) -- (0.775, 0.8) node[midway, left] {$Jet \, inlet$};
                \draw[thick, ->] (0.1, 0.1) -- (0.1, 0.2) node[above] {$y$};
                \draw[thick, ->] (0.1, 0.1) -- (0.2, 0.1) node[right] {$x$};
            \end{scope}
            \end{tikzpicture}%
        }{
        \includegraphics{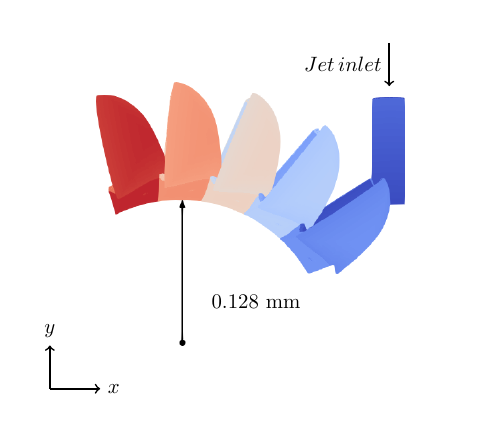}
    }
}

\caption{Schematic of the initial 3D Pelton runner configuration  which consist of the jet and five buckets. The position of the nozzle and the incoming jet are displayed on the right figure (b). The left figure (a) shows the plane of symmetry. }
\label{fig:peltonSchematic}
\end{figure}

The final application examines a Pelton turbine, i.e., a hydraulic impulse machine adapted for high head and low discharge. Its main components are the distributor, the injectors, the runner, and the housing. Mesh-based methods are well-suited to deal with the confined flow in the distributor and injectors. However, the runner and housing are characterized by complex free surface flow, where SPH is an attractive alternative. The simulated configuration adopts a set-up similar to that described in a previous study \cite{andritzoverview}. As illustrated in Fig. \ref{fig:peltonSchematic}, it refers to a runner sector comprising five buckets fed with a single water jet. It employs a symmetry condition in the mid-plane of the runner. 

Following the approach outlined in \cite{andritzoverview}, the simulation includes all possibly wetted surfaces and is carried out in a pure Lagrangian mode, discretizing solely the liquid fluid phase and the solid bucket model, excluding the air phase. The box-shaped physical domain has a height of $0.55$ m, a width of $0.55$ m, and a depth of $0.068$ m. The runner is centered in the x-y plane, and the runner center is positioned at $(0,0,0)$. The buckets are modeled by (solid) particles with prescribed motion. Fluid particles are injected with a speed of $44.30$ m/s, and the initial particle distance reads $\Delta x = 0.001$ m. The speed of sound is assigned to $c = 443.00$ m/s, i.e., $v_{in}/c = 0.1$, and the influence of gravity is neglected. The simulation is performed over $N_s= 30 000$ time steps and terminates when all buckets have interacted with the jet. Based on the results obtained for the previous test case in Sec. \ref{sec:jet}, this application is simulated using the hybrid mean/block-mean imputation $h_{MI-BMI}$  using $N_w=15$ windows and a bunch width of $b=250$.

Regarding SVD, the case shares many properties with the example discussed in Sec. \ref{sec:jet}, particularly the presence of input and output boundaries where particles can enter and exit the domain. However, unlike the previous scenario, no fully developed state is reached where the number of active liquid particles remains virtually constant, but the total number of simultaneously active particles at the end of the simulation grows to over 620 000. As a result, the reduced system exhibits a fully unsteady behavior. The 3D test case is also significantly more computationally intensive and is therefore computed using a thread-parallel approach on one GPU.

\begin{figure}
\centering
\captionsetup[subfloat]{margin=10pt, format=hang, singlelinecheck=false, justification=centering}
\subfloat[\; FOM time step = 6000]{%
    \includegraphics[width=0.5\textwidth]{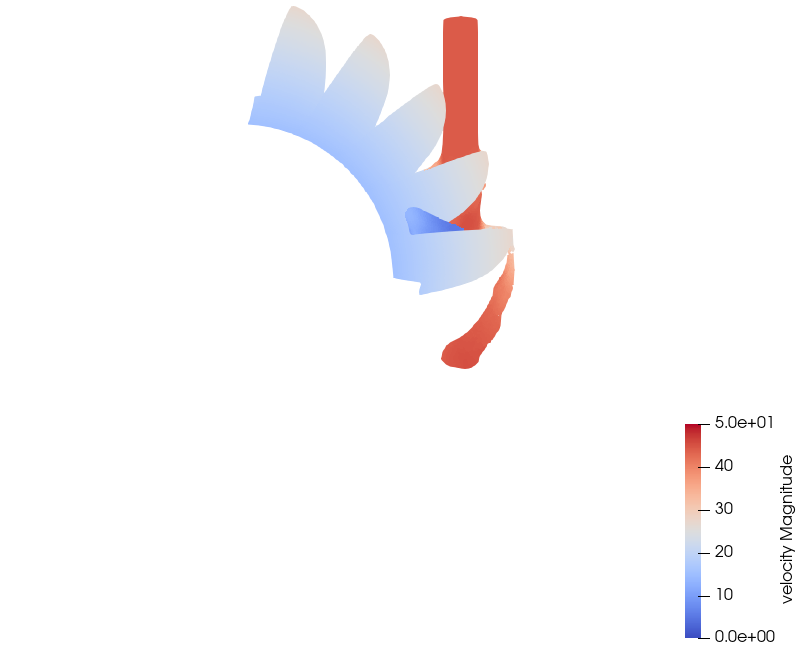}%
}
\subfloat[\; ROM time step = 6000]{%
    \includegraphics[width=0.5\textwidth]{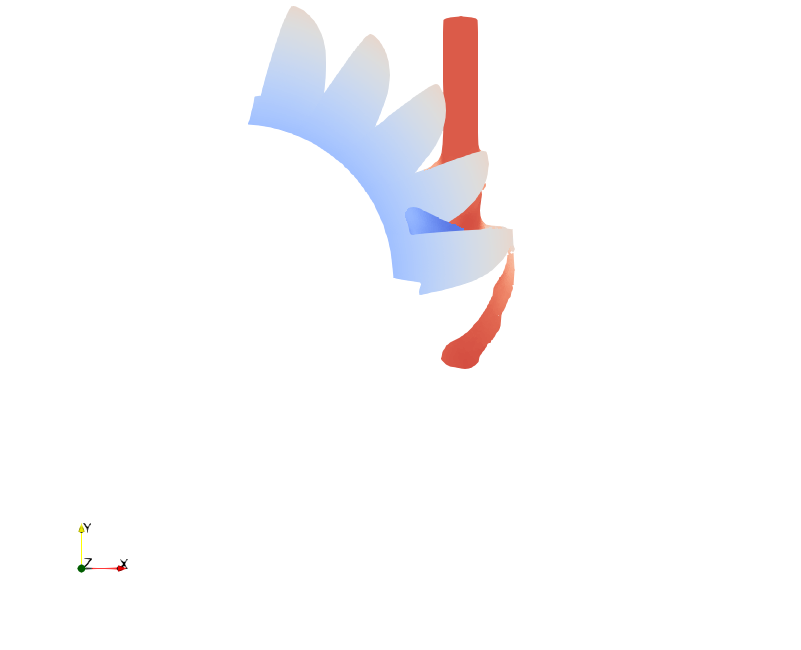}%
}
\hfill
\subfloat[\; FOM time step = 16000]{%
    \includegraphics[width=0.5\textwidth]{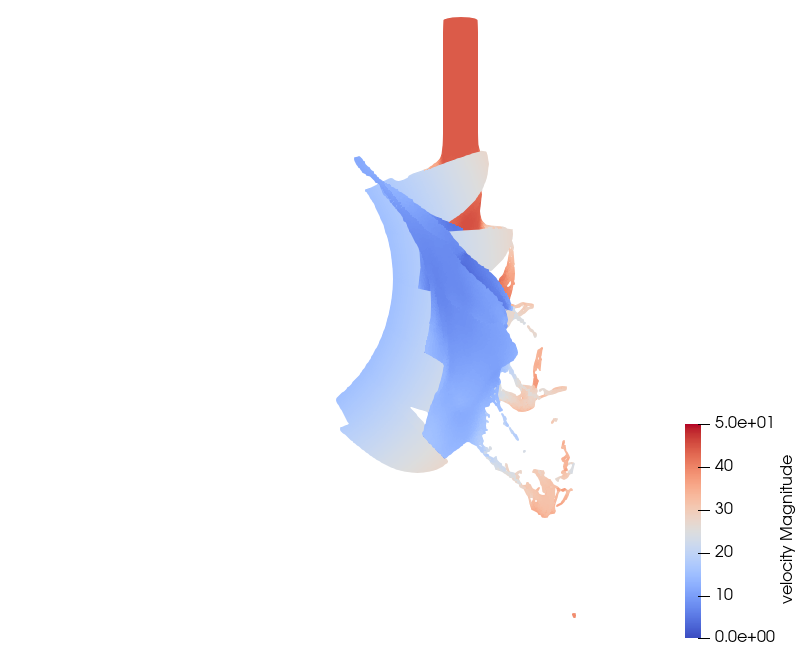}%
}
\subfloat[\; ROM time step = 16000]{%
    \includegraphics[width=0.5\textwidth]{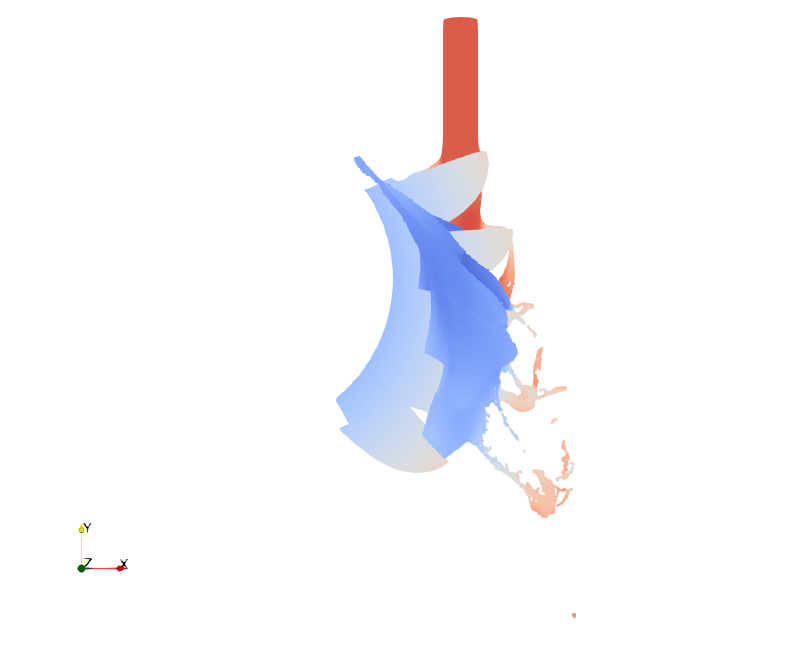}%
}
\hfill
\subfloat[\; FOM time step = 26000]{%
    \includegraphics[width=0.5\textwidth]{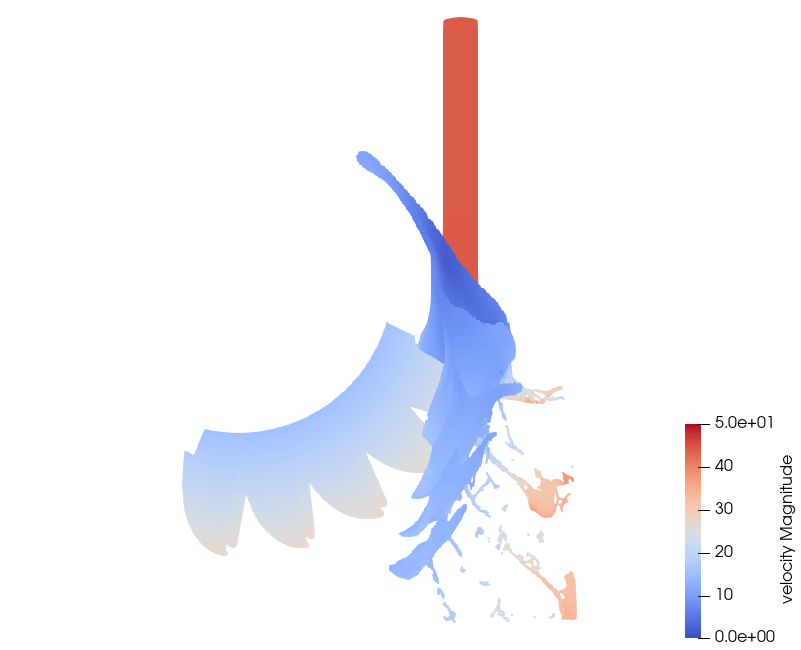}%
}
\subfloat[\; ROM time step = 26000]{%
    \includegraphics[width=0.5\textwidth]{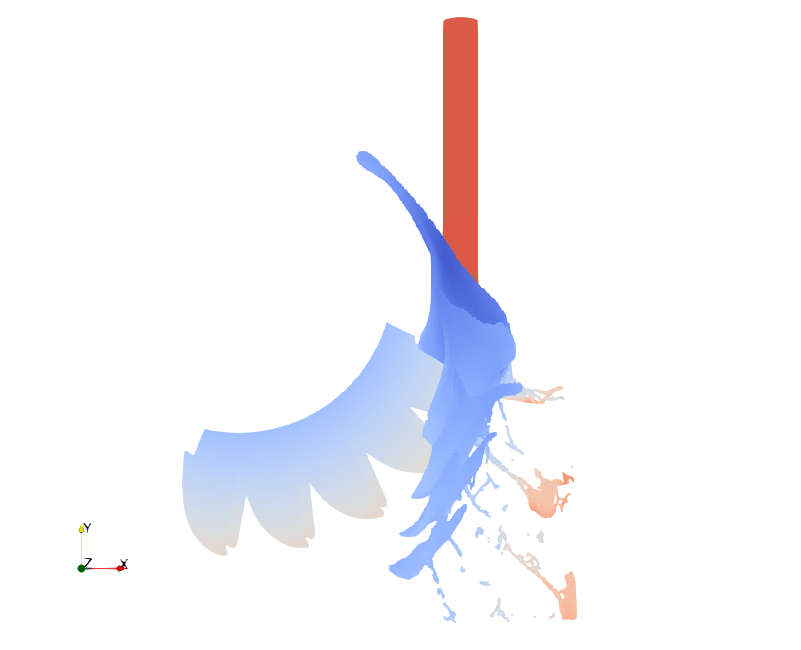}%
}
\caption{Snap-shots of the particle positions colored by velocity magnitude obtained from the original FOM (left; SPH) and the SVD-based reconstruction (right) of the 3D Pelton runner flow at three different time instants.}
\label{fig:peltonqualitative}
\end{figure}

Figure \ref{fig:peltonqualitative} displays a visual comparison of the full-order SPH data (left) and the reconstructed jet (right) extracted at three different exemplary time instants. This qualitative comparison shows an excellent agreement. Note that only the fluid particles are reconstructed from the SVD and no penetration by the reconstructed fluid particles into the non-reconstructed/prescribed bucket region is observed.

\begin{figure}
\centering
\includegraphics[width=\textwidth]{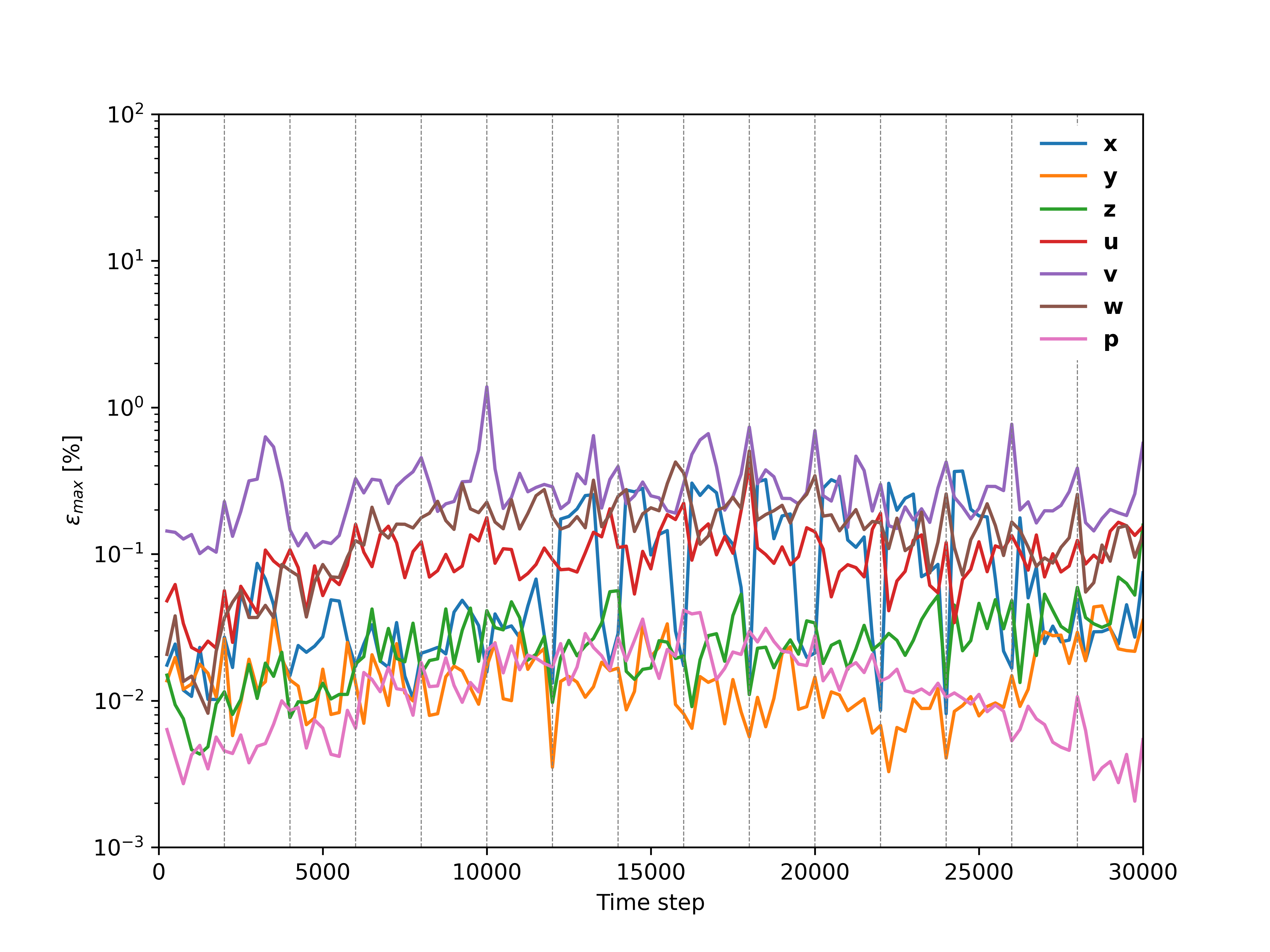}%
\caption{3D Pelton runner: Temporal evolution of the maximum reconstruction error of fluid field data in percent obtained with adaptive truncation, $h_{MI-BMI}$ imputation, employing a bunch width of $b=250$ and $N_w=15$ (equally sized) time windows/SVDs indicated by the dashed vertical lines. }
\label{fig:peltonerror}
\end{figure}

The quality of the reconstruction by the ROM is assessed according to Eqn. (\ref{eq: max reconstruction error}) and shown in Fig.  \ref{fig:peltonerror}. Attention is confined to the error obtained for reconstructing the pressure, velocity, and position fields. Since the pressure and the density are related according to the algebraic equation of state (\ref{eq:Tait}), storing both fields in the snapshot matrix is strictly speaking not necessary. 
Figure \ref{fig:peltonerror} reveals an excellent agreement between FOM and ROM, as expected from the visual comparison of the reconstructed and the simulated water body in Fig \ref{fig:peltonqualitative}. With only a few exceptions, maximum relative error magnitudes inside the fluid remain below one per cent.

With regard to pressure, however, another detail comes to the fore, which concerns the solid pressures and their reconstruction. The  solid pressures in SPH are usually determined from the fluid properties using boundary-specific formulae and the algebraic equation of state (\ref{eq:Tait}). The  pressure levels can quickly be in the range of several million Pascals in wetted regions,   
but fall back to zero Pascal in (temporarily) unwetted solid areas. The reconstruction of the pressure field by the SVD is therefore more challenging for the solid than in the fluid particles. In essence, the snapshot matrix for the pressure of a wall particle resembles a random matrix with many zero entries, and even when maintaining a large number of modes, it is difficult to maintain the same accuracy as for the fluid particles. For this reason, the adaptive truncation approach would lead to storing almost all the available modes. Therefore, we decided to use a fixed truncation rank for the pressure field, which allows assigning the degree of compression but not the resulting accuracy. The accuracy of the reconstruction is evaluated based on the maximum and mean error, as well as for the integral quantities force and moment. Alternatively, analogous to the procedure for the FOM SPH method, the reconstructed fluid pressure can be determined in a post-processing step, which appears to be much simpler.
Nonetheless,  we obtain the ROM pressure field for the solid from the SVD of the FOM solid pressures in the present study, to evaluate the forces and torques on the turbine. Mind that only a subset of solid particles interacts with water during the simulation; for those that do not interact, the pressure value remains zero. Solid particles are included in their corresponding bunch matrix only if they have interacted with fluid particles at least once. This avoids storing irrelevant data. Solid particles are then handled similarly to newly injected fluid particles, with the distinction that imputation is unnecessary since no data is missing.

\begin{figure}
\centering
\includegraphics[width=0.5\textwidth]{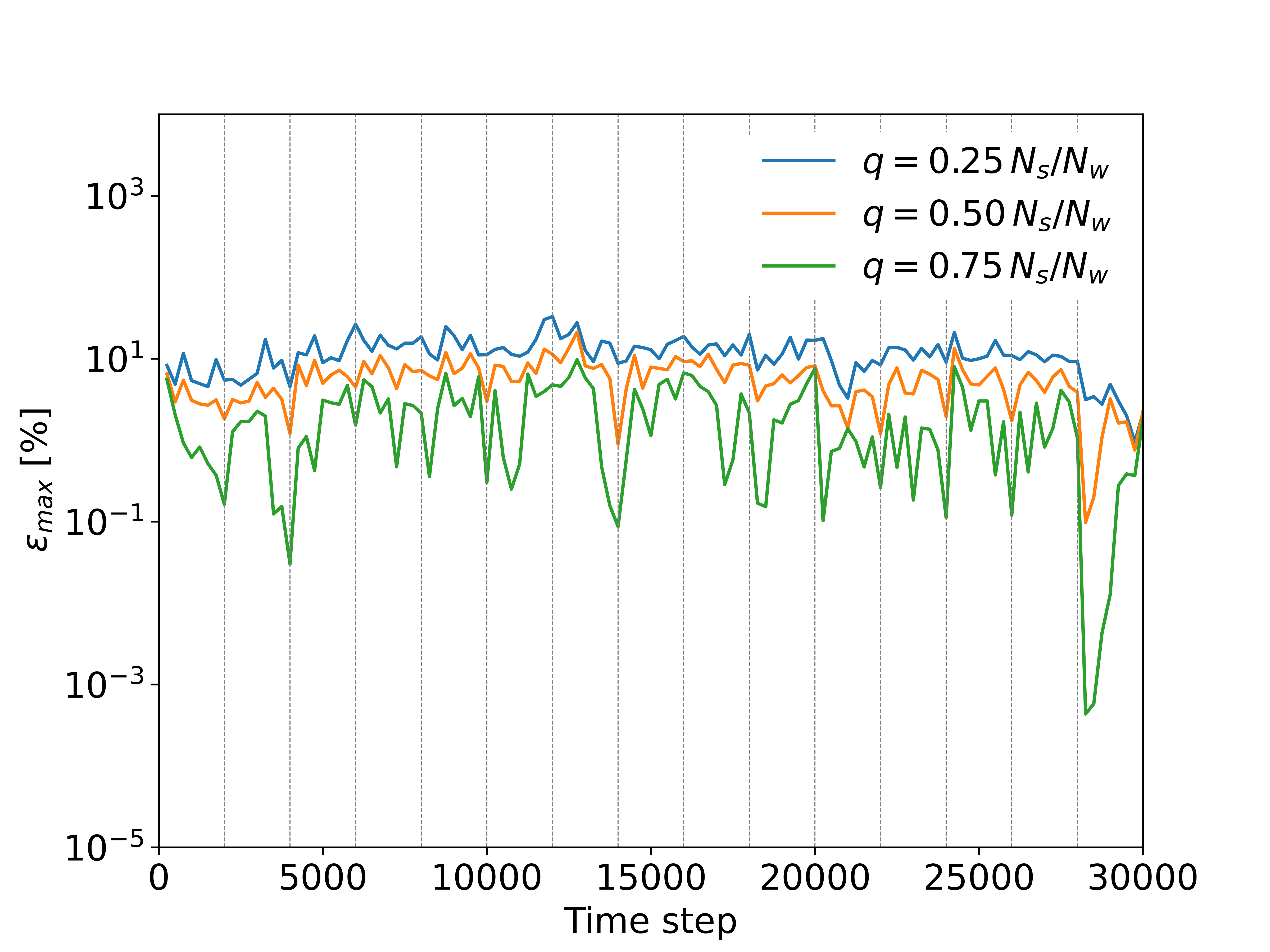}%
\hfill
\includegraphics[width=0.5\textwidth]{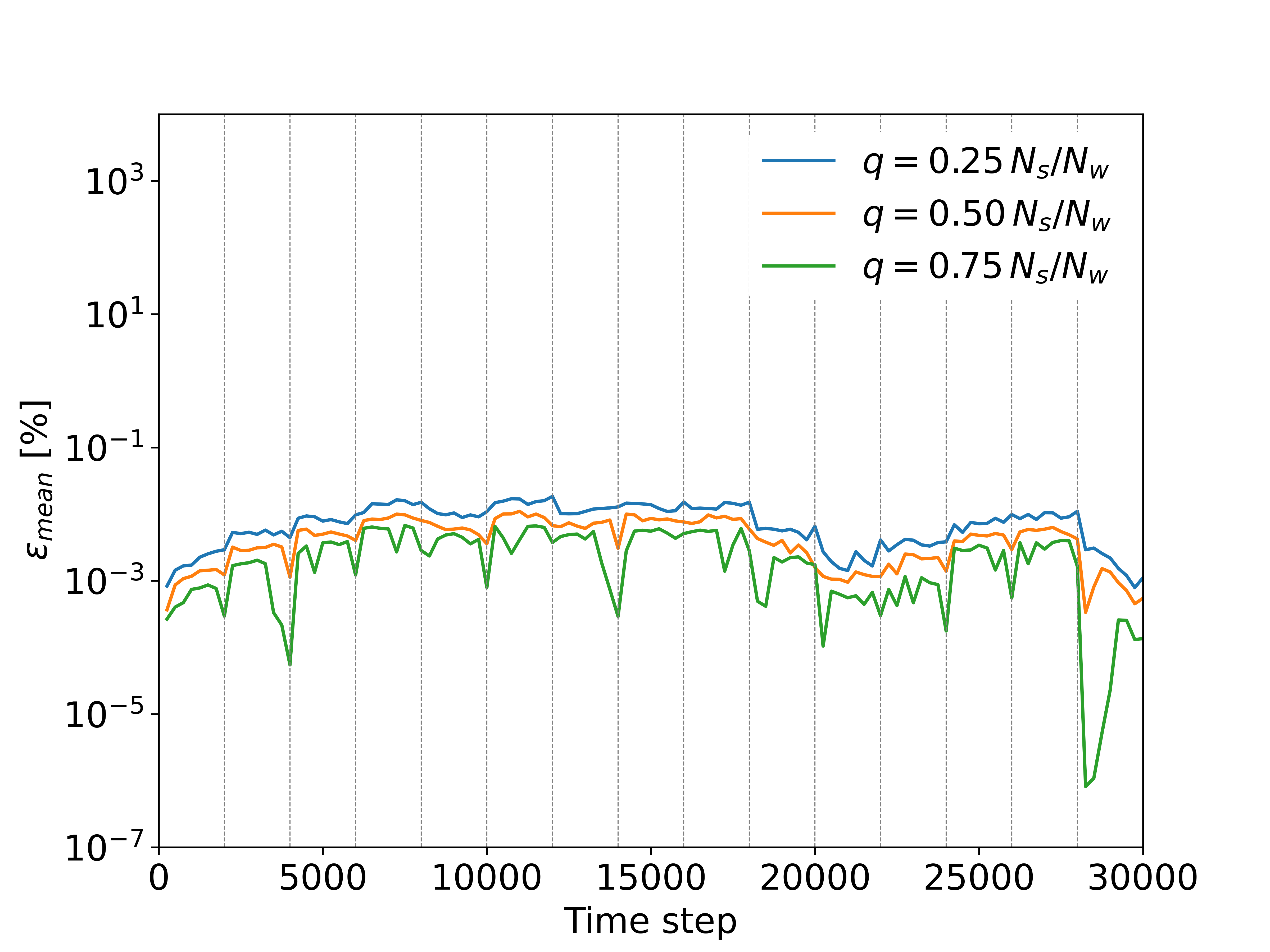}
\caption{3D Pelton runner: Temporal evolution of the maximum (left) and mean (right) reconstruction error of the solid particle pressures in percent obtained with three fixed truncation ranks $q$, employing a bunch width of $b=250$ and $N_w=15$ (equally sized) time windows/SVDs.}
\label{fig:peltonwallerror}
\end{figure}

\begin{figure}
\centering
\includegraphics[width=\textwidth]{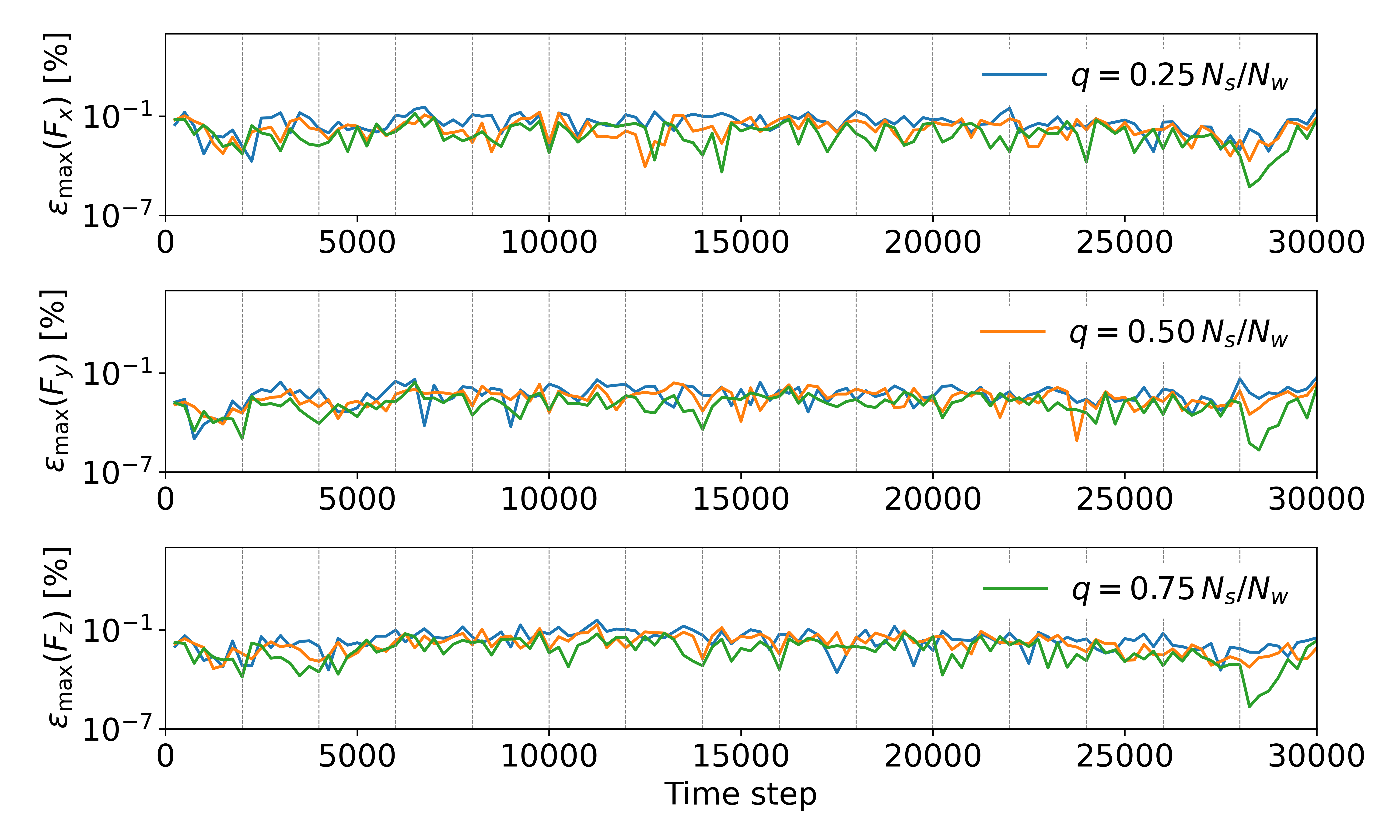}%
\caption{3D Pelton runner: Temporal evolution of the reconstruction error of the force in percent obtained with three fixed truncation ranks $q$, employing a bunch width of $b=250$ and $N_w=15$  (equally sized) time windows/SVDs.}
\label{fig:force}
\end{figure}

\begin{figure}
\centering
\includegraphics[width=\textwidth]{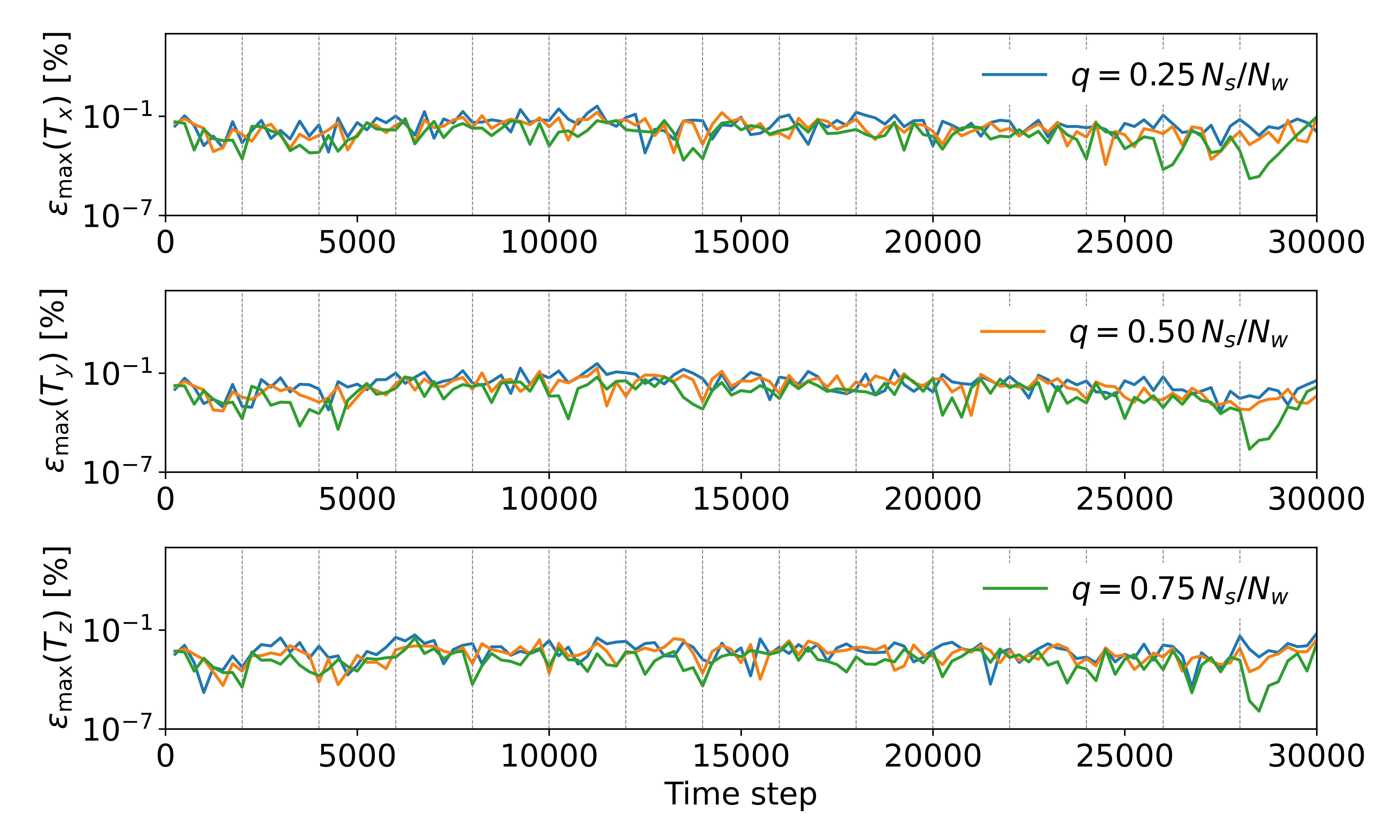}%
\caption{3D Pelton runner: Temporal evolution of the reconstruction error of the torque in percent obtained with three fixed truncation ranks $q$, employing a bunch width of $b=250$ and $N_w=15$  (equally sized) time windows/SVDs indicated by the dashed vertical lines.}
\label{fig:torque}
\end{figure}

Figure \ref{fig:peltonwallerror} depicts the maximum (left) and mean (right) reconstruction errors obtained for the solid pressure using three different (fixed) truncation ranks in the range of 25\%-75\% of the maximum possible rank. 
The irregularity of the snapshot matrix of the solid pressures mainly concerns the maximum error. The mean error, however, remains relatively small, which suggests that the rank truncation primarily impairs the local accuracy of the wall pressure. The latter is confirmed by evaluating the reconstruction of integral quantities, i.e., forces and torques, as shown in Figs. \ref{fig:force}-\ref{fig:torque}, revealing an error is in the order of per mille.

The attained compression ratio (CR) for the Pelton runner case is evaluated against the full-order full-storage effort in Table  \ref{table:pelton-CR}. This assessment is performed separately for the fluid and solid particles and subsequently collectively. The tabulated values for the fluid data are of the same order of magnitude as those in the 2D cases, cf. for example Table \ref{table:IJ-Wind-CR}, and the experienced CR reduction can be attributed the ROM of the solid pressure. Mind that post-processing the ROM for the solid pressures from the ROM fluid data in line with the employed boundary condition could represent a more efficient strategy. The computational overhead is $t^{SVD} / t^{SPH} \approx 7.26\%$ for the considered simulation which  agrees with best practice results obtained for the 2D impinging jet case depicted in Table \ref{table:IJ-Wind-Time}. 

\begin{table}[ht!]
\centering
\begin{tabular*}{\textwidth}{@{\extracolsep{\fill}} c c c c}
\hline
 & Fluid & Solid & Total  \\
\hline
CR & 9.82 & 3.42 & 7.55 \\
\hline  
\end{tabular*}
\caption{Compression ratio (CR) of the reduced-order model obtained for the 3D Pelton runner case normalized with the full-order full-storage effort. Result were obtained with adaptive truncation and $h_{MI-BMI}$ imputation for the fluid, fixed truncation $q=0.25N_s/N_w$ for the solid, employing a bunch width of $b=250$ and $N_w=15$ equally sized time windowed SVDs.}
\label{table:pelton-CR}
\end{table} 

\section{Conclusions}
\label{sec:conclusion}
An incremental SVD-based strategy has been developed to reduce transient SPH results. The approach is based on an existing incremental SVD algorithm that was originally devised for mesh-based regular snapshot data matrices, and is further developed here for 
processing highly irregular data matrices associated with Lagrangian particle simulation methods. 

The system of equations from the Full-Order Model (FOM) does not influence the development of the Reduced-Order Model (ROM) itself 
and therefore allows its application to other numerical methods characterized by irregular spatio/temporal data.
The method has been embedded into a parallel, industrialized 
in-house SPH software ({\it Asphodel}) and was successfully 
validated for a sequence of relevant 2D and 3D test cases.
The suggested approach reduces storage requirements by $\mathcal{O} (90 \%)$ with good agreement between ROM and FOM data while maintaining reasonable computational overhead in the order of $\mathcal{O}(10\%)$. For the final water turbine application, the temporal evolution of force and torque values returned by the ROM is in excellent agreement with FOM recordings. 

Various aspects important for accuracy, computational effort, and memory requirements were addressed, such as an adaptive rank truncation approach, the imputation strategy for gaps in the snapshot matrix caused by temporarily inactive particles and the sequencing of the data history into temporal windows as well as the bunching of the SVD updates.
To ensure accuracy, an appropriate threshold was identified from the decrease in singular values, where the SVD was truncated when the values decreased by five orders of magnitude compared to the dominant singular value.
  Different imputation strategies have been tested for missing data in the data matrices, and a hybrid mean/block-mean method has been found to be the most effective approach that outperforms the gappy POD method. The suggested windowing strategy supports the effective management of scenarios with variable dynamics. 
  
\section{CRediT Authorship Contribution Statement}
\textbf{Eduardo Di Costanzo}: Conceptualization, Methodology, Software, Validation, Formal analysis, Investigation, Writing - original draft, Visualization.
\textbf{Niklas K{\"u}hl}: Conceptualization, Software, Resources, Writing - review \& editing, Supervision, Project administration.
\textbf{Jean-Christophe Marongiu}: Resources, Writing - review, Project administration, Funding acquisition.
\textbf{Thomas Rung}: Conceptualization, Resources, Writing - review \& editing, Supervision, Project administration, Funding acquisition.

\section{Data Availability Statement}
Data sharing not applicable, since no new data was generated, but only standard 2D testcases were used to scrutinze the method. For the final 3D testcase, the employed geometry is confidential, but the method essentially works for any irregular data matrix.

\section{Acknowledgments}
The first author was supported by the State Secretariat for Education, Research and Innovation (SERI) of Switzerland and Co-Funded by the European Union. Views and opinions expressed are however those of the author(s) only and do not necessarily reflect those of the European Union or European Research Executive Agency (REA). Neither the European Union nor the granting authority can be held responsible for them.

\end{document}